\documentclass[useAMS,usenatbib]{mn2e}

\usepackage{graphicx}

\title[Dynamical and chemical effects of FR~II radio sources on the intra-cluster 
medium]{Dynamical and chemical effects of FR~II radio sources on the intra-cluster 
medium}
\author[M. Huarte-Espinosa et al.]{M. 
Huarte-Espinosa,$^{1}$\thanks{E-mail: mh475@mrao.cam.ac.uk} 
M. Krause,$^{1,2,3}$ P. Alexander,$^{1}$ and C. R. Kaiser$^{4}$\\ 
$^1$Astrophysics Group, Cavendish Laboratory, 19~J.~J.~Thomson~Ave., Cambridge CB3 0HE, UK \\
$^2$Max-Planck-Institut f\"ur Extraterrestrische Physik, Giessenbachstrasse, 85748 Garching, Germany \\
$^3$Universit\"atssternwarte M\"unchen, Scheinerstr.~1, 81679 M\"unchen, Germany \\
$^4$School of Physics and Astronomy, University of Southampton, Southampton SO17 1BJ, UK}

\begin{document}

\date{Received \today}

\pagerange{\pageref{firstpage}--\pageref{lastpage}} \pubyear{2007}

\maketitle

\label{firstpage}

\begin{abstract}
We investigate the effects of intermittent strong jets 
from an Active Galactic Nuclei (AGN) of a massive galaxy in the 
core of a cool core galaxy cluster, on the 
dynamics and metal distribution of the intra-cluster medium (ICM). 
We use a simple model for the metal distribution within the host galaxy
which includes metal injection via star formation.
We carry out 2D axisymmetric hydrodynamic simulations of these systems. After 
having established a cooling flow, two light jets are injected 
in opposite directions with a range of (intermittent) active phases. We follow the time
evolution of the system from the jets' active phases up to \hbox{3\,Gyr}. 
The general metallicity evolution for all our simulations is very similar on large-scales.
The convective flows driven by the jets advect gas and metals from the central galaxy to 
distances beyond \hbox{1.5\,Mpc} within the cluster. Intermittent jets 
are able to distribute metals to grater radii. The metal injection has 
effects on the general metal abundances present in the ICM,
the faster the metal replenishment time the higher the metal abundance.
Although metallicity gradients in the very central regions of cool core clusters
are likely to be shaped by less-energetic phenomena, we find evidence in our simulations 
for metallicity gradients similar to those observed out \hbox{to~$\sim$\,400\,kpc} radius. 
The metal distribution details of the central galaxy have no effects on the dynamical
evolution of the ICM metals.
\end{abstract}


\section{Introduction}
\label{intro}

The abundance and distribution of heavy elements (metals) in 
the intra-cluster medium 
(ICM) holds information about the history of galaxy clusters 
(clusters from now on). Metals formed as a result of star formation in the cluster
galaxies must be transported into the ICM and a variety of mechanisms are possible, 
for example galactic-scale winds, or outflows driven by Active Galactic Nuclei (AGN).
X-ray spectroscopic observations provide a direct prove of metallicity gradients in 
the ICM. Many clusters show an approximately flat metallicity distribution, of about 
a third of the solar ($Z_{\odot}$), out to megaparsec distances from the
cluster core  \citep{degrandi01}. Moreover, the metallicity profiles of cool core 
intermediate redshift (\hbox{$\sim$\,0.1\,--\,0.35}) clusters, 
particularly those that show no evidence of shocks associated with mergers, 
have an abundance excess of the order \hbox{0.45\,--\,0.55\,Z$_{\odot}$} at their cores 
\citep{degrandi01,baldi07,leccardi08}. High resolution Chandra X-ray observations 
of the core of Hydra~A, show that metals in the central regions were formed by 
different stellar populations \citep{david01}, with a metal replenishment  
time of the order of \hbox{60\,M$_{\odot}$\,yr$^{-1}$} over a period of 1\,Gyr. 
Similar results are found for other clusters as well \citep{molendi04}.

Over recent years there has been a growing consensus that radio galaxies have
a significant impact on the cluster environment (see \citealt{mcnamara07} 
for a recent review). Jets from radio-loud AGN provide an ideal mechanism for coupling
the energy output from the AGN to the surrounding gas, both via heating by the bow shock
(\citealt{binney95,kaiser99}; Reynolds, Heinz \& Begelman 2001; \citealt{alexander02,VLJI}), 
but also by driving gas from the core
during both the active phase of the jet (e.g., Alexander 2002), but also as the buoyant
remnant -- bubbles -- of the radio-source cocoons rise though the cluster gas 
(e.g., \citealt{churazov01,blanton01,reynolds01}). The effects can be very long lasting
as a large-scale convective flow can be established by the rising bubbles 
\citep{basson}. These flows have important effects on the entropy and energy of the
central ICM (see e.g., \citealt{kaiser05} and references there in), and can transport
metals from the core to radii of order a megaparsec within the cluster 
(\citealt{renzini04}; Heath, Krause \& Alexander 2007). Moreover, the bubbles are inferred to produce hydrodynamic
turbulence in the ICM \citep{rebusco06}. These processes may well be linked to the 
cluster-wide feedback with infall of cooling gas triggering the AGN activity 
(e.g., \citealt{binney95,kaiser03,shabala08}). The observed lack of evolution of the ICM 
metallicity out to redshift of order~1 \citep{carilli01,mushotzky97,leccardi08},
suggests that metal advection took place at or before a redshift of unity.
Furthermore, the metallicity gradients in the ICM also suggest that the metals were 
expelled from the dominant central cluster galaxy rather than from ram-pressure 
stripping of cluster galaxies in general \citep{renzini04}. Supernovae do not provide 
sufficient energy for that process and an attractive alternative are powerful radio 
galaxies \citep{heath07}.

In this paper we examine the advection of metals driven by radio-loud AGN. We 
investigate numerically the effects 
that the powerful jets and continuous metal formation, of a massive galaxy in the core 
of a cool core cluster, have on the hydrodynamics and chemistry of the ICM. 
We explore the consequences of varying the jet power, the galaxy metal 
distribution profile and the metal formation rates. Our implemented cluster
and galaxy models are described in Section~2 along with our numerical setup. 
We describe and discuss our results in Section~3 and give 
our conclusions in Section~4. In this paper we extend the work of \citet{heath07}. 
In particular, we use an improved numerical scheme (see below) which enables us to 
model intermittent jets as well as injection of metals from star formation.
 
\section[]{The simulations}
\label{simul}

\subsection[]{Numerical setup and initial conditions}
\label{init}

We solve the equations of hydrodynamics using the Flash code 
\citep{fryxell00} assuming axisymmetry in a 
2D sphere. Given this geometry we use spherical polar 
coordinates (r,$\theta$). Flash is a well
tested \citep{calder02} parallel modular hydrodynamics code with an adaptive mesh. 
It solves the Riemann problem using a Piecewise-Parabolic Method able 
to work well with contact discontinuities, making it well-suited to our current
problem.

To implement the galaxy cluster we use a computational domain 
with ranges \hbox{r\,$\in$\,[10,3000]\,kpc}, with outflow
and hydrostatic boundary conditions at the inner and outer radii respectively, and 
\hbox{$\theta$\,$\in$\,[0,$\pi$]\,rad} both with reflective boundary conditions 
to establish the rotational symmetry of the system. Our initial grid
setup consisted of \hbox{512x16\,zones} which were adaptively refined by a half,
attaining a maximum resolution of 2.92\,kpc in the radial direction and 
1.27\,kpc in the angular one. Each simulation took 5 hours (on average) to run on 
20 processors of the CamGrid\footnote{www.escience.cam.ac.uk/projects/camgrid/} 
cluster of the University of 
Cambridge, and the production runs were made at the Darwin\footnote{ 
http://www.hpc.cam.ac.uk/darwin.html} cluster of the University. 

The initial conditions of our ICM gas were simulated as a monoatomic 
hydrogen plasma with an ideal gas equation 
of state ($\gamma=5/3$), an initial constant temperature of \hbox{10$^7$\,K} and a King density 
profile \citep{king72}:
 
\begin{equation}
\rho_{g}(r) = \frac{\rho_{0}}{(1+(r/a_{0})^2)^{3\beta/2}}.
\label{icm}
\end{equation}
We set the cluster's core density, $\rho_0$, and radius, $a_{0}$, 
to be \hbox{10$^{-23}$\,kg\,m$^{-3}$} and 100\,kpc respectively, and take 
\hbox{$\beta$\,$=$\,2$/$3}. For this initial gas distribution to be in hydrostatic
equilibrium, the gravitational acceleration of the dark matter potential well must 
satisfy: 

\begin{equation}  
g_{r} =  -\frac{    3 \beta k_{B} T }{  
	\mu m_{H} a_{0}^2 } \frac{r}{( 1 + (r/a_{0})^2) },
\end{equation}  
where $\mu$\,$=$\,1$/$2 and $m_{H}$ is the proton mass. We neglect the self-gravity of the 
cluster gas and take this as a fixed background gravitational field.

We include an energy loss term, $dE/dt$, to account for the ICM radiative 
cooling due to Bremsstrahlung:

\begin{equation}  
dE/dt = - 1.722 \times 10^{-40} \left( \frac{n}{m^{-3}} \right)^2
	\sqrt{\left( \frac{T}{K} \right) } \, W\,m^{-3}.
\end{equation}  

Our simulations consist of three phases: (1) The cooling flow phase, 
in which we let the initial ICM to evolve subject to Bremsstrahlung radiation 
and gravity only, until it reaches a quasi-steady state. The system then 
develops a nearly spherically symmetric inflow. 
(2) The radio galaxy active phase(s), when we use 
the \hbox{$\theta$\,$\notin$\,[0.5,$\pi-$0.5]\,rad} region of the inner radial 
boundary (with \hbox{r\,=\,10\,kpc}) to inject two light jets into the cluster 
with a velocity of a third of the speed of light (Lorentz \hbox{factor\,$=$\,~1.048}). 
We set the density of the jets to be~10$^{-3}$ times smaller than the 
initial central ICM density,
and the pressure of the jets was adjusted to that of the cooled central ICM of 
\hbox{2.2\,$\times$\,10$^{-17}$\,N\,m$^{-2}$}. This simulates a powerful FR~II 
\citep{FR} radio source with a jet power of \hbox{8.8\,$\times$\,10$^{39}$\,Watts}. 
(3) The remnant phase, during which the ICM, the metals (see the following Section) and 
the convective flows left by the 
jets evolve passively, under the influence of gravity and Bremsstrahlung
radiation, for 3\,Gyr. Our different simulations and parameters are summarised in Table~1.

\begin{table*}
 \centering
 \begin{minipage}{120mm}
  \caption{Simulations and parameters. The names of the simulations contain the information about
the time between the jet outburst (interludes, I), the metal injection replacement time (R) and
a letter P when the central galaxy metal distribution is implemented with a Plumer profile.}
  \begin{tabular}{@{}lccccc@{}}
  \hline
   Simulation 	&Number    &Duration  of   & Time between    &Injection re-   &    Metal    \\
      name      &of jet    &each outburst  &  outbursts      &placement time  &distribution \\
                &episodes  & [Myr]        &   [Myr]       & $t_c$ [Myr]   &   profile   \\
 \hline
 I0-RInf 	   & 1 	   & 30 	   & 0   	& $\infty$  	& flat		\\
 I0-R1000    	   & 1 	   & 30 	   & 0   	& 1000	   	& flat		\\
 I0-R500    	   & 1 	   & 30 	   & 0   	& 500	   	& flat		\\
 I10-RInf	   & 3 	   & 10 	   & 10  	& $\infty$ 	& flat		\\
 I100-RInf	   & 3 	   & 10 	   & 100 	& $\infty$  	& flat		\\
 I100-R1000 	   & 3 	   & 10 	   & 100 	& 1000      	& flat		\\
 I100-R500 	   & 3 	   & 10  	   & 100 	& 500       	& flat		\\
 I100-RInf-P	   & 3 	   & 10 	   & 100 	& $\infty$  	& Plumer	\\
 I100-R1000-P     & 3 	   & 10 	   & 100 	& 1000      	& Plumer	\\
 I0-R500-P 	   & 1 	   & 30 	   & 0   	& 500       	& Plumer	\\
 I100-R500-P 	   & 3 	   & 10 	   & 100 	& 500       	& Plumer	\\
\hline
\end{tabular}
\end{minipage}
\end{table*}

\subsection{The tracer fields, conservation and injection}
\label{conserv}

We implement the distribution and evolution of metals using Flash's tracer fields 
(fields from now on). The metallicity, $Z$, is the ratio of the metal density 
to the ICM gas density, $Z( \vec r ) = \rho_m( \vec r ) / \rho_{g}( \vec r  )$, 
and the field is defined so that:

\begin{equation}
f( \vec r ) = N Z( \vec r ), 
\label{Z-N}
\end{equation}
where $N$ is a constant factor required by Flash to keep \hbox{$f\,\in[$0,1$]$}.
Flash uses two fields: $f$, representing the metal distribution, 
and  $\bar{f}$, the non-metal gas fraction. Then, fields are conserved in the sense
that $f + \bar{f} = 1$ in every zone of the simulation grid. Thus,
normalisation parameters are required to fulfil this condition.

We use two galaxy metal density distributions: a flat distribution:

\begin{equation}
\begin{array}{c l}
    \rho_{_{0f}}  & \rmn{for} \,\, r \in [10,50] \rmn{kpc}; \\ 
   	0	  & \rmn{elsewhere}, 
\end{array}
\label{rhom}
\end{equation}

and a Plumer-like distribution (see Table~1):

\begin{equation}
\rho_{mP}(r) = \frac{\rho_{_{0P}} }{ (1 + ( r / r_c )^2)^{5/2} },
\label{rhop}
\end{equation}
where $r_c$ in the denominator of equation~(\ref{rhop}) is the 
core radius of the galaxy metal distribution, which we set to 15\,kpc. 
The metal density, $\rho_{_{0P}}$, in equation~(\ref{rhop}) was adjusted to be equal to 
$\rho_0$ (equation~\ref{icm}) making the metal density, $\rho_{mP}(r)$, a reasonable 
and arbitrary fraction of the ICM gas. 
On the other hand, $\rho_{_{0f}}$, in equation~(\ref{rhom}), was adjusted so that the 
initial metal mass in the cluster is the same for all our simulations.

Metals are injected via star formation which we consider only in the central dominant
cluster galaxy. To simulate this process (from phase 2 of the simulation onwards) 
we add metals to the ICM gas, with a radial dependence following the initial metal 
distribution --\,either $\rho_{_{0f}}(r)$ or $\rho_{mP}(r)$ 
(equations~\ref{rhom} or~\ref{rhop} respectively)\,-- 
with a metal replacement time, $t_c$, so that:

\begin{equation}
\frac{ d\rho_m(r,t) }{ dt } = \frac{\rho_m(r)}{t_c}, \qquad 
\frac{ d\rho_g(r,t) }{ dt } = \frac{\rho_m(r)}{t_c}. 
\label{update1}
\end{equation}
Values for $t_c$ (see Table 1), are based on previous work 
which consider a constant star formation rate \citealt{david01};
Gaibler, Krause \& Camenzind 2005. 
From equations~(\ref{update1}) the computational Euler step for these densities 
is given by:

\begin{equation}
\begin{array}{ll}
\rho_m^{n+1}(r,t) = &\rho_m^n(r,t) + \rho_m(r) \, (dt/t_c), \\
\rho_g^{n+1}(r,t) = &\rho_g^n \, (r,t) + \rho_m(r) \, (dt/t_c), 
\end{array}
\label{update2}
\end{equation}
where $n+1$ denotes the densities at the current timestep, $dt$, and $n$, the densities
at the previous timestep. The evolution of the metallicity is hence given by 
equations~(\ref{Z-N}) and~(\ref{update2}):

\begin{equation}
Z^{n+1}(r,t) = \frac{\rho_m^{n+1}}{\rho_g^{n+1}} = \frac{ \rho_m^n + \rho_m(r) \, (dt/t_c)}{ 
\rho_g^n + \rho_m(r) \, (dt/t_c)},
\label{Z(t)}
\end{equation}
\noindent and thus the evolution of the fields follows so that: 

\begin{equation}
f^{n+1} = f^n \left( \frac{ \rho_g^n }{\rho_g^n + \rho_m(r) \, (dt/t_c)} \right) +
\frac{N \rho_m(r) \, (dt/t_c)}{\rho_g^n + \rho_m(r) \, (dt/t_c)}, \\
\label{f(t)}
\end{equation}
where $f^{n+1}$ and $f^n$ represent the fields for the current and the previous
timesteps respectively.

We tested our implementation and conservation of the fields by following the 
evolution of an annulus sector with a constant field distribution, during a cooling 
flow phase (see Section~2). We found a loss of fields due to 
advection of \hbox{2 parts in 10$^6$} over a Hubble time. The inner computational 
boundary has outflow boundary conditions (except for the jets' nozzles during the 
active phases) and thus fields (metals) can be lost only if they flow through this 
boundary.  
A similar test for our implementation of fields injection 
resulted in a metal loss of not more than \hbox{1 \%}, over a Hubble time 
--\,since our simulations last for \hbox{3\,Gyr}, 
we confidently neglect these numerical effects.

\section[]{Results and Discussion}
\label{results}

\begin{figure*}
  \includegraphics[scale=0.35,bb=0.0in 9.5in 6.5in 1in,clip=]{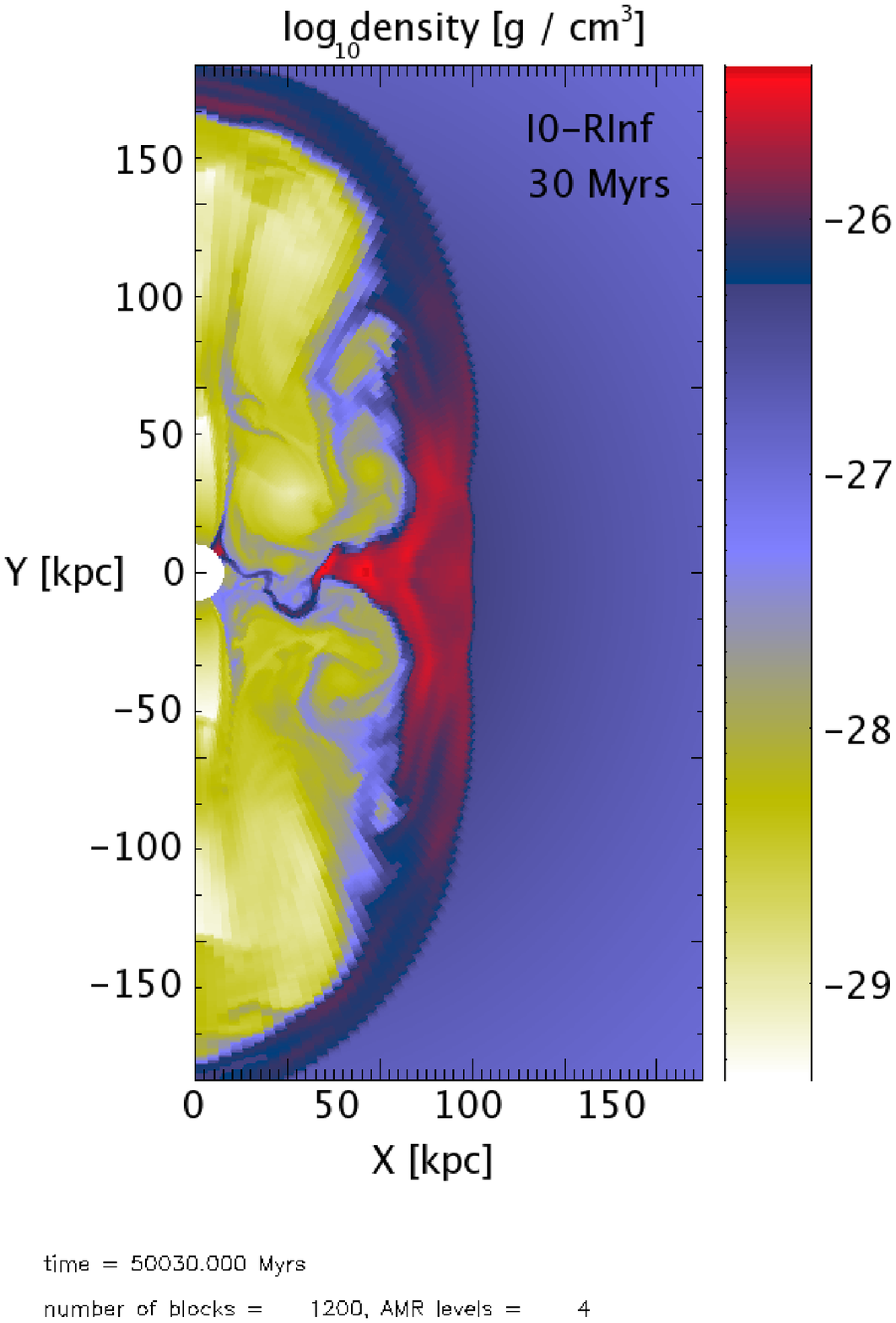} 
  \includegraphics[scale=0.35,bb=0.0in 9.5in 6.5in 1in,clip=]{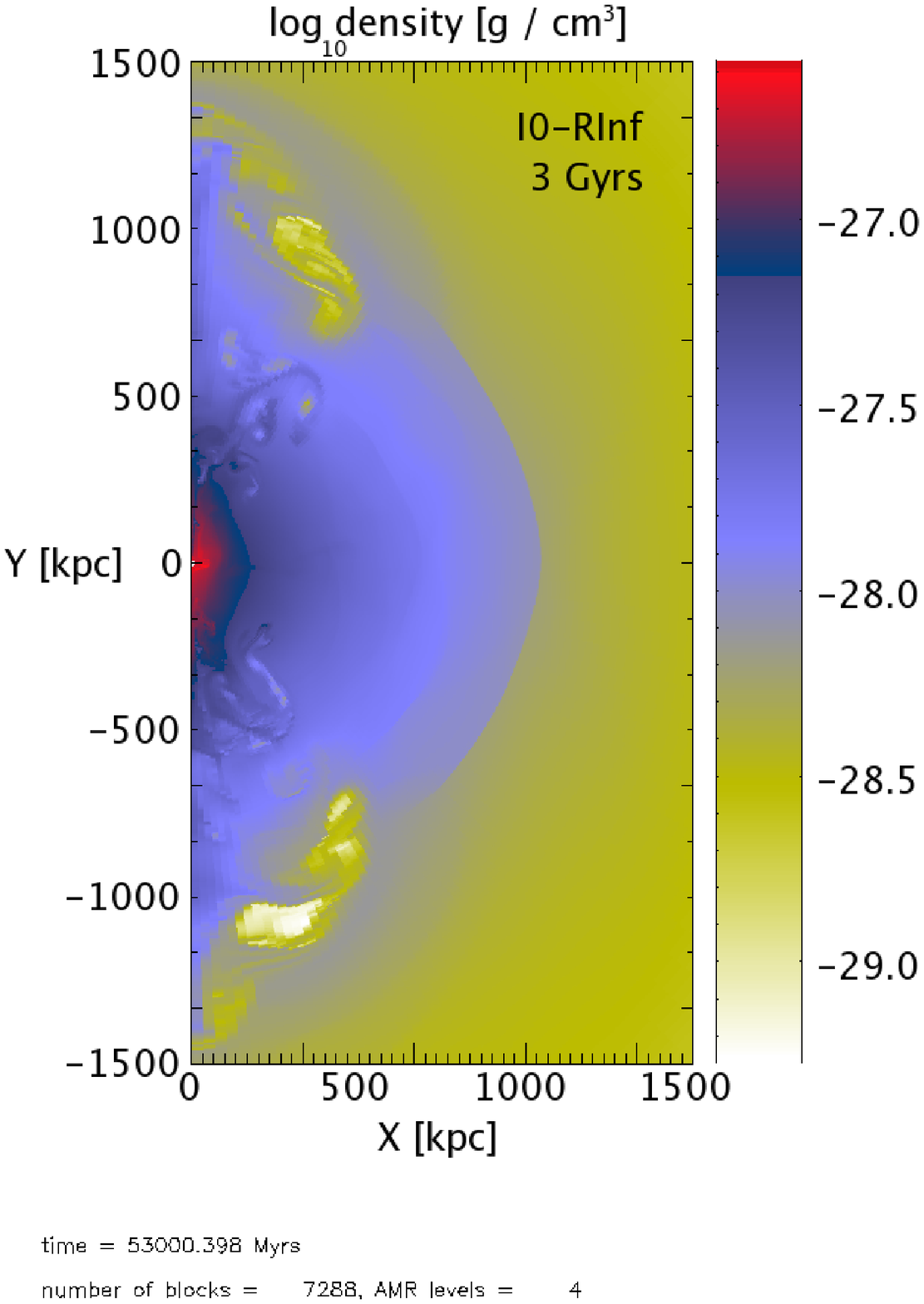} 
  \includegraphics[scale=0.35,bb=0.0in 9.5in 6.5in 1in,clip=]{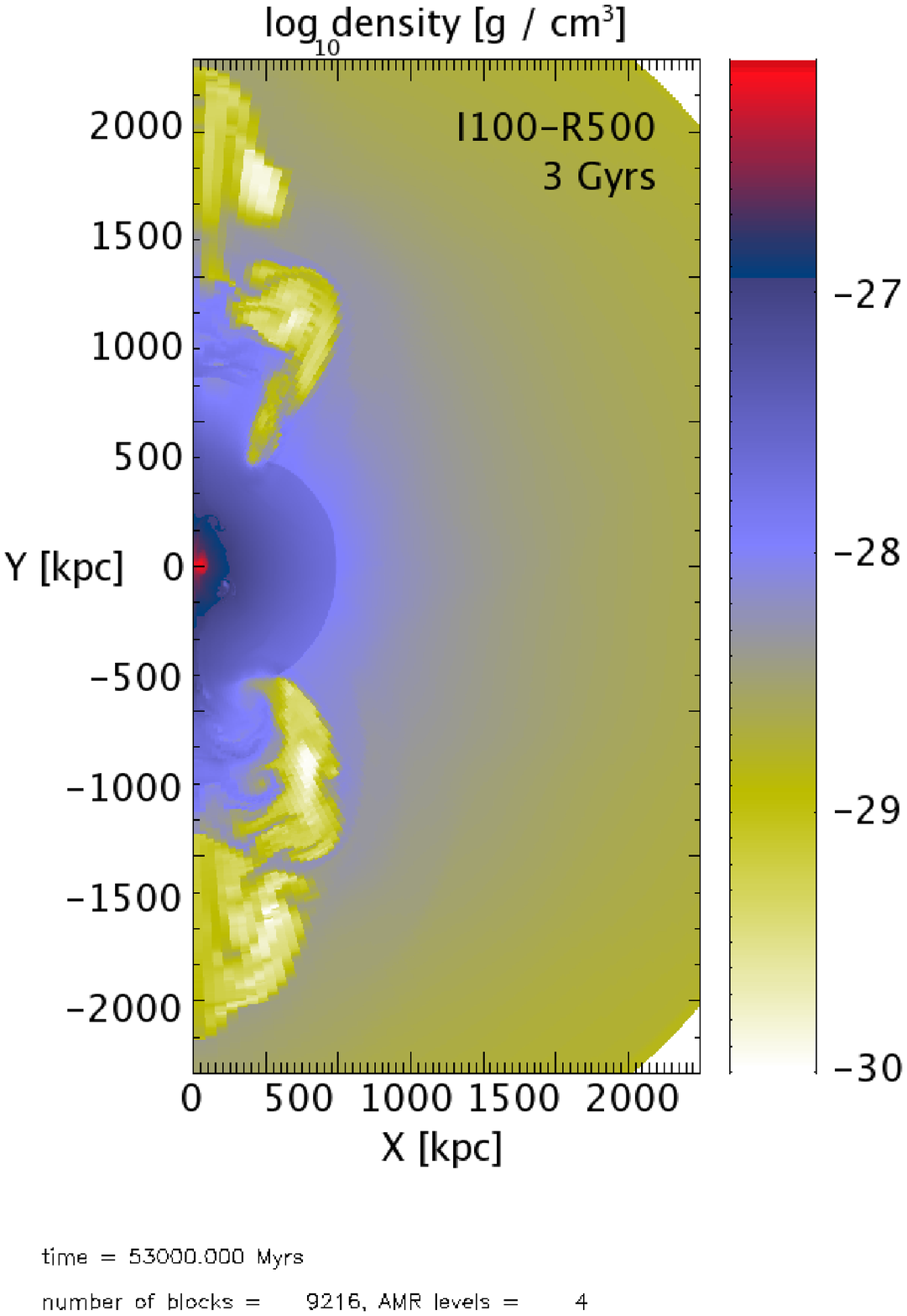}\\
\vskip7cm 
  \includegraphics[scale=0.35,bb=0.0in 9.5in 6.5in 1in,clip=]{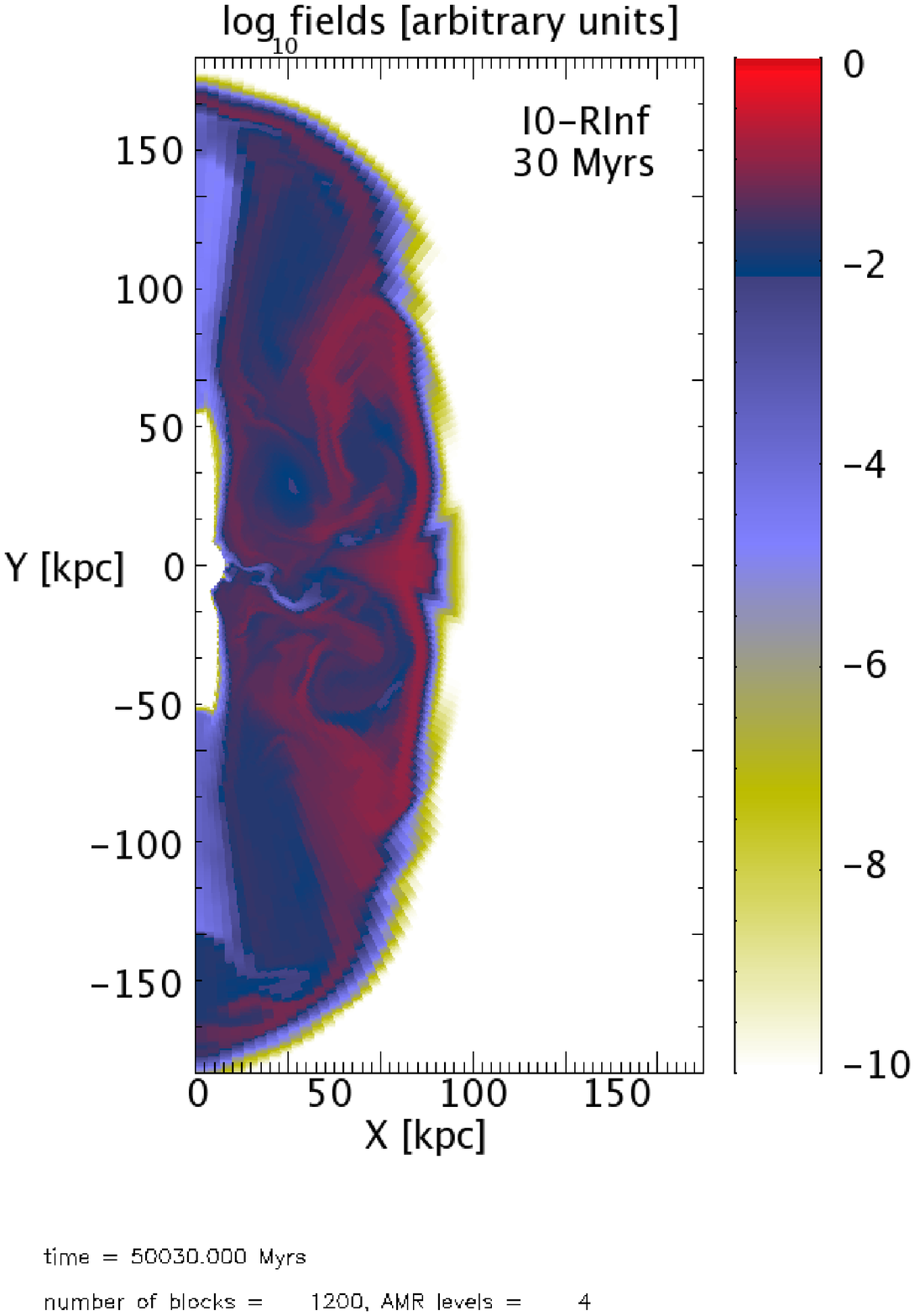} 
  \includegraphics[scale=0.35,bb=0.0in 9.5in 6.5in 1in,clip=]{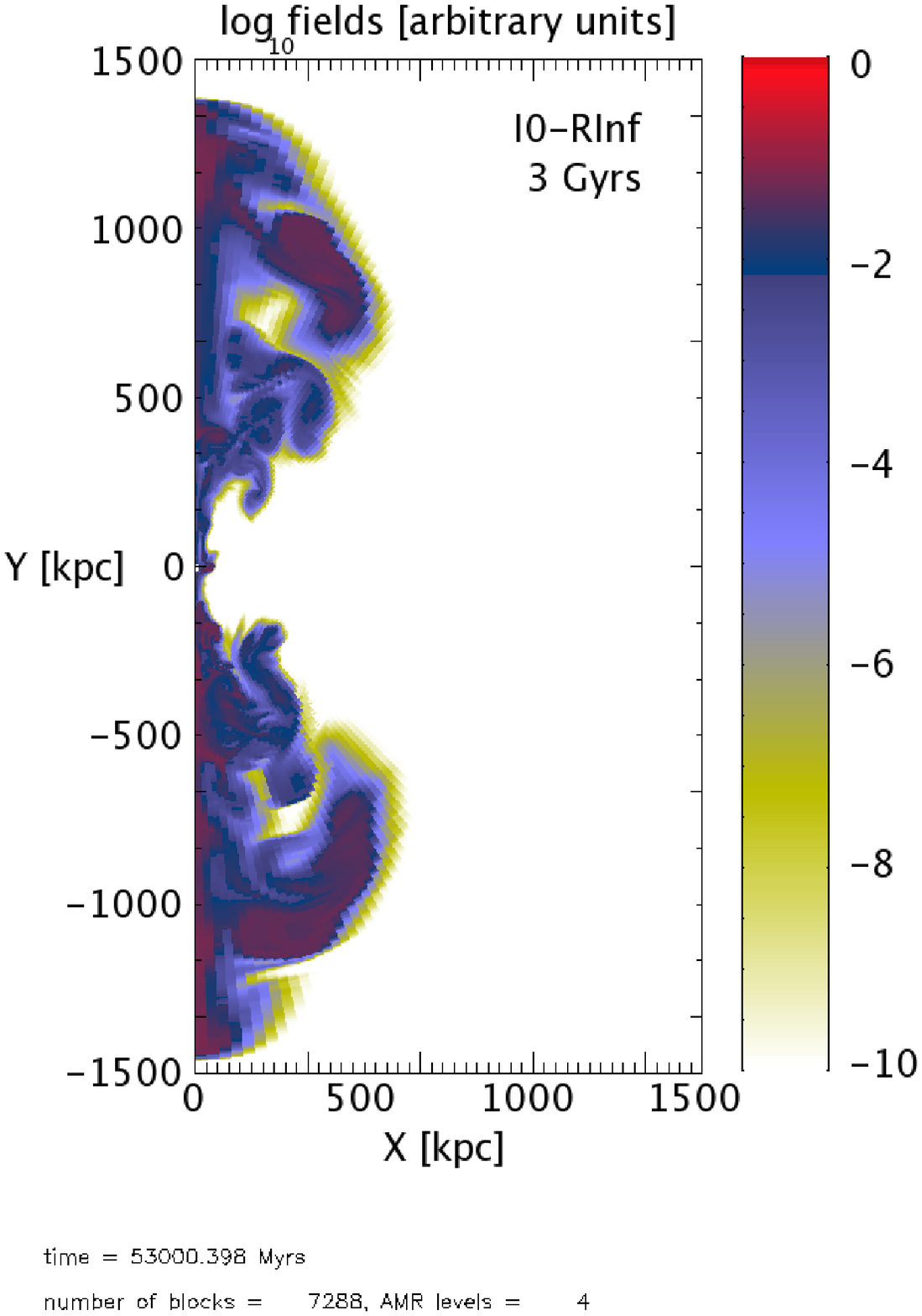} 
  \includegraphics[scale=0.35,bb=0.0in 9.5in 6.5in 1in,clip=]{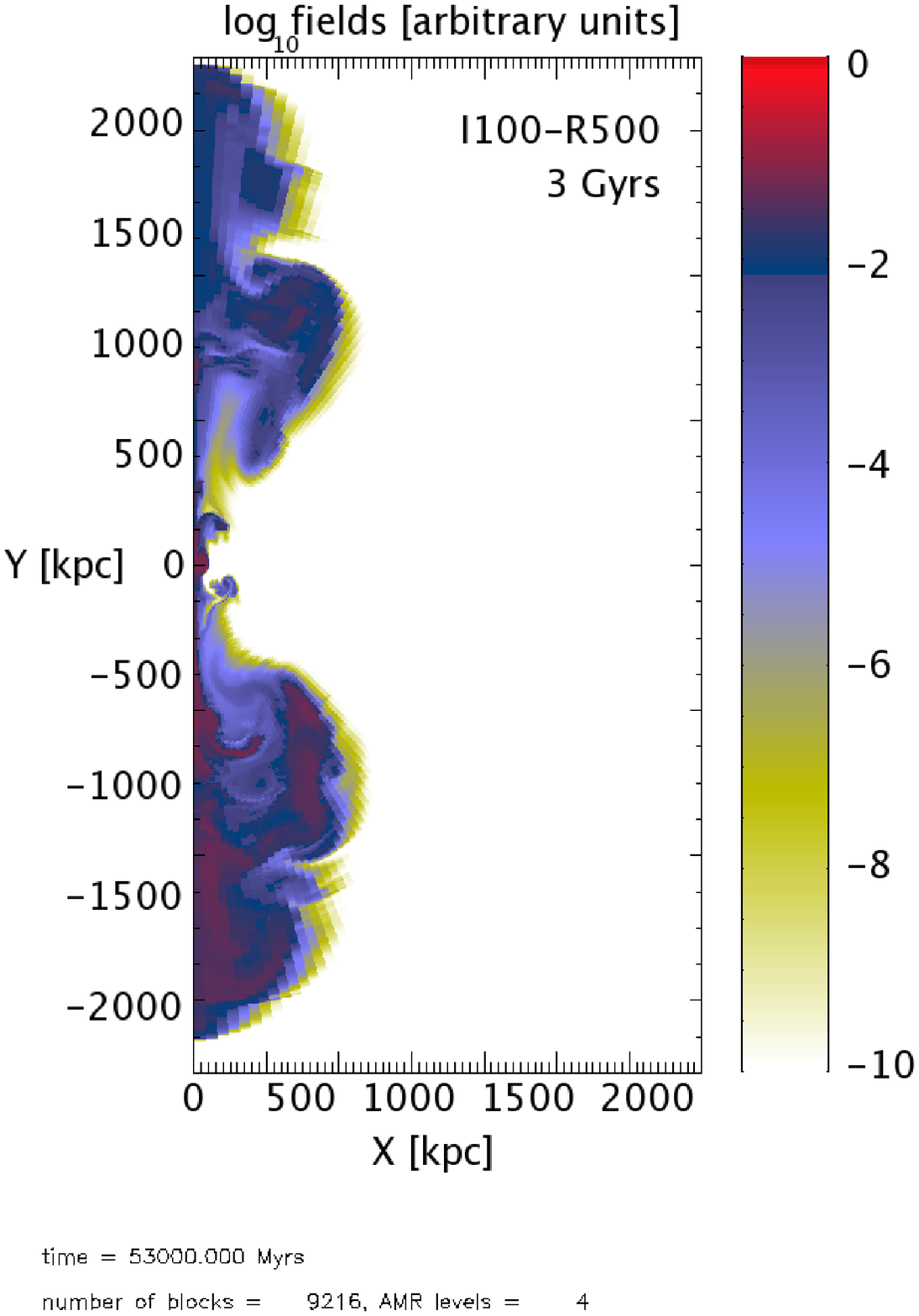}\\
  \vspace*{210pt}
 \caption{Plots (xy plane) of the logarithms of the ICM gas density 
(top) and the tracer fields (bottom). Left panel shows run I0-RInf (see Table~1) at 
the end of the jets' active phase. The jets and the cocoon are visible.
The middle and right panels show runs I0-RInf and I100-R500 respectively, at the end 
of the simulation (3\,Gyr).}
\label{fig1}
\end{figure*}

The basic dynamical evolution observed in the simulations is as follows.
During phase 1 (see~Section~2) radiative cooling leads to the formation of a 
cooling flow. We run the simulation until the cooling flow reaches a 
steady state and develops a core with 
cool inflowing gas and a radial gradient in the velocity -- this gradient is very steep 
near to the core and flat away from it. The initial 
hydrostatic equilibrium conditions prevail
away from the core in agreement with canonical models (e.g., see \citealt{fabian94}).
During phase~2, the initial hydrodynamical structure of powerful jets develops. 
The jets drive a supersonic bow shock from the centre of the galaxy through the ICM,
but at this stage the fields are not removed from the core.
Behind the shock, the under-dense jets form a low density cocoon 
bounded by a contact surface. The jets push aside the 
central cool gas and metals in their way. 
The cocoon grows roughly self-similarly in accordance with the models of 
\citet{K-A} and \citet{alexander06}. 
The left panels in Figure~\ref{fig1} show  
the distribution of the logarithms of the density (top) and the fields (bottom),
at the end of the active phase of \hbox{run I0-RInf} (30\,Myr).
The jets, the cocoon and ambient medium are clearly visible. 

Once the cycle of the jets is finished, the system has developed several gas layers
with different densities and velocities. The expansion of the cocoon decelerates 
gradually. 
Hydrodynamic instabilities grow and disrupt the cocoon producing 
turbulent flows in the central ICM. The remnant cocoon forms two hot buoyant
bubbles that propagate away from the core, in agreement with previous 
3D~hydrodynamical simulations (e.g., see \citealt{churazov01,basson}).

The fields mimic the motion of the gas flows closely and mixing of
gas with low and high metal concentrations occurs, particularly
close to the most prominent turbulent vortices. The later effects are more
important when intermittent jets are present (see Table~1). 
The convective flows drag the metals very efficiently --\,$\sim$\,95\%\,of
the initial galaxy metals in general\,-- to distances up \hbox{to\,2\,Mpc} 
(see Figures~\ref{fig1} and~\ref{fig2}). 
The bow shock continues propagating, decelerates gradually and reaches 
the outer computational boundary (3\,Mpc) at the end of our simulations  
(\hbox{3\,Gyr}). The middle and right panels in Figure~\ref{fig1} show 
the distribution of the logarithms of the density (top) and the fields 
(bottom), of these convective flows, at the end of the simulation for
Runs~I0-RInf and~I100-R500.

\subsection{The time evolution of the metal distribution}

\begin{figure*}
  \includegraphics[bb=65 20 483 355, scale=0.34, clip=]{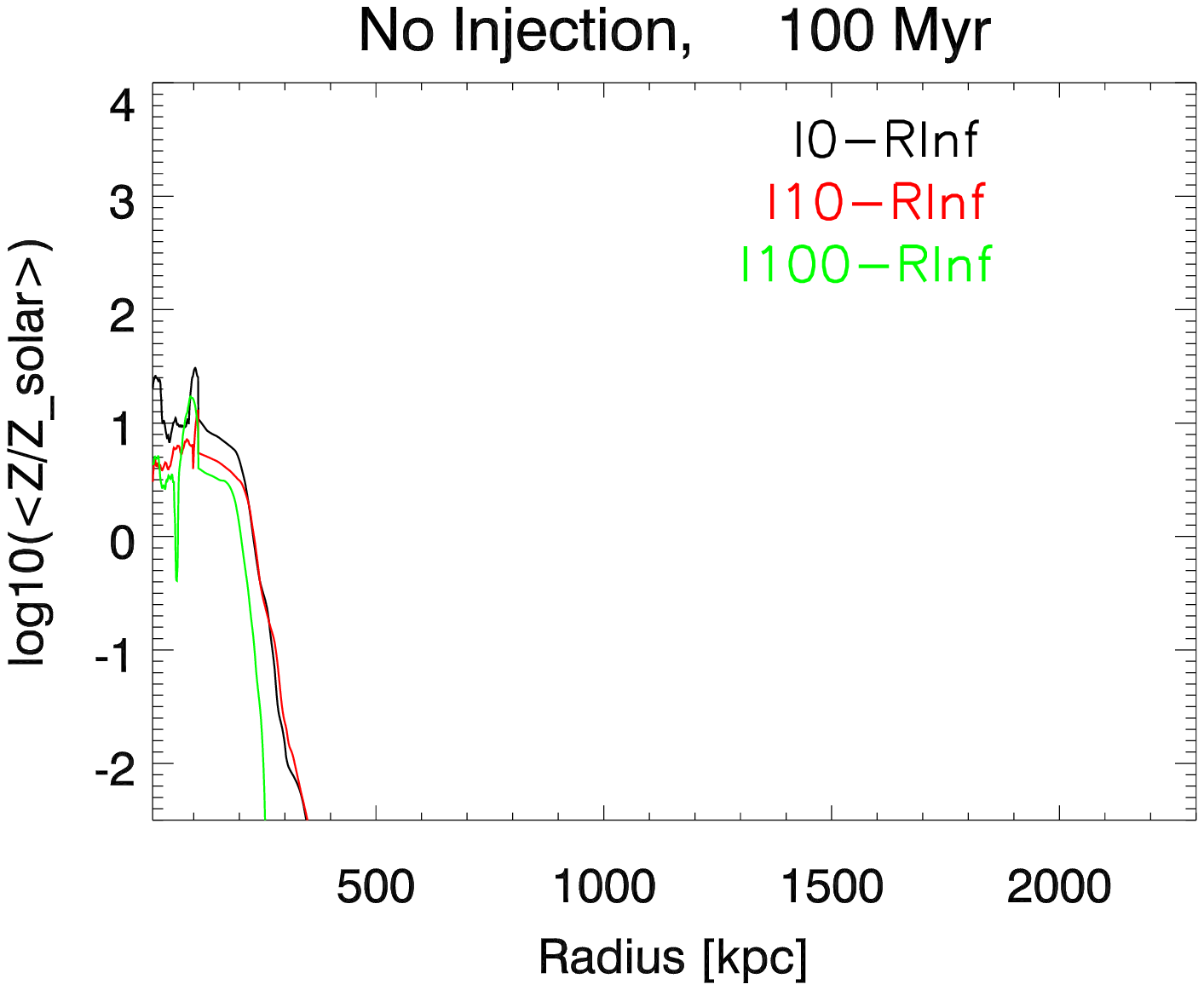}
	\hspace*{\columnsep}%
  \includegraphics[bb=65 20 483 355, scale=0.34, clip=]{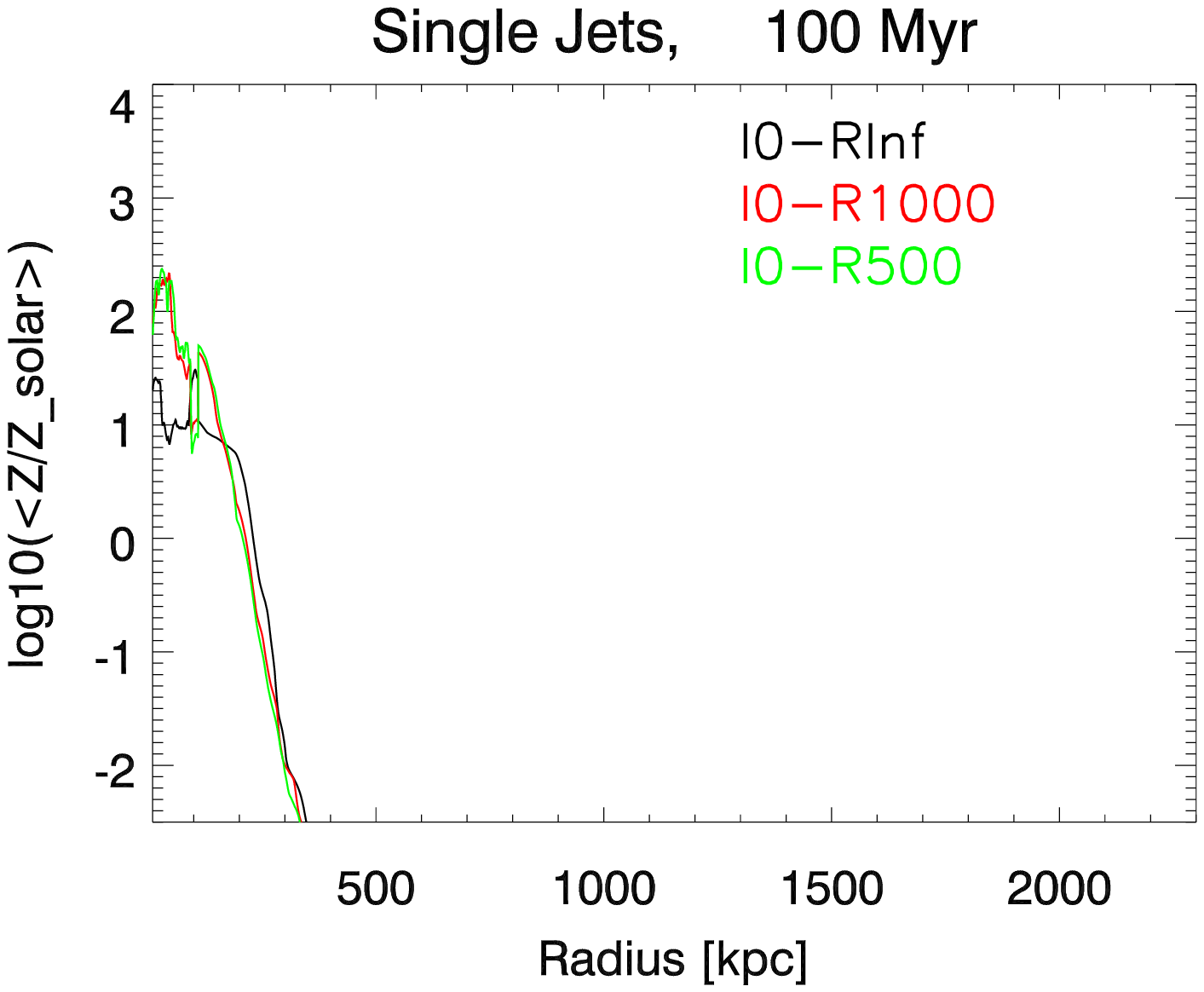} 
	\hspace*{\columnsep}%
  \includegraphics[bb=65 20 483 355, scale=0.34, clip=]{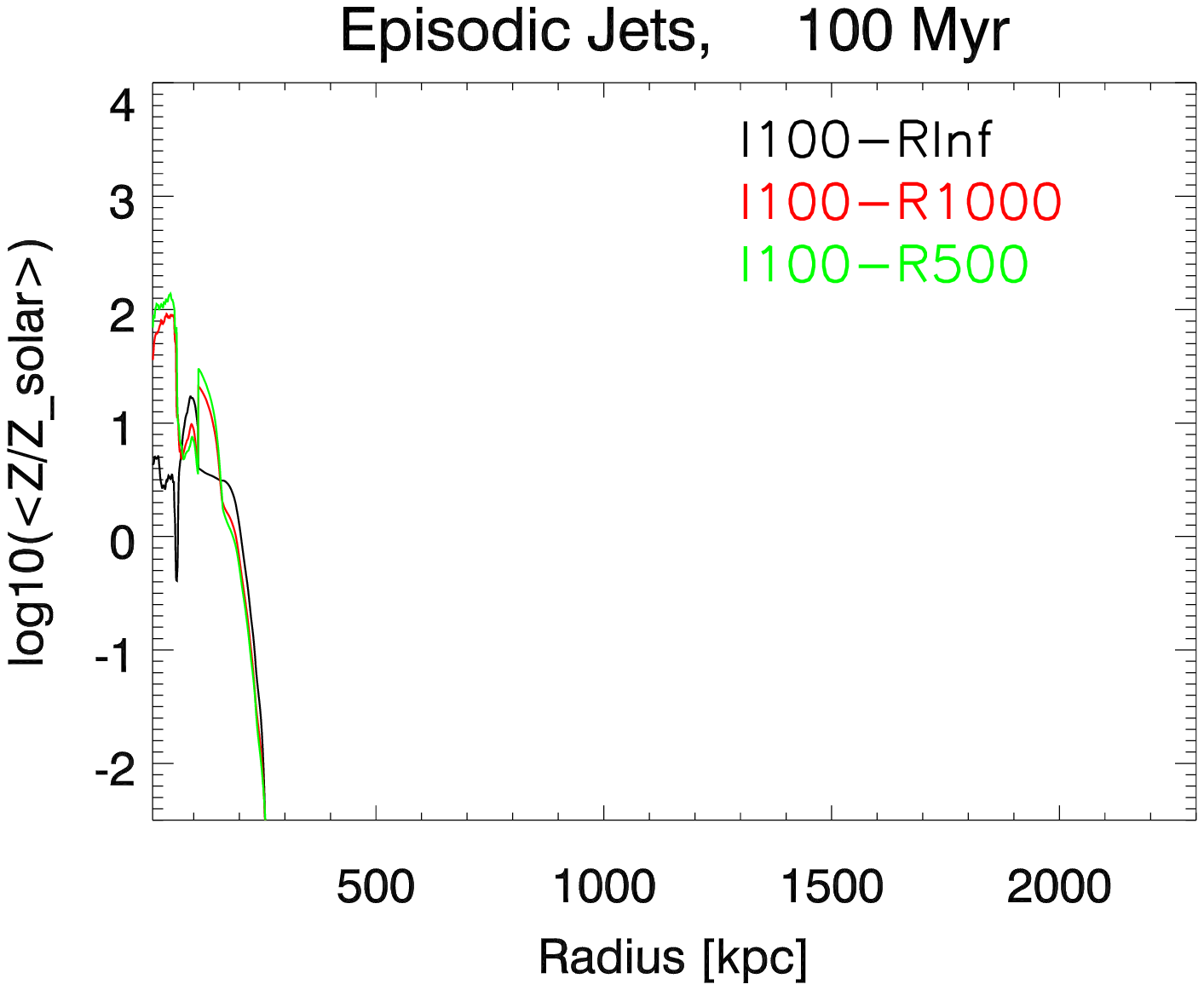} \\
  \includegraphics[bb=65 20 483 355, scale=0.34, clip=]{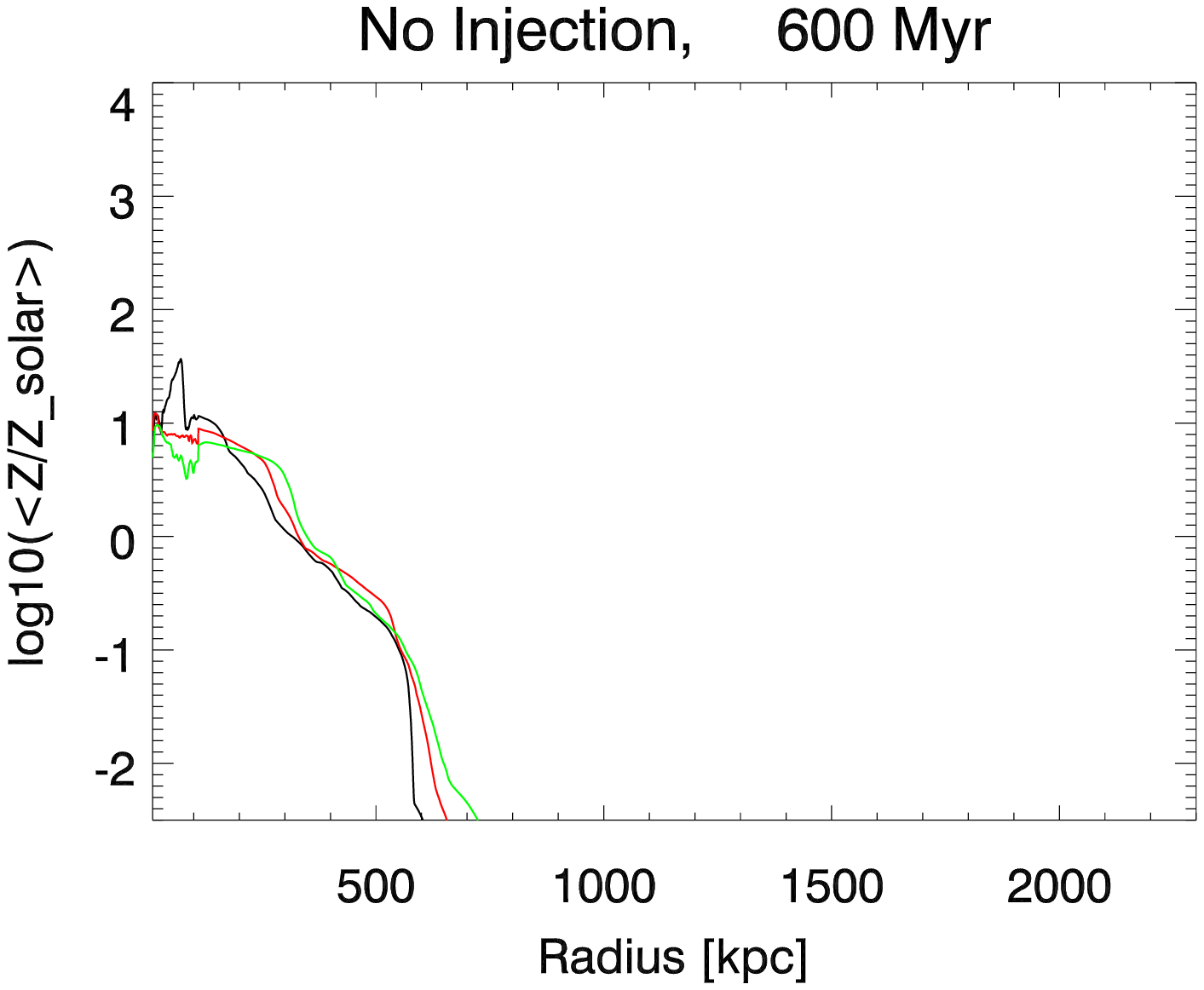} 
	\hspace*{\columnsep}%
  \includegraphics[bb=65 20 483 355, scale=0.34, clip=]{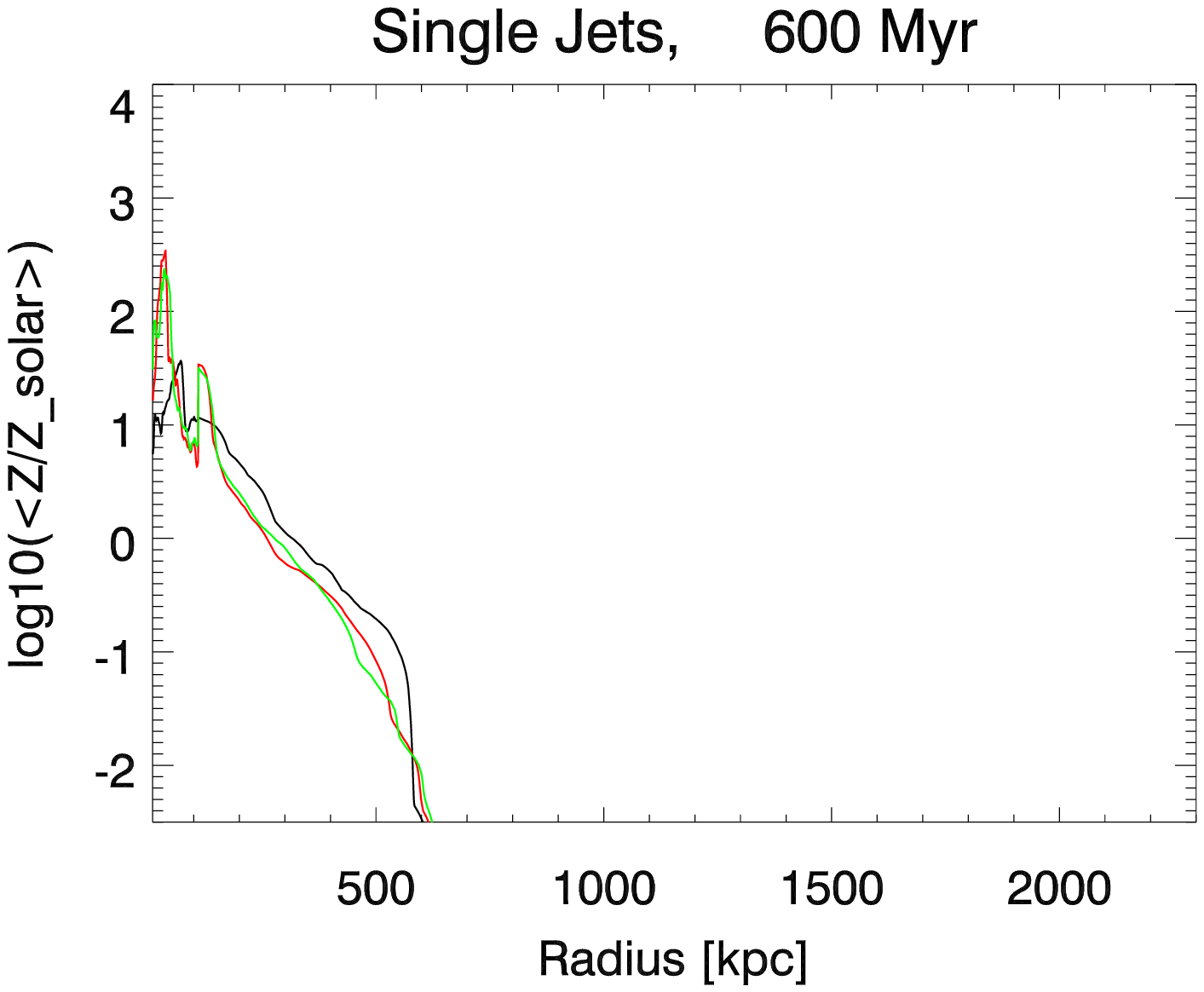} 
	\hspace*{\columnsep}%
  \includegraphics[bb=65 20 483 355, scale=0.34, clip=]{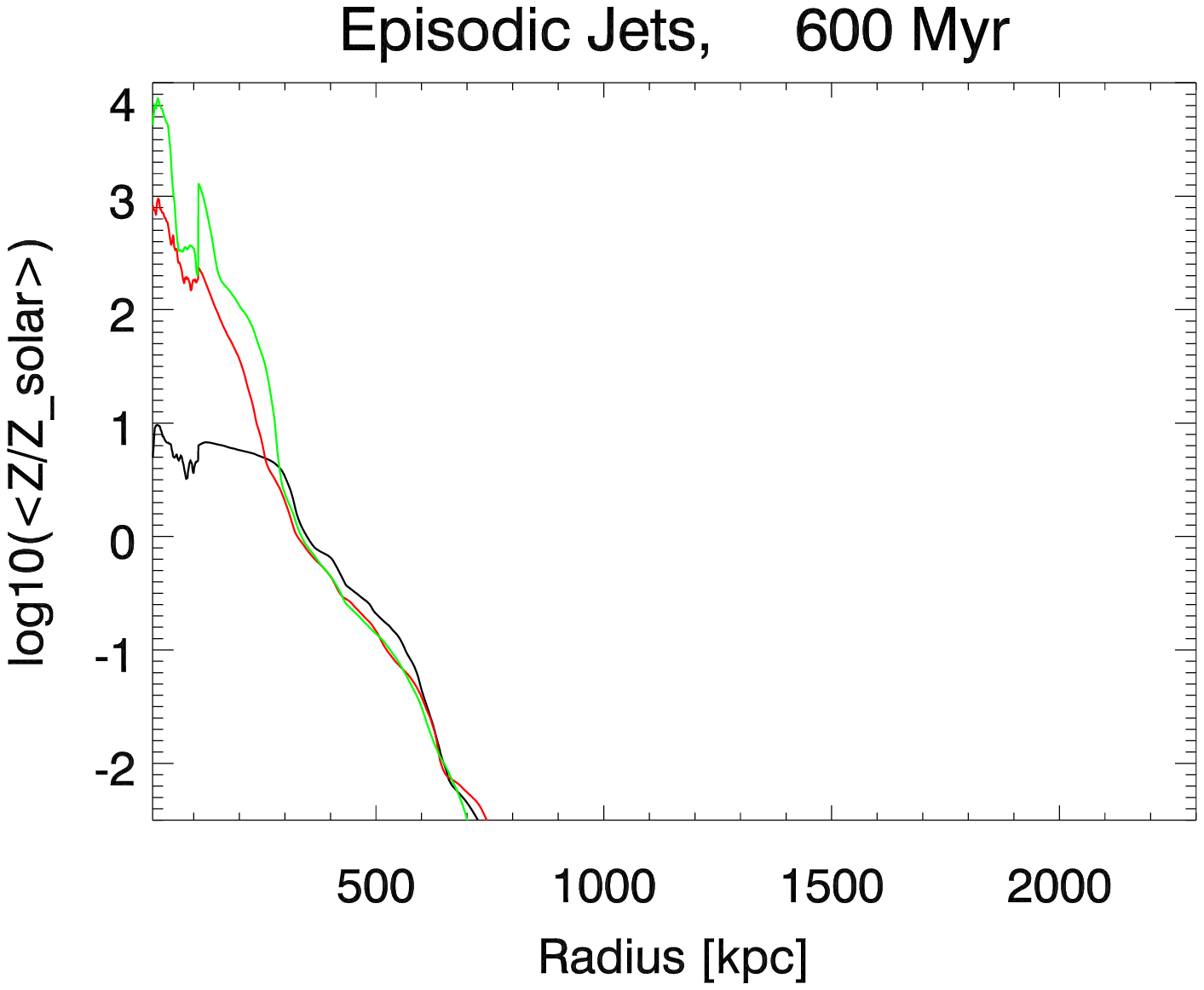} \\
  \includegraphics[bb=65 20 483 355, scale=0.34, clip=]{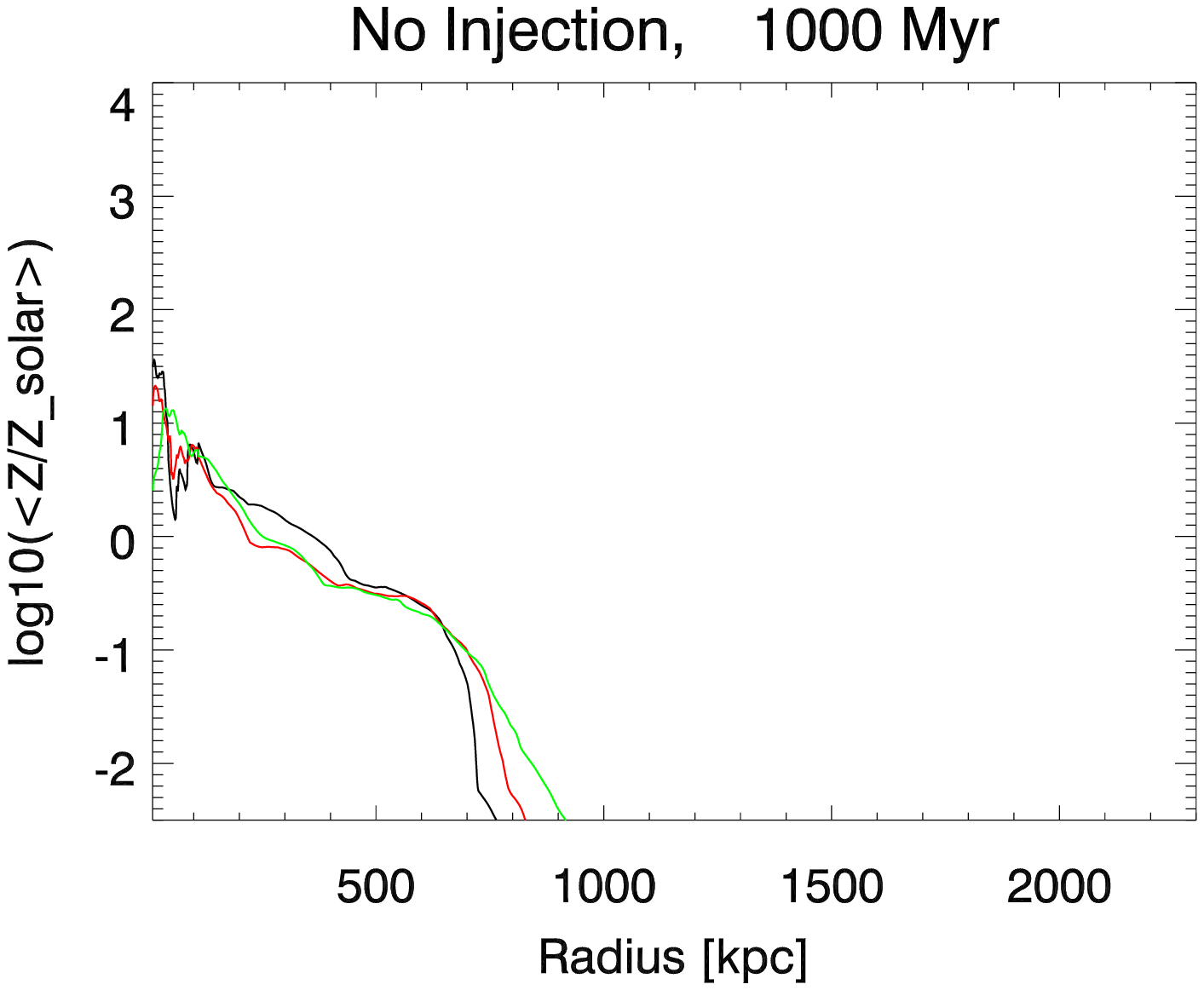} 
	\hspace*{\columnsep}%
  \includegraphics[bb=65 20 483 355, scale=0.34, clip=]{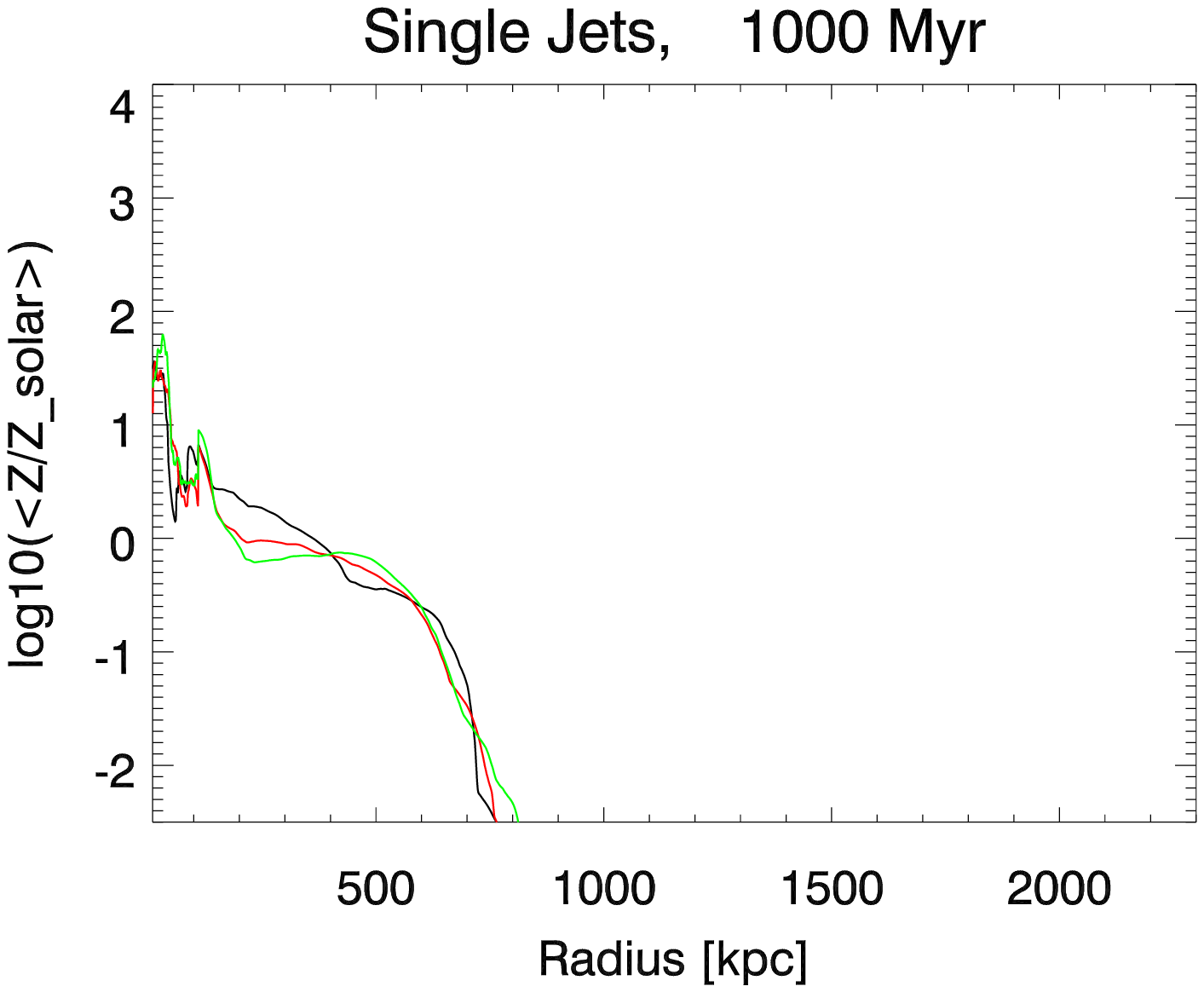} 
	\hspace*{\columnsep}%
  \includegraphics[bb=65 20 483 355, scale=0.34, clip=]{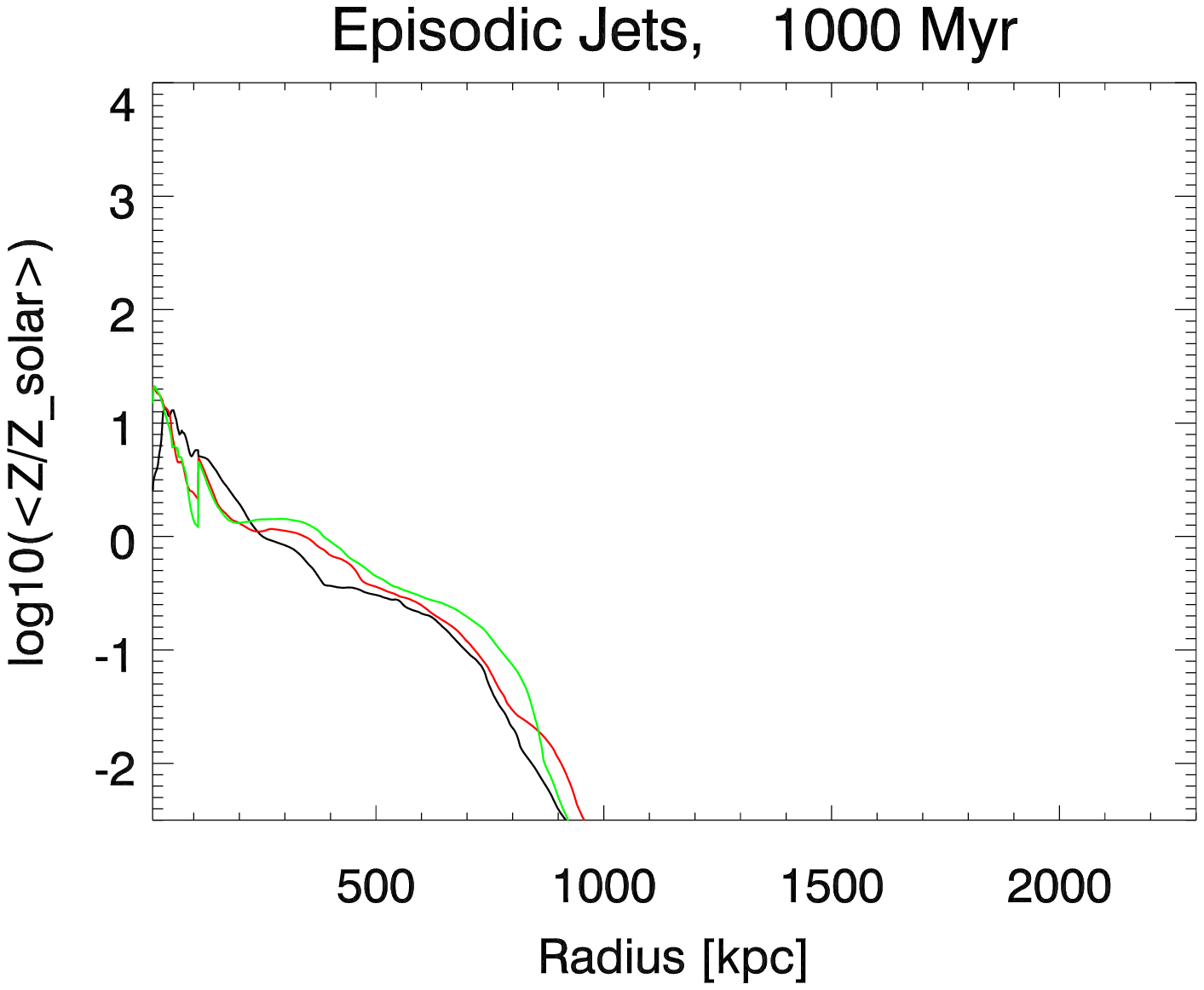} \\
  \includegraphics[bb=65 20 483 355, scale=0.34, clip=]{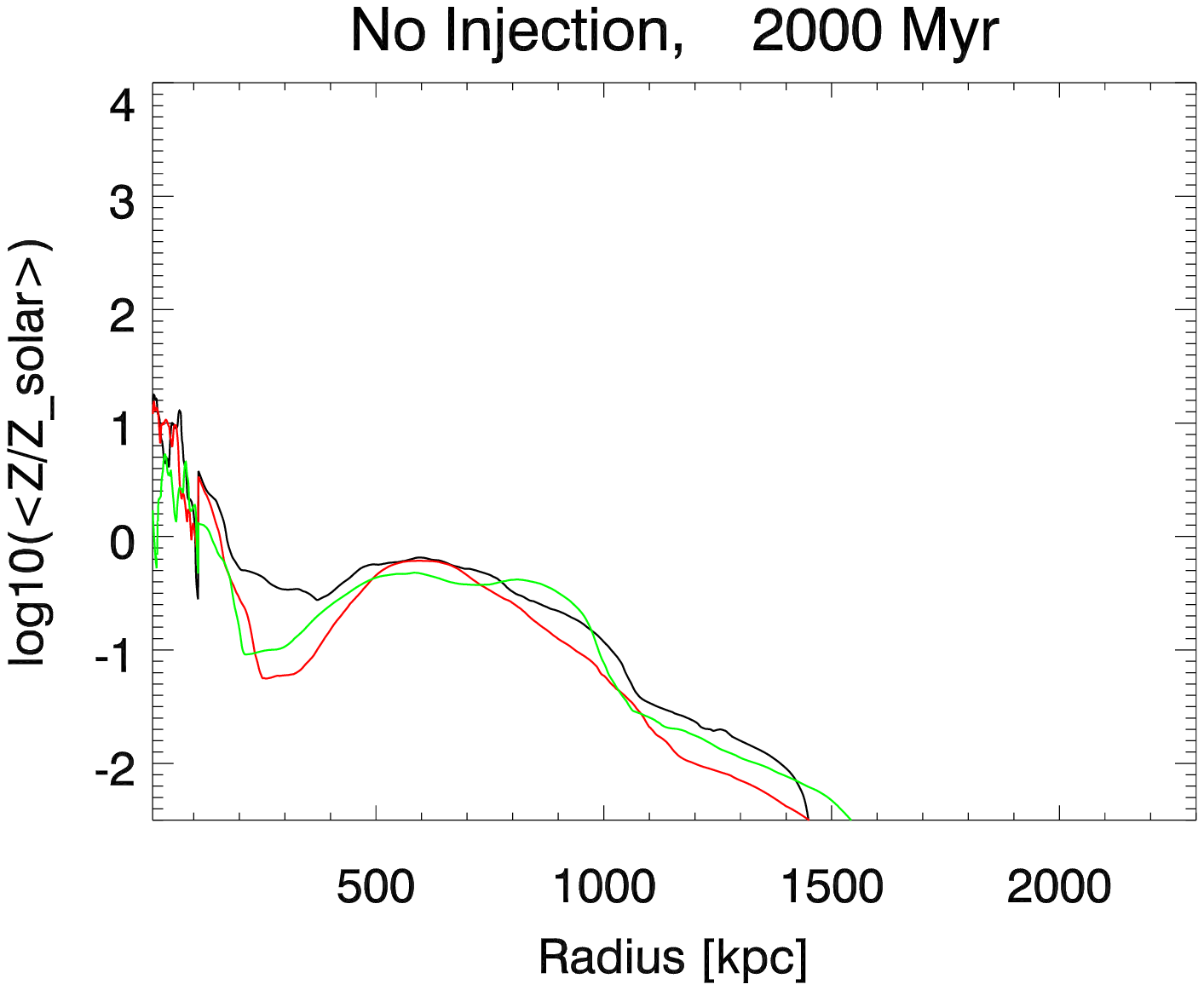} 
	\hspace*{\columnsep}%
  \includegraphics[bb=65 20 483 355, scale=0.34, clip=]{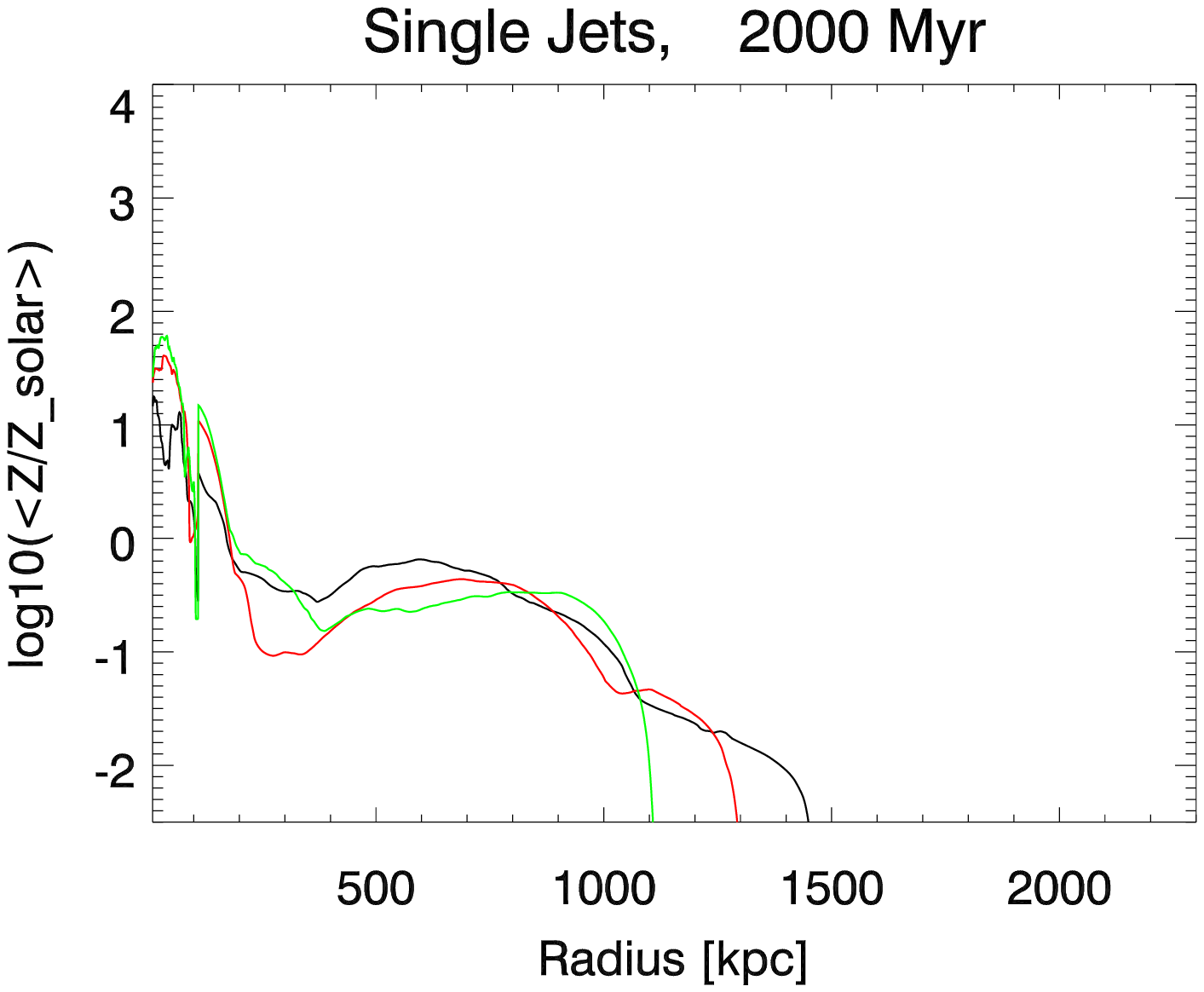} 
	\hspace*{\columnsep}%
  \includegraphics[bb=65 20 483 355, scale=0.34, clip=]{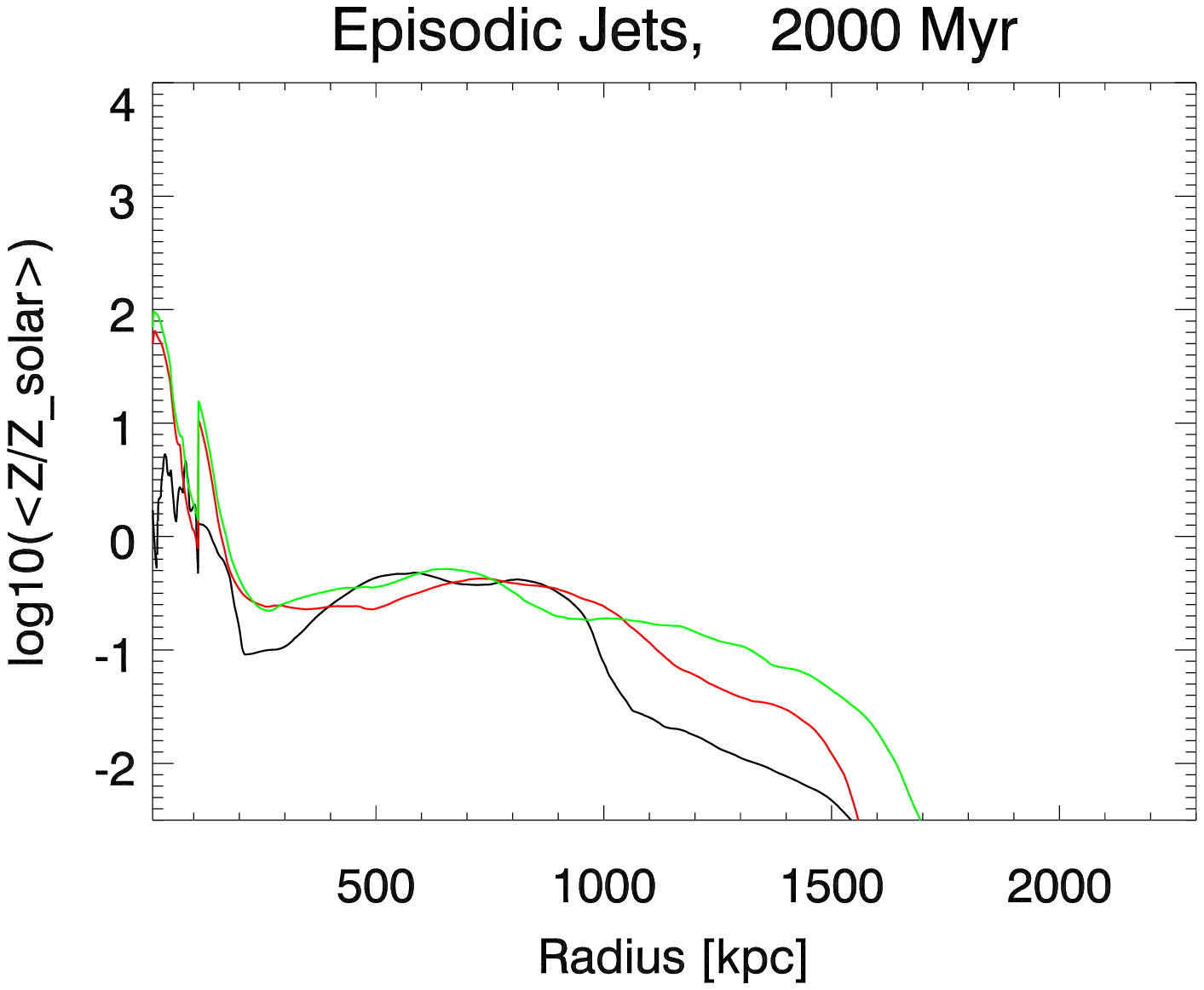} \\
  \includegraphics[bb=65 20 483 355, scale=0.34, clip=]{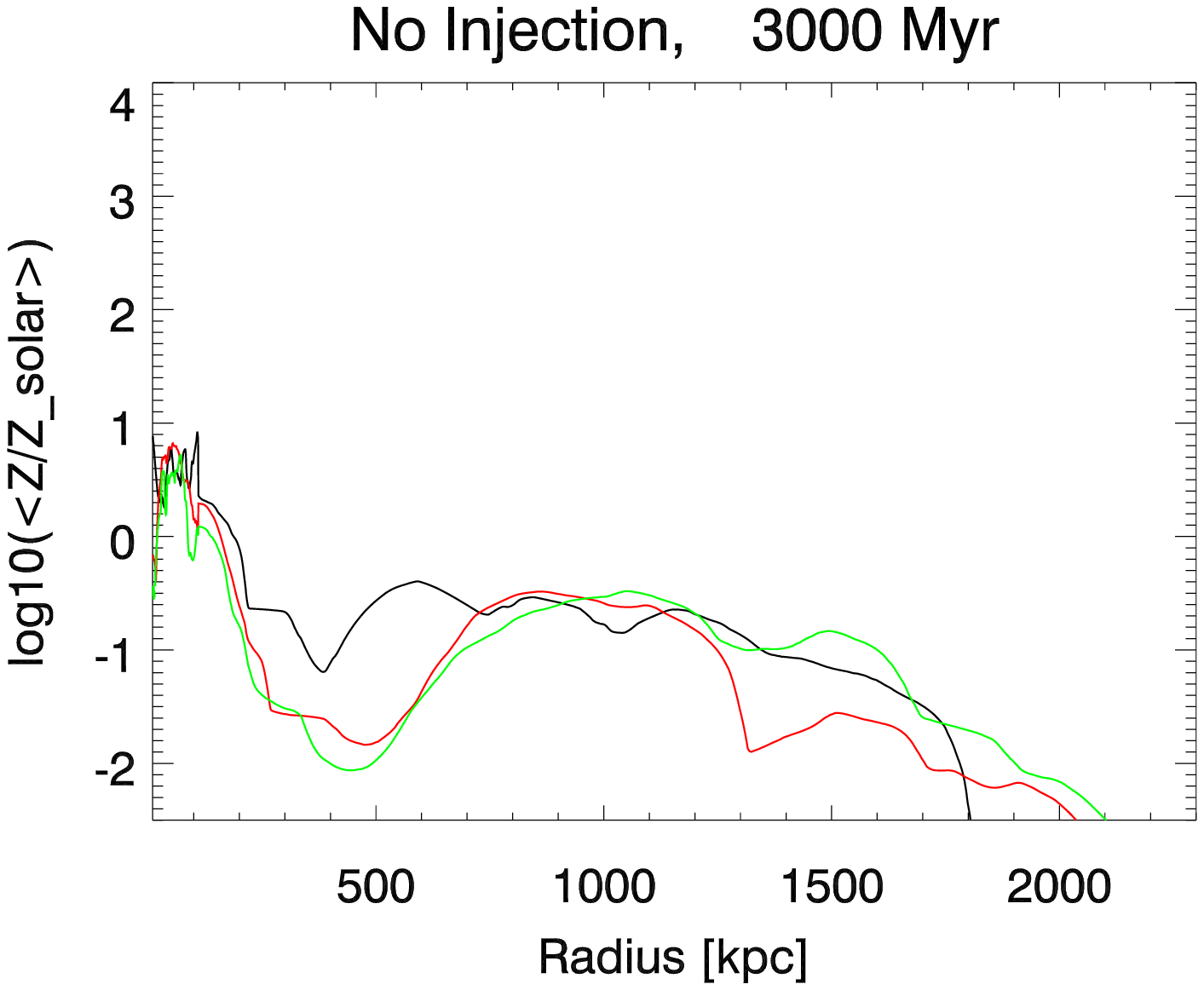} 
	\hspace*{\columnsep}%
  \includegraphics[bb=65 20 483 355, scale=0.34, clip=]{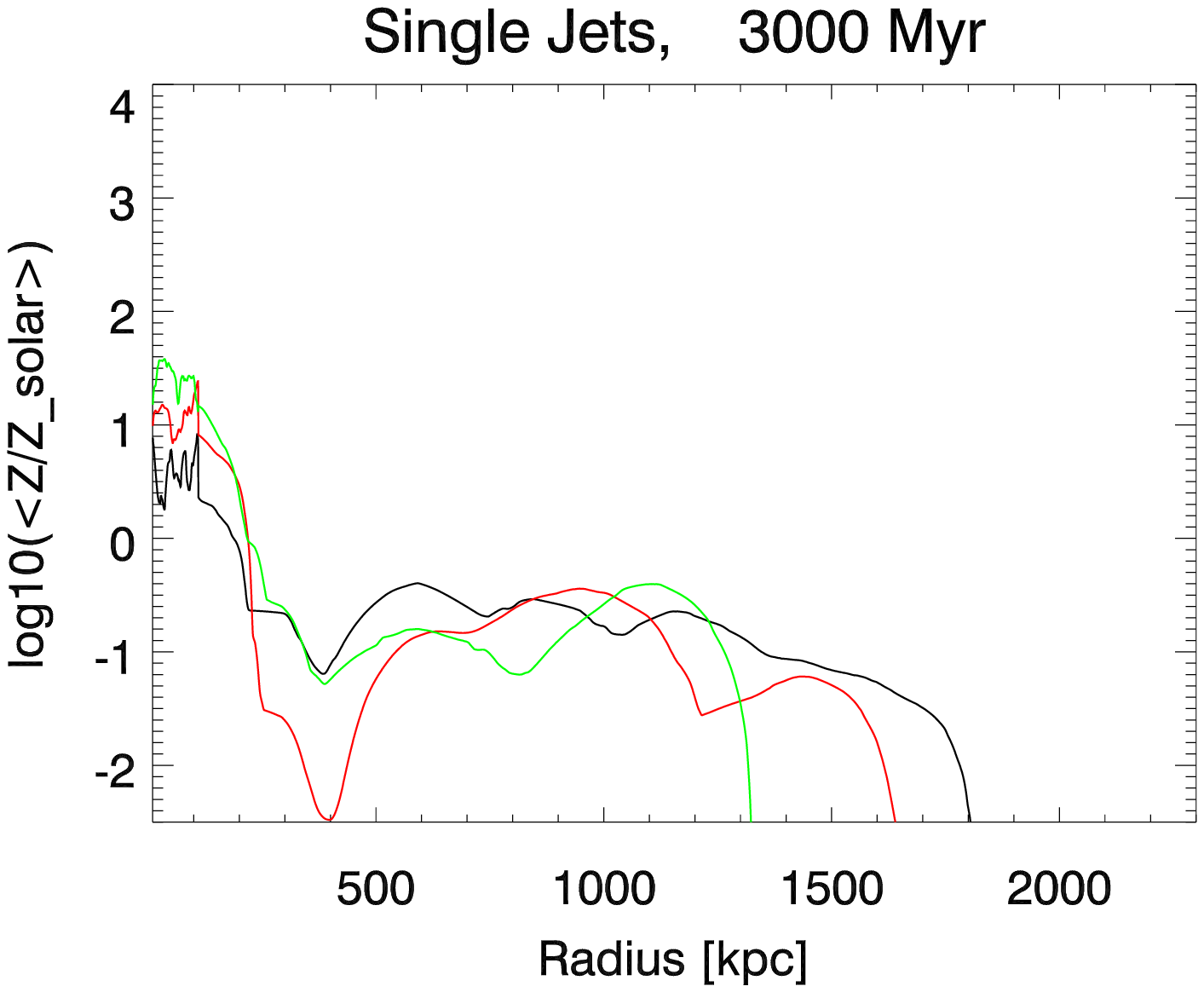} 
	\hspace*{\columnsep}%
  \includegraphics[bb=65 20 483 355, scale=0.34, clip=]{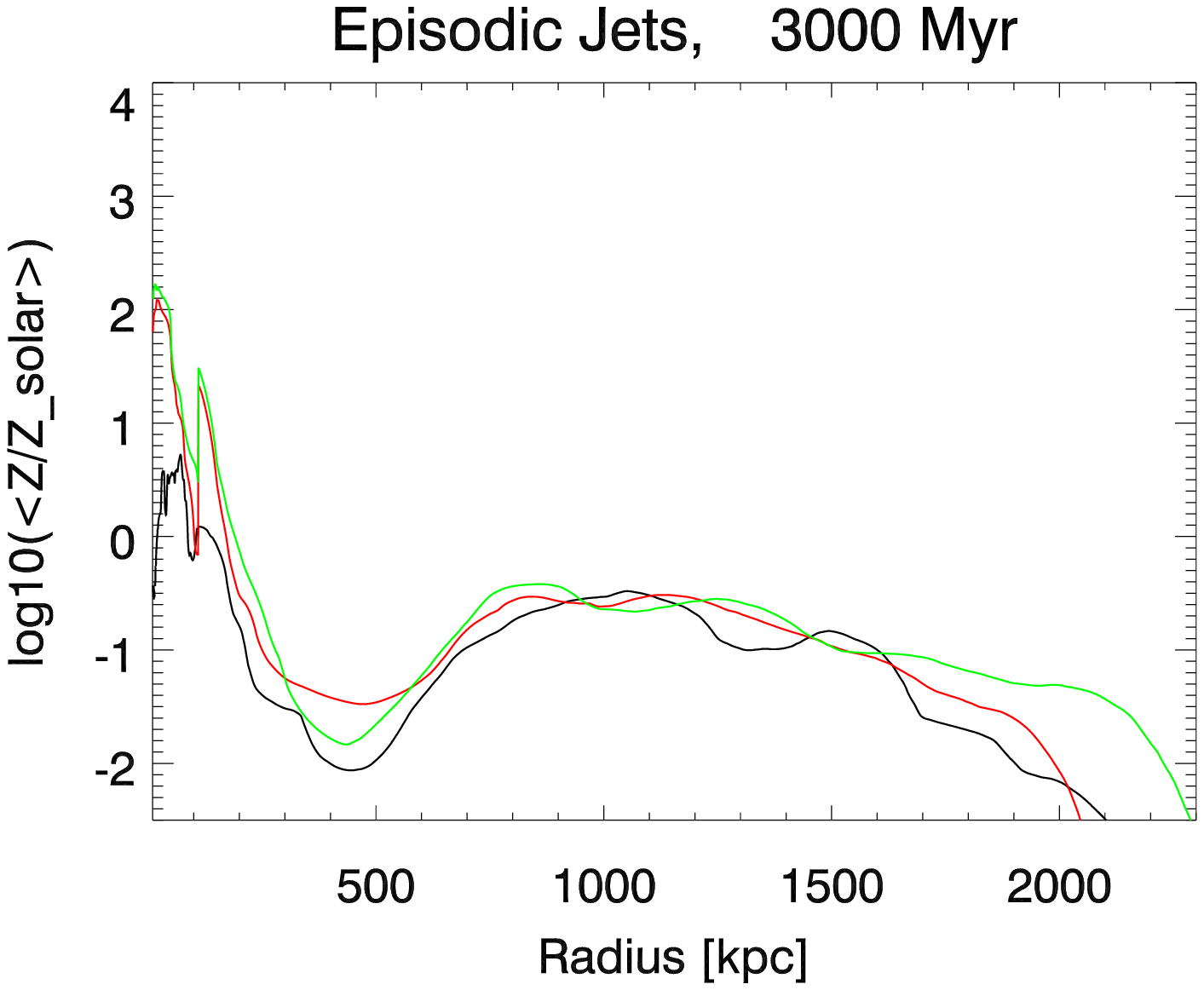} \\
  \vspace*{0pt}
 \caption{Time evolution of the mean relative metallicities as functions
of the radius from cluster centre. Each point shows the logarithm of the 
azimuthal integration 
(spherical shells) of the ratio of the metal density over the ICM gas density 
(see equation~\ref{plots}). i.e., the mean metallicities are normalised in order to be
compared with the observations (see Section~\ref{data}). Column~1, comparison 
of simulations of galaxies with no metal injection and different jets' duty cycles. 
Column~2, comparison of Runs with a single 30\,Myr outburst and 
different matter injection rates. Column~3, comparison of simulations with three 
10\,Myr outbursts (and two 100\,Myr interludes) and different mater injection 
rates (see Table~1).}
\label{fig2}
\end{figure*}

\begin{figure*}
  \vspace*{0pt}
  \includegraphics[bb=65 20 483 355, scale=0.34, clip=]{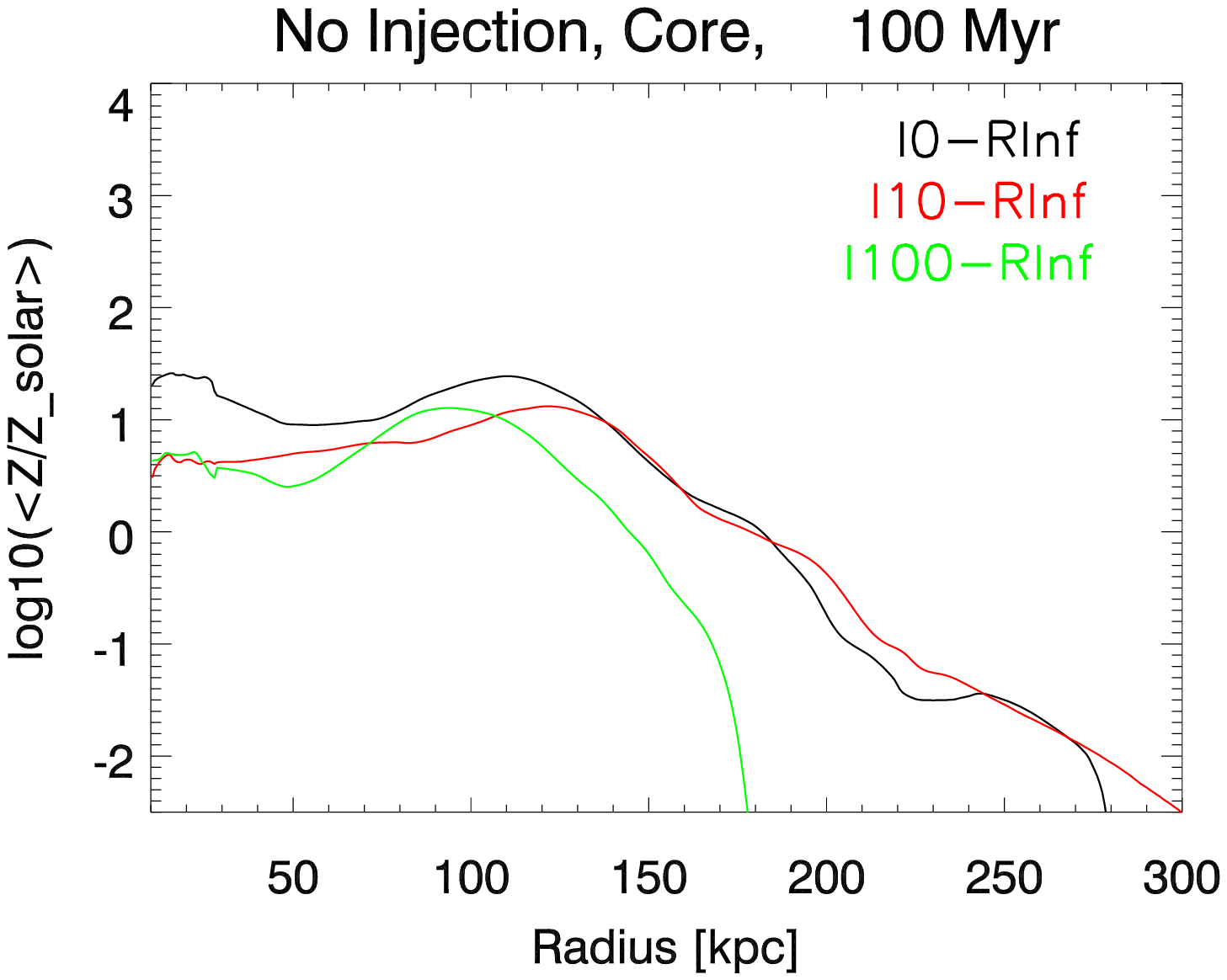} 
	\hspace*{\columnsep}%
  \includegraphics[bb=65 20 483 355, scale=0.34, clip=]{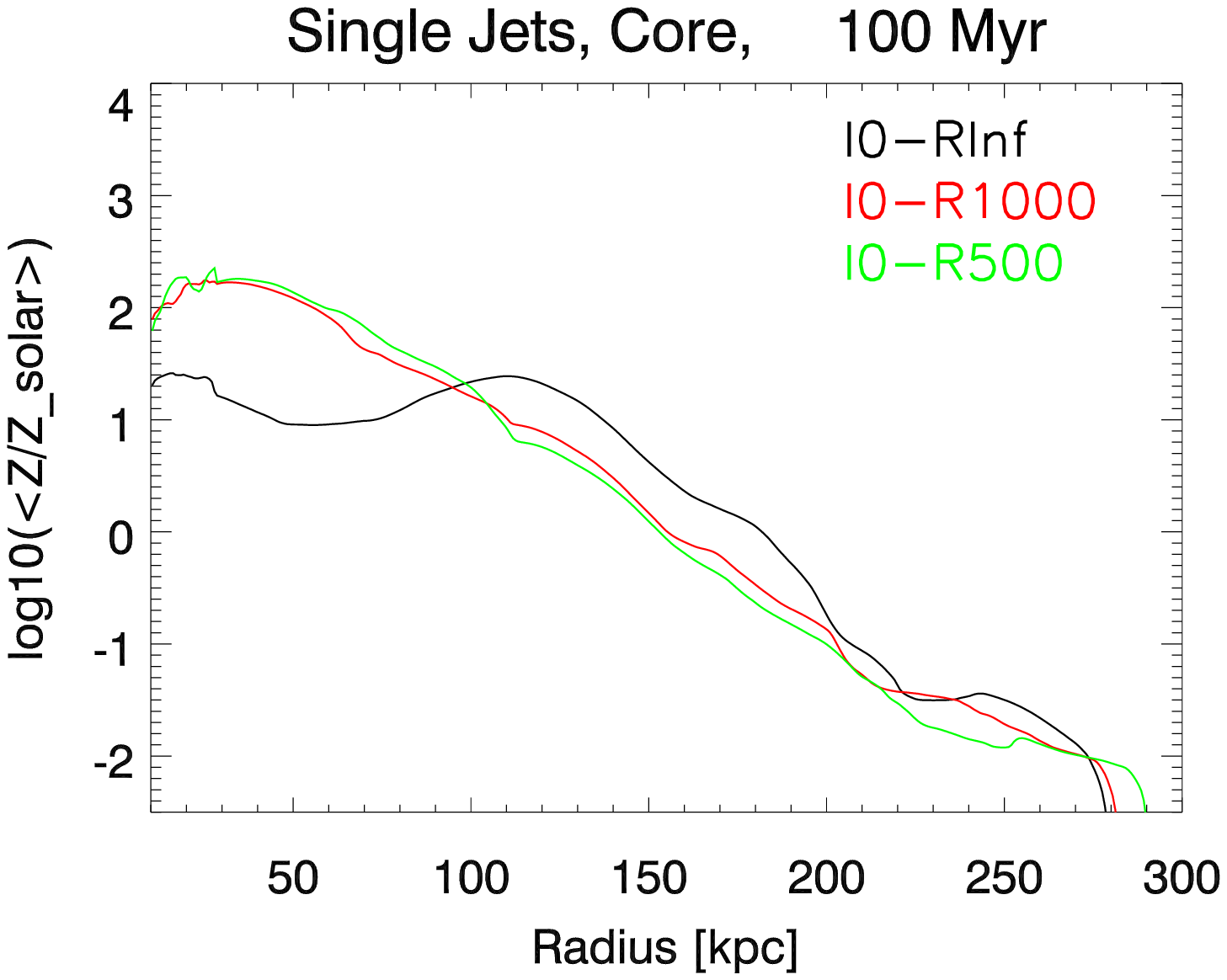} 
	\hspace*{\columnsep}%
  \includegraphics[bb=65 20 483 355, scale=0.34, clip=]{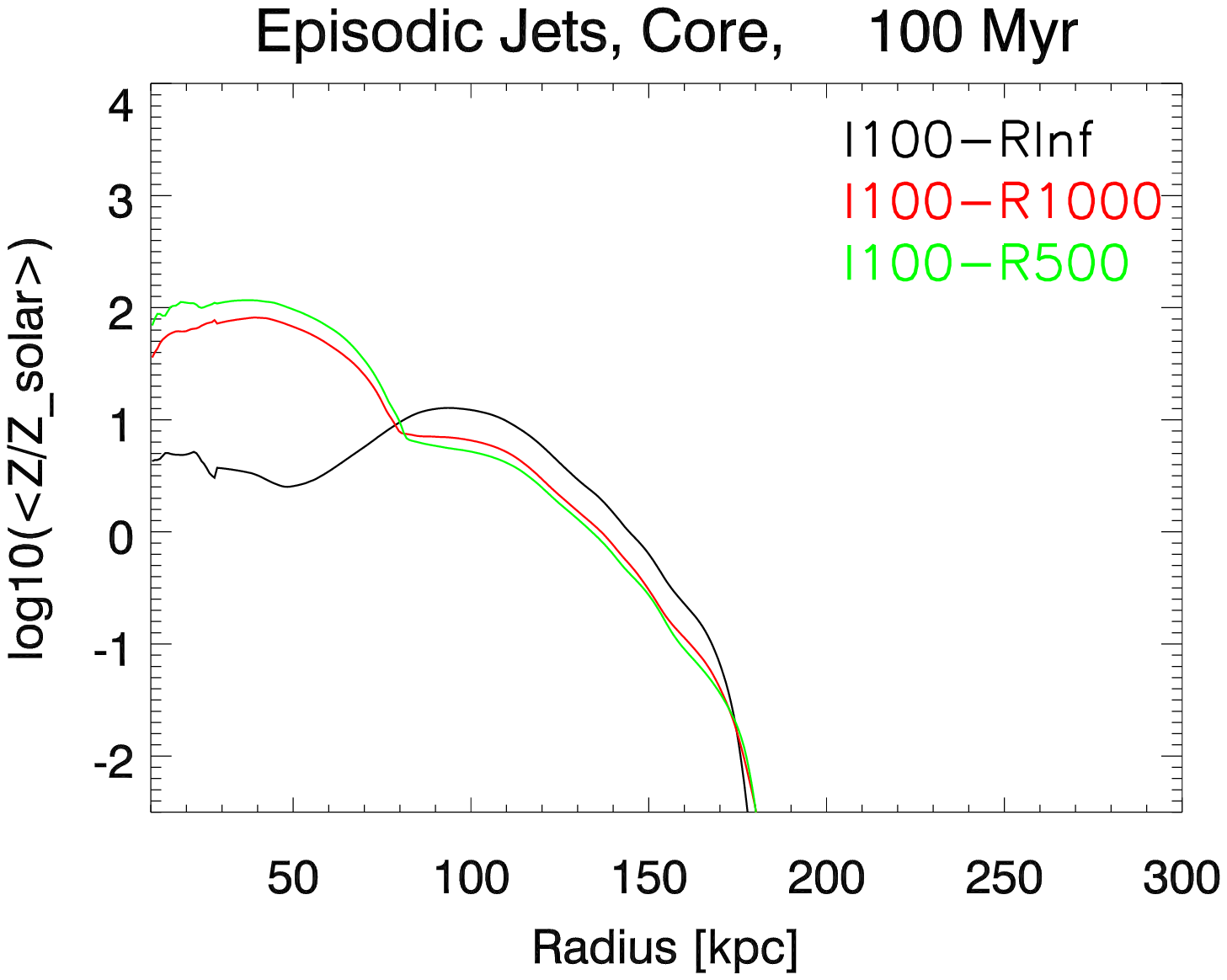} \\
  \includegraphics[bb=65 20 483 355, scale=0.34, clip=]{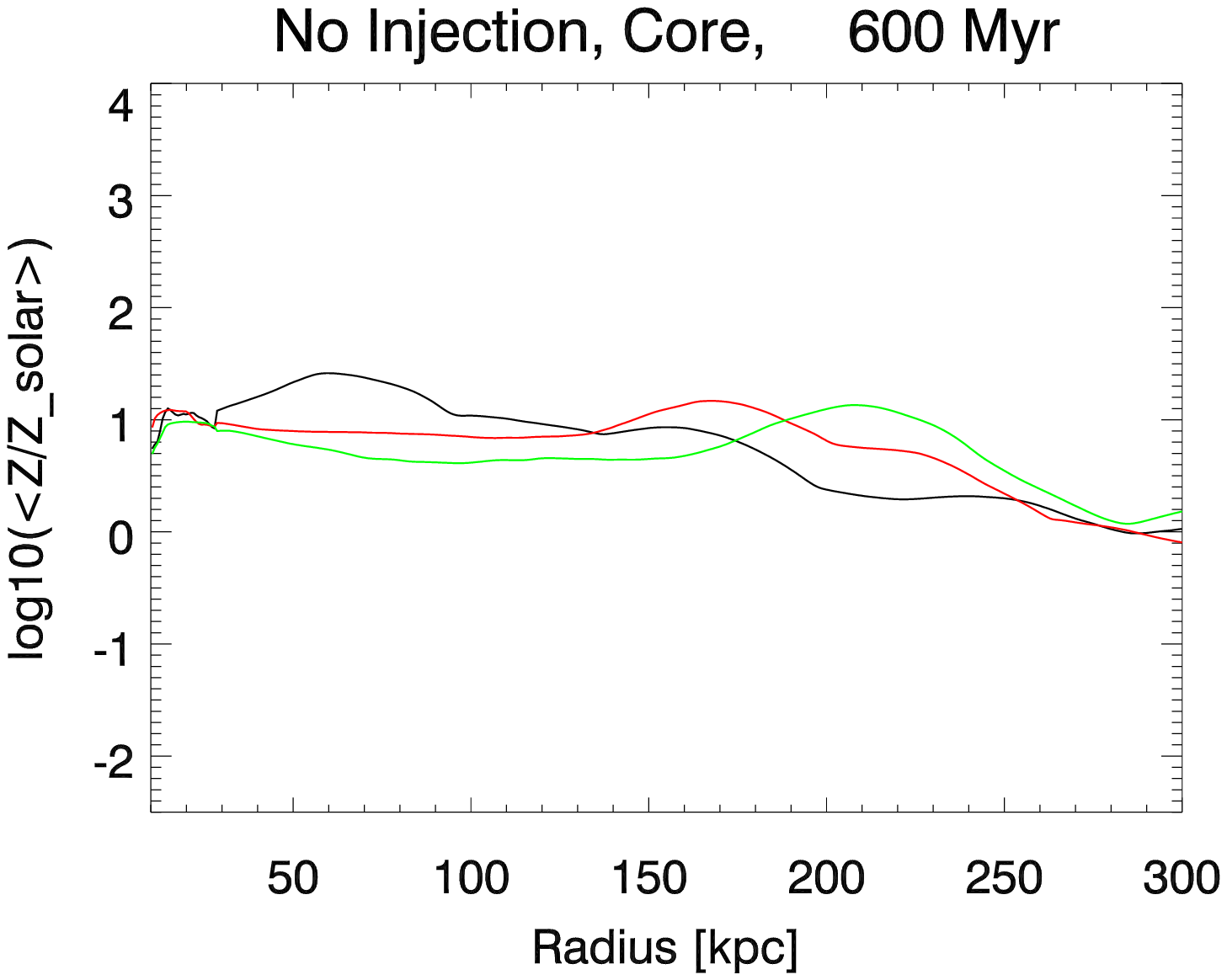} 
	\hspace*{\columnsep}%
  \includegraphics[bb=65 20 483 355, scale=0.34, clip=]{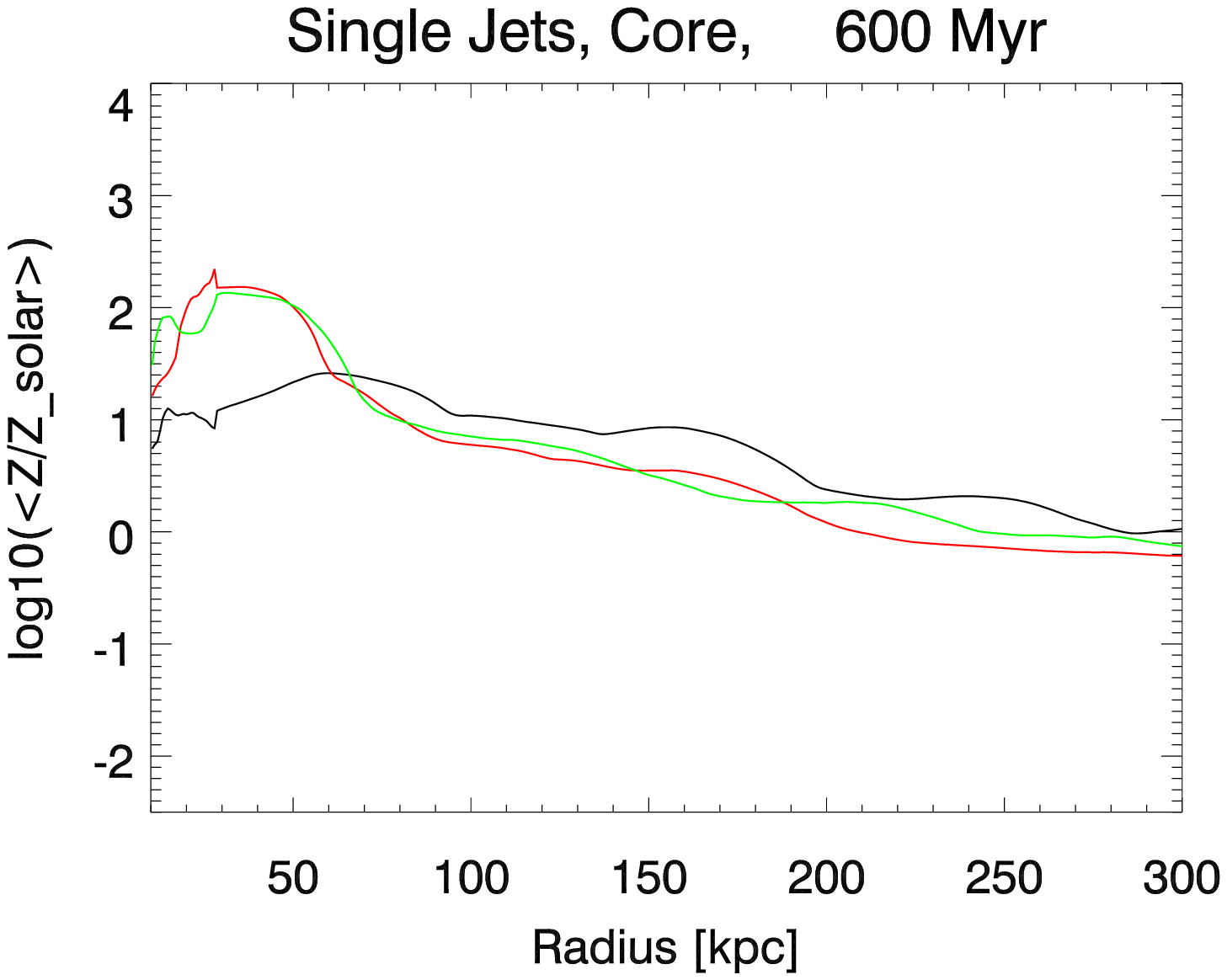} 
	\hspace*{\columnsep}%
  \includegraphics[bb=65 20 483 355, scale=0.34, clip=]{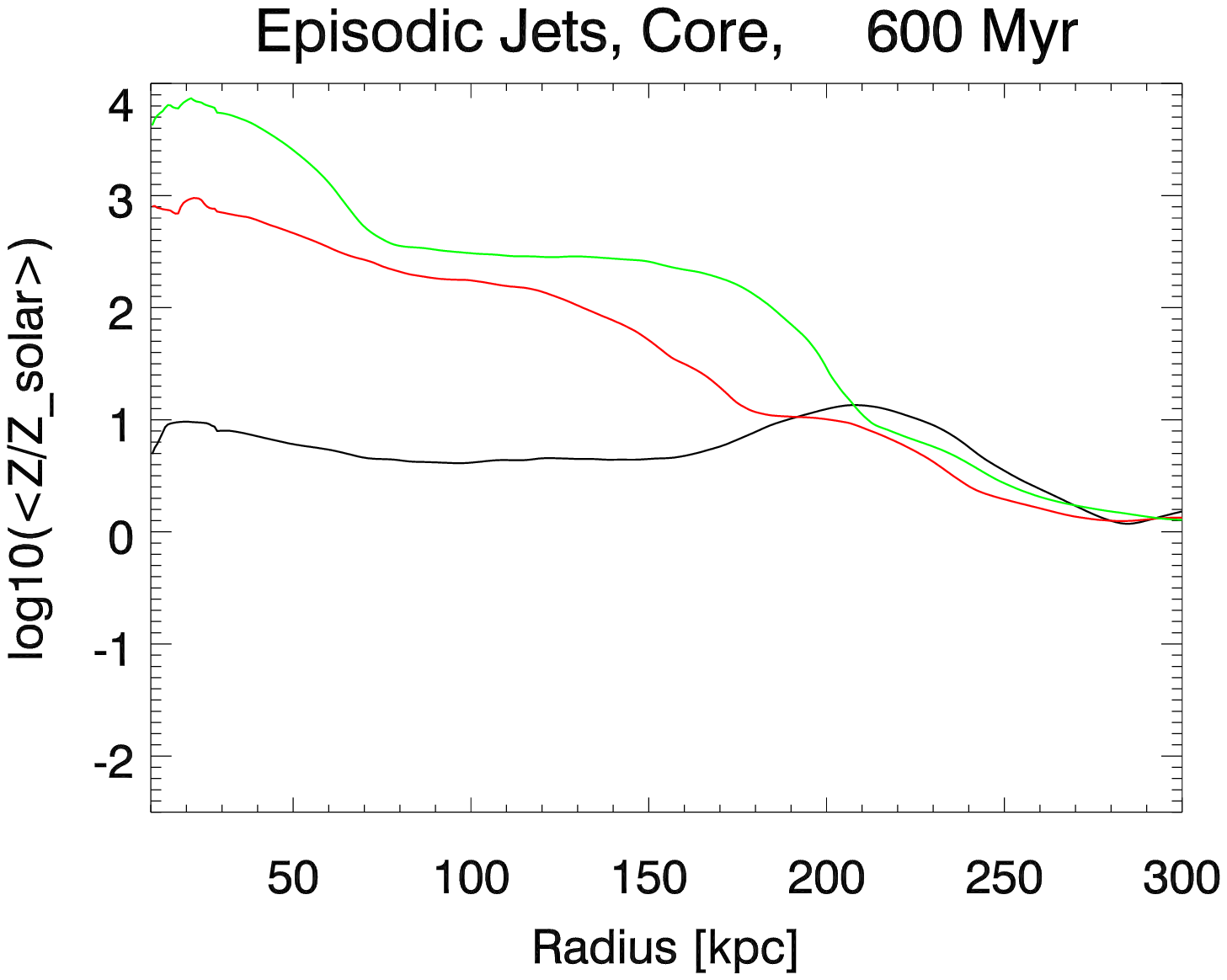} \\
  \includegraphics[bb=65 20 483 355, scale=0.34, clip=]{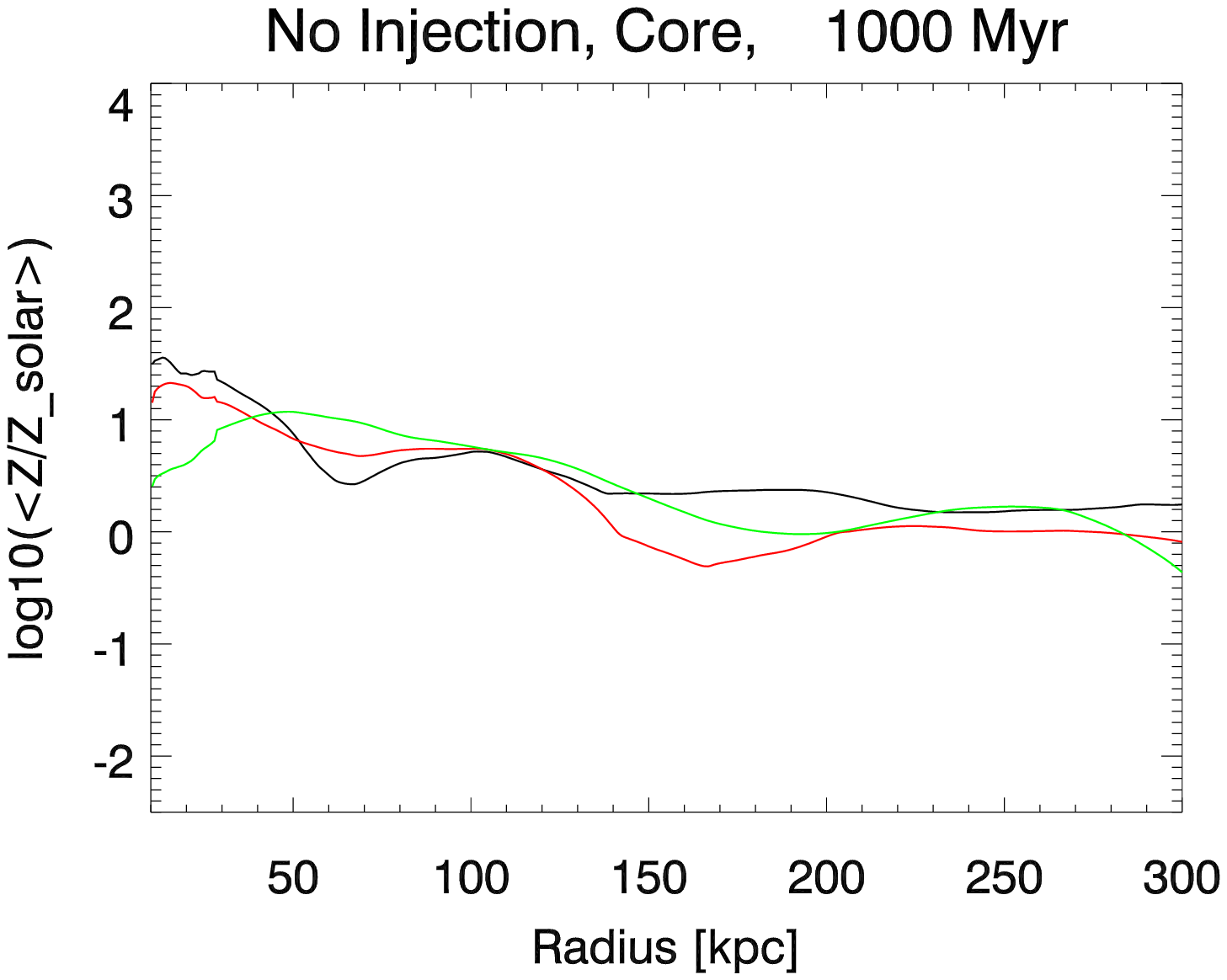} 
	\hspace*{\columnsep}%
  \includegraphics[bb=65 20 483 355, scale=0.34, clip=]{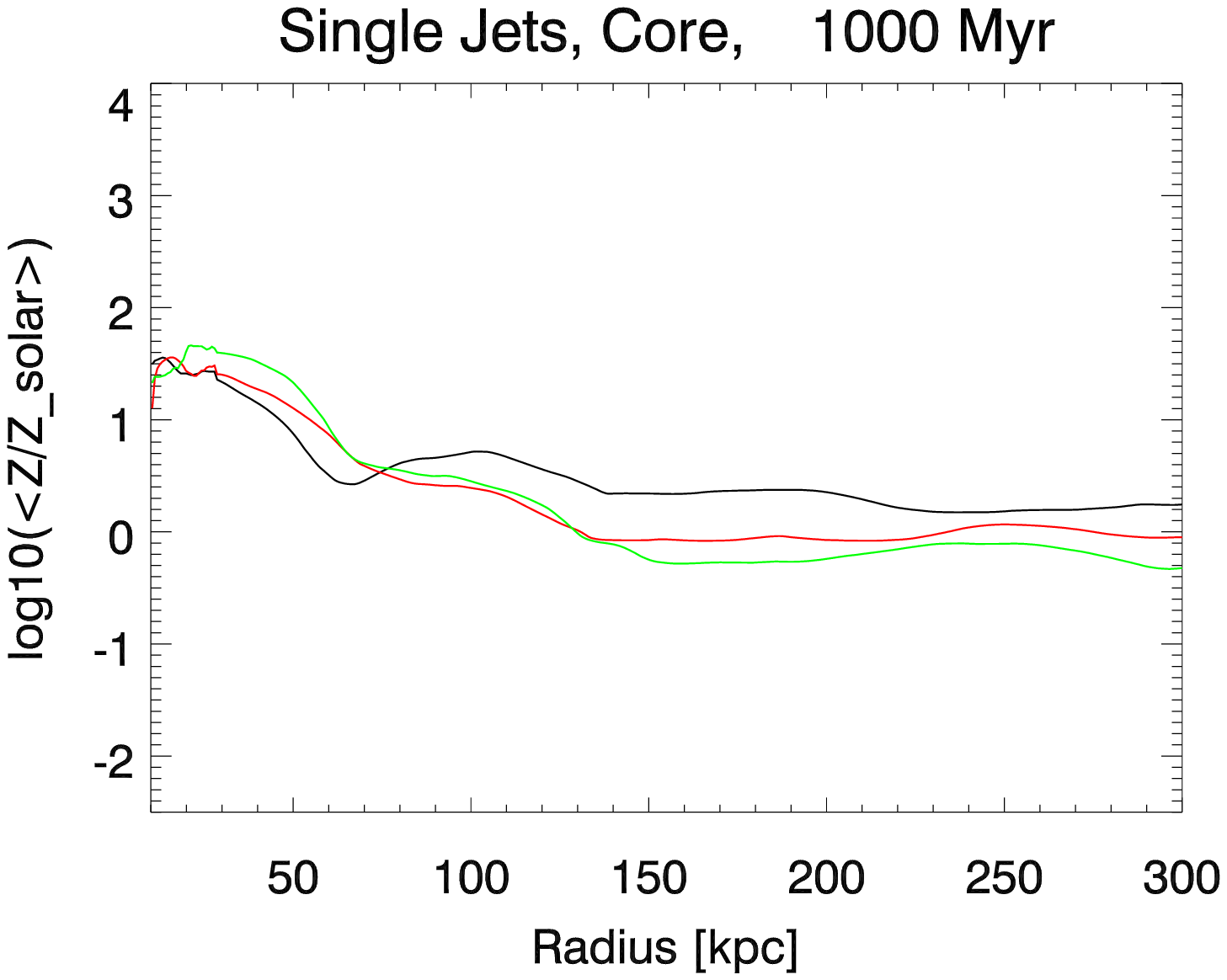} 
	\hspace*{\columnsep}%
  \includegraphics[bb=65 20 483 355, scale=0.34, clip=]{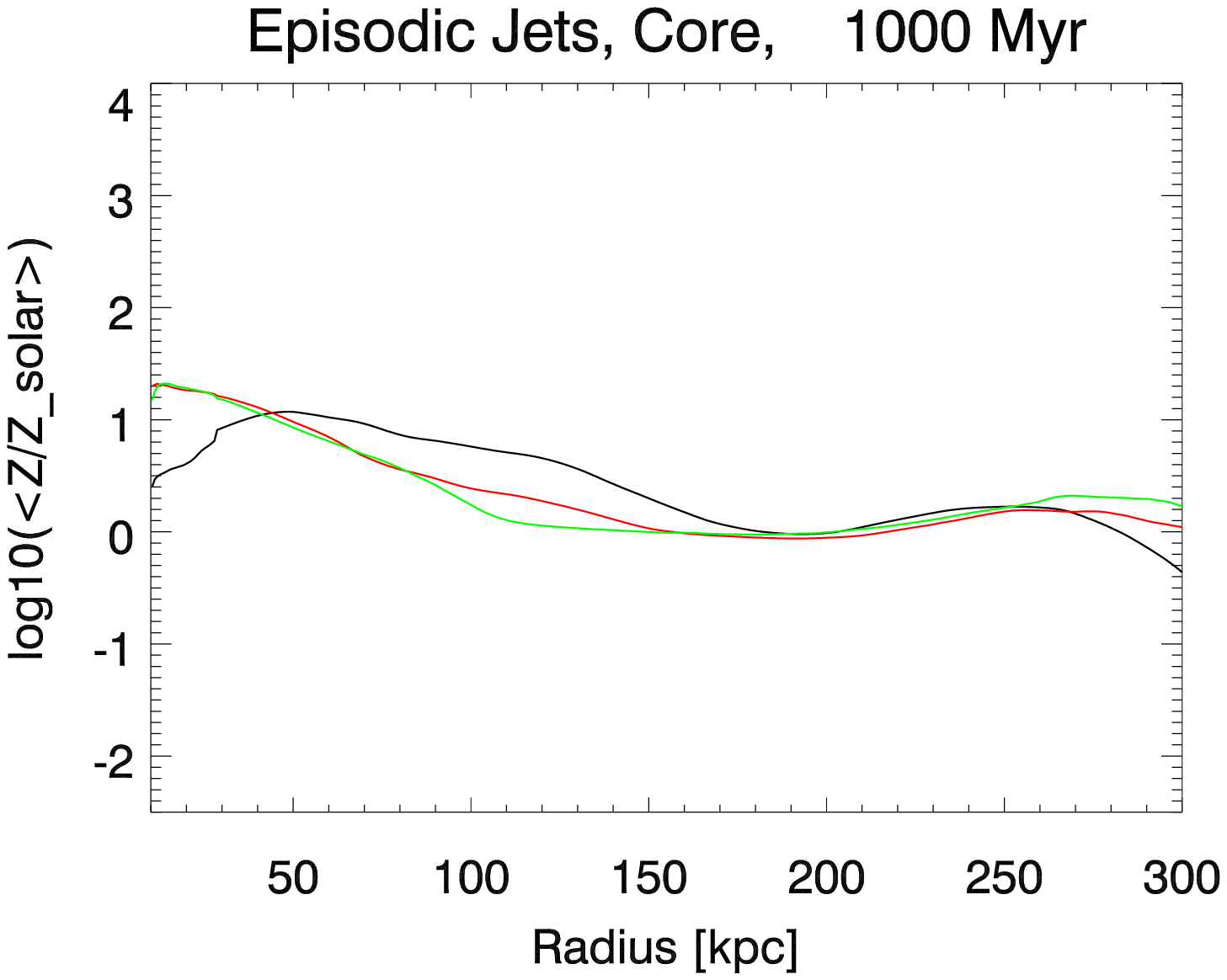} \\
  \includegraphics[bb=65 20 483 355, scale=0.34, clip=]{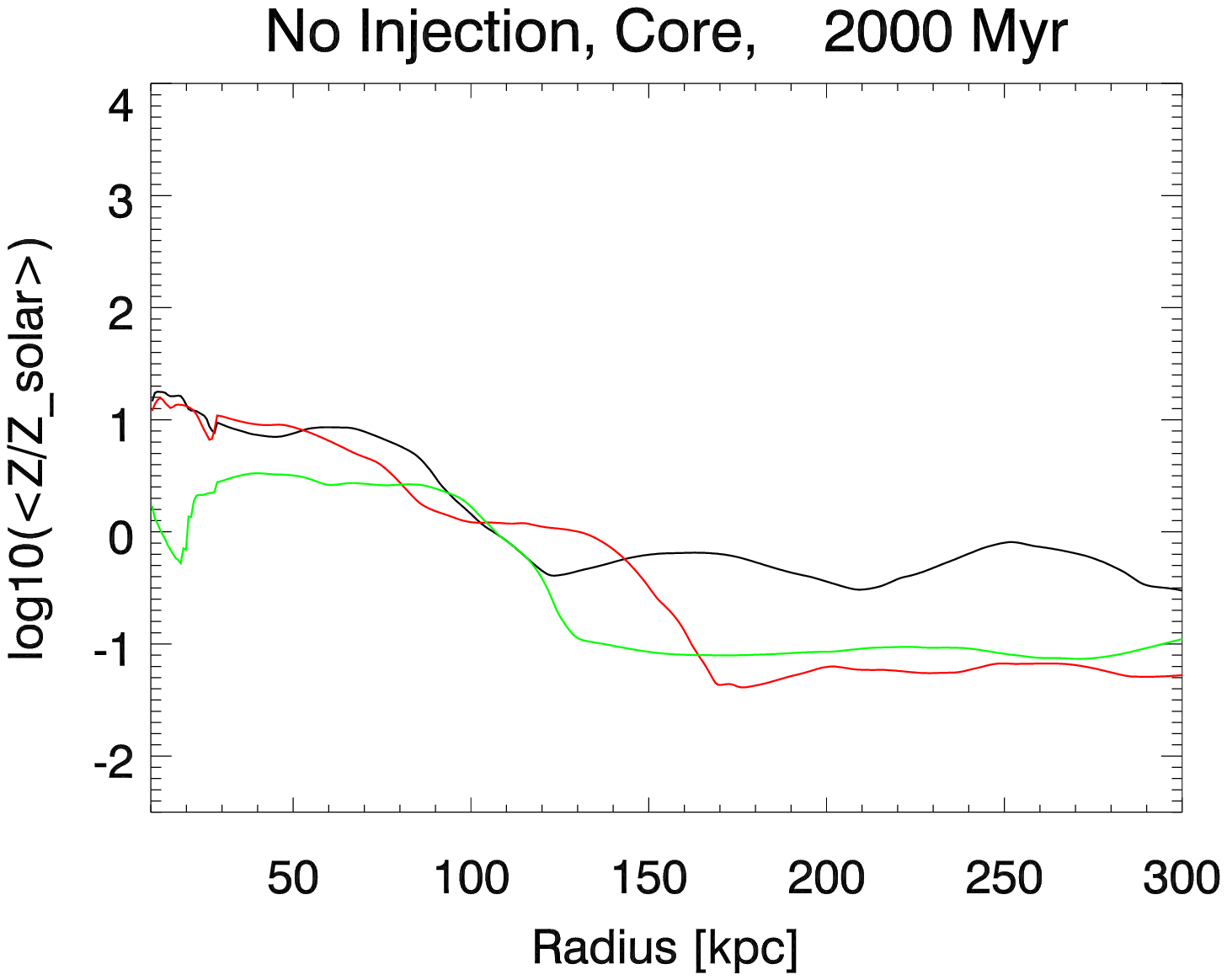} 
	\hspace*{\columnsep}%
  \includegraphics[bb=65 20 483 355, scale=0.34, clip=]{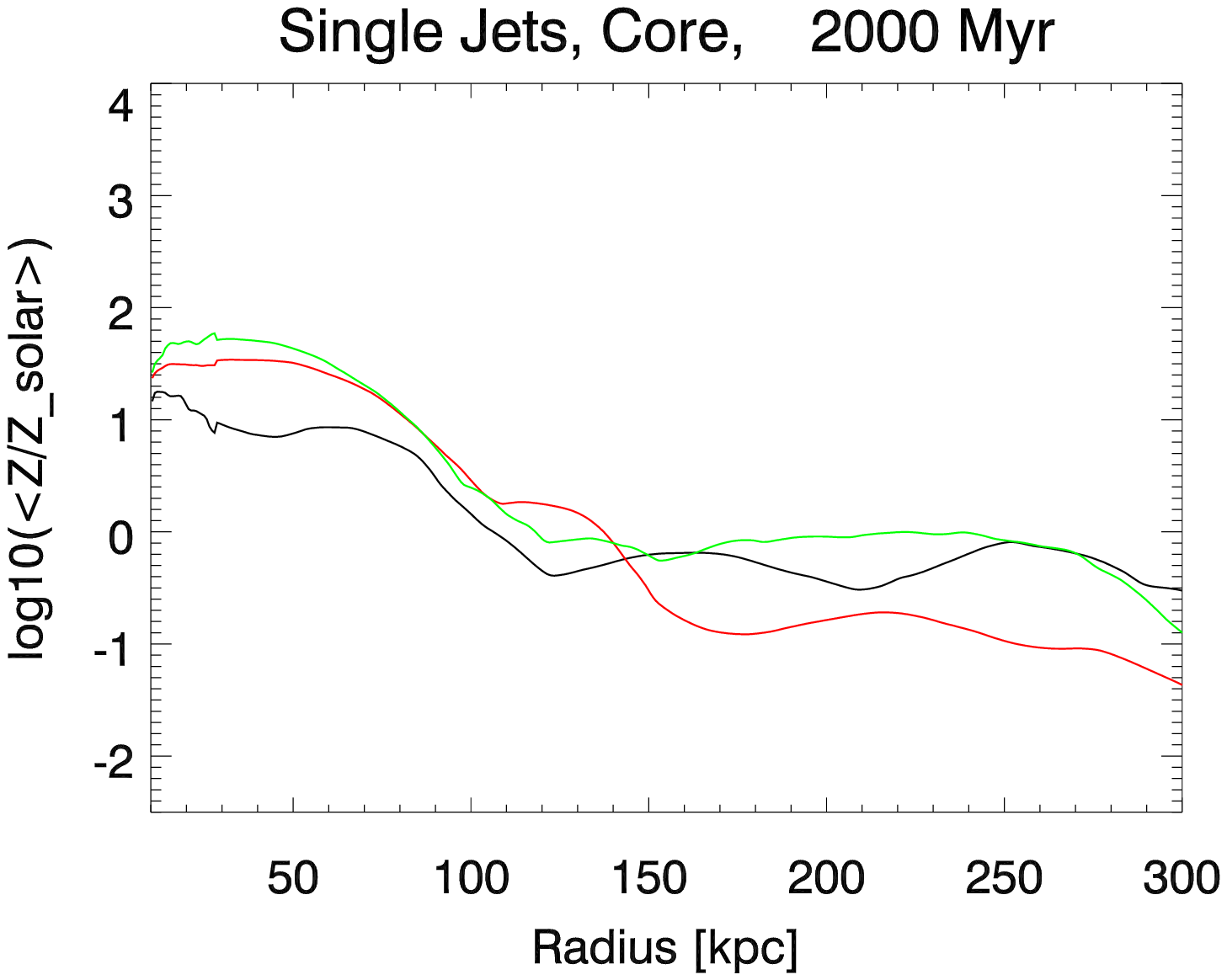} 
	\hspace*{\columnsep}%
  \includegraphics[bb=65 20 483 355, scale=0.34, clip=]{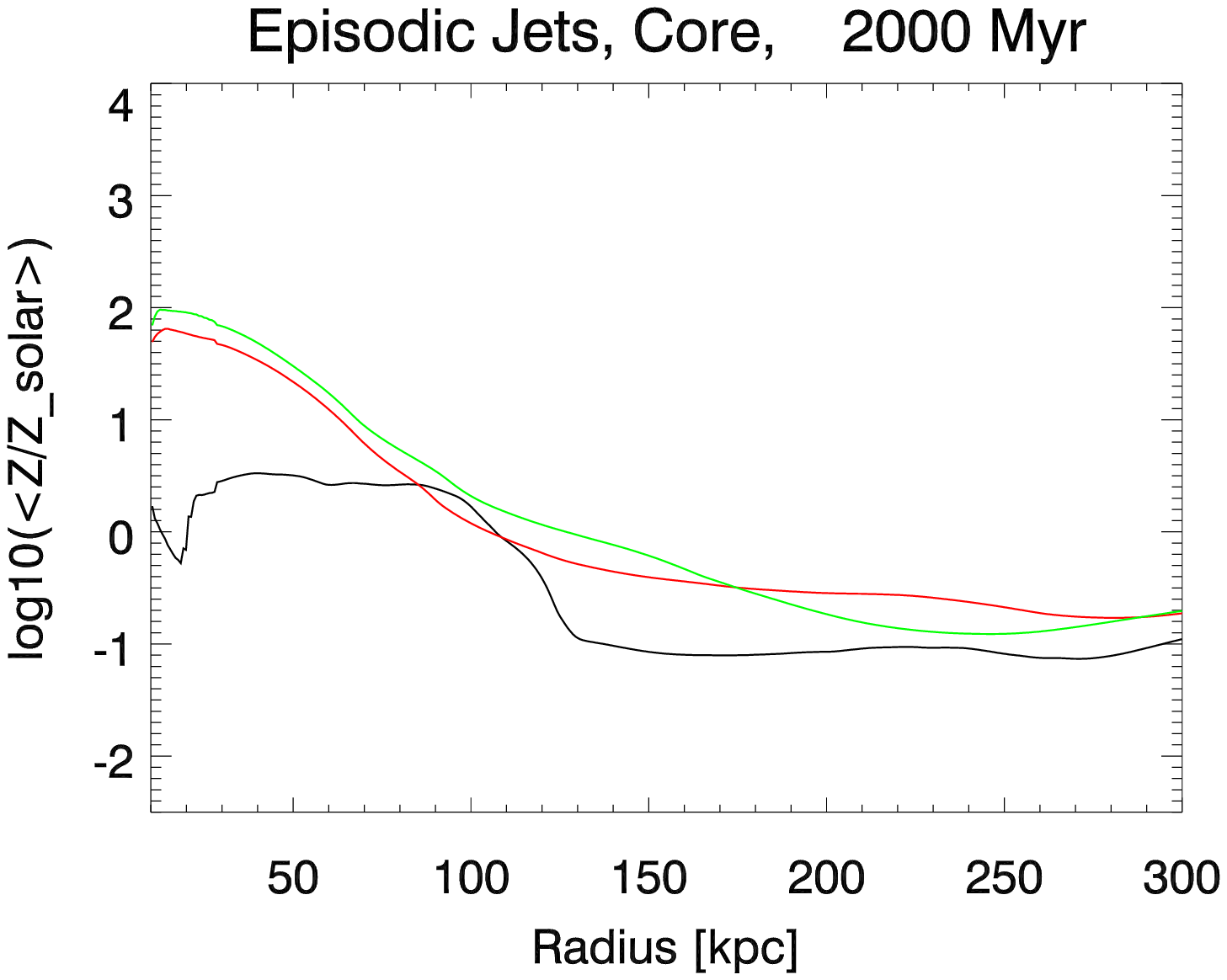} \\
  \includegraphics[bb=65 20 483 355, scale=0.34, clip=]{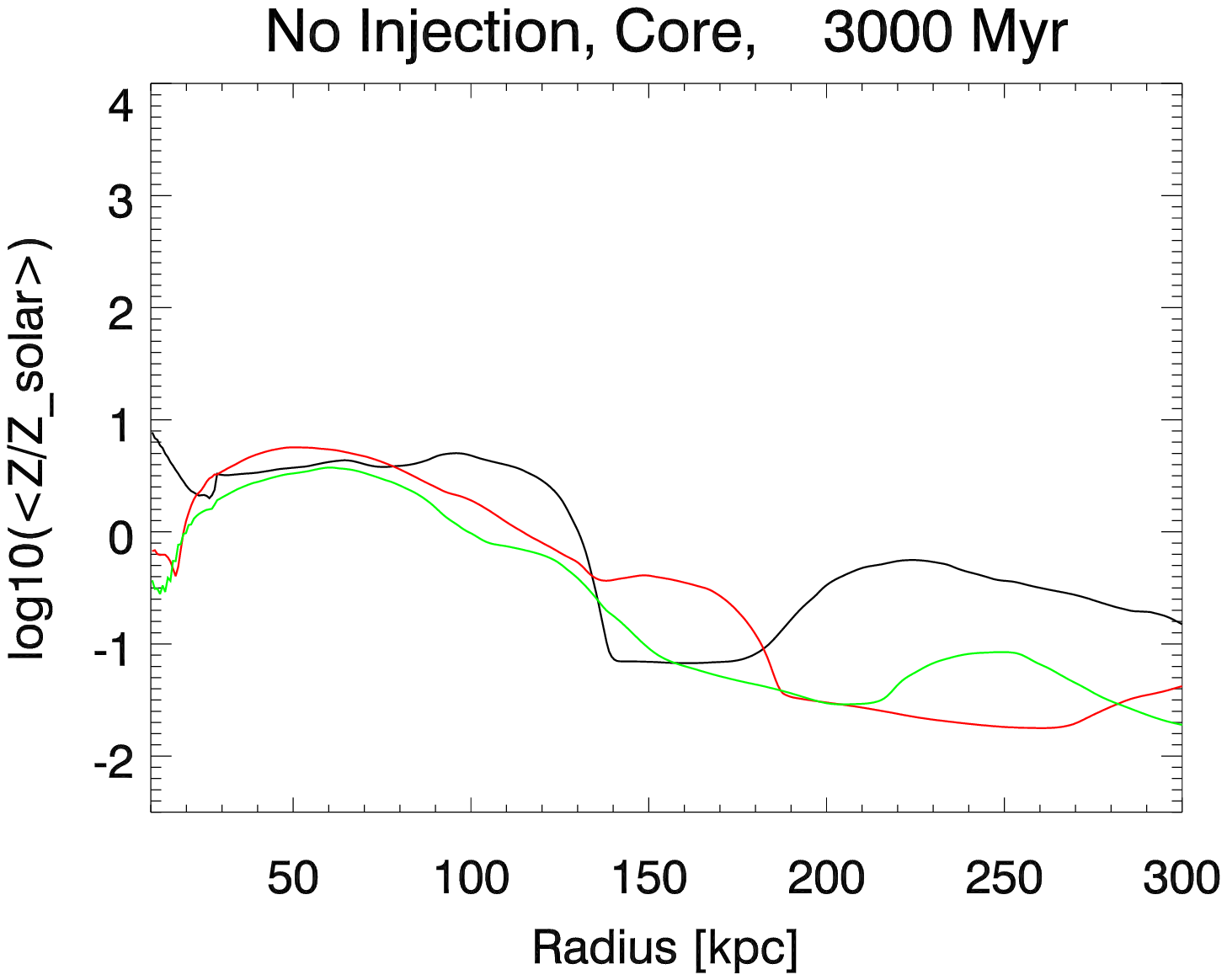} 
	\hspace*{\columnsep}%
  \includegraphics[bb=65 20 483 355, scale=0.34, clip=]{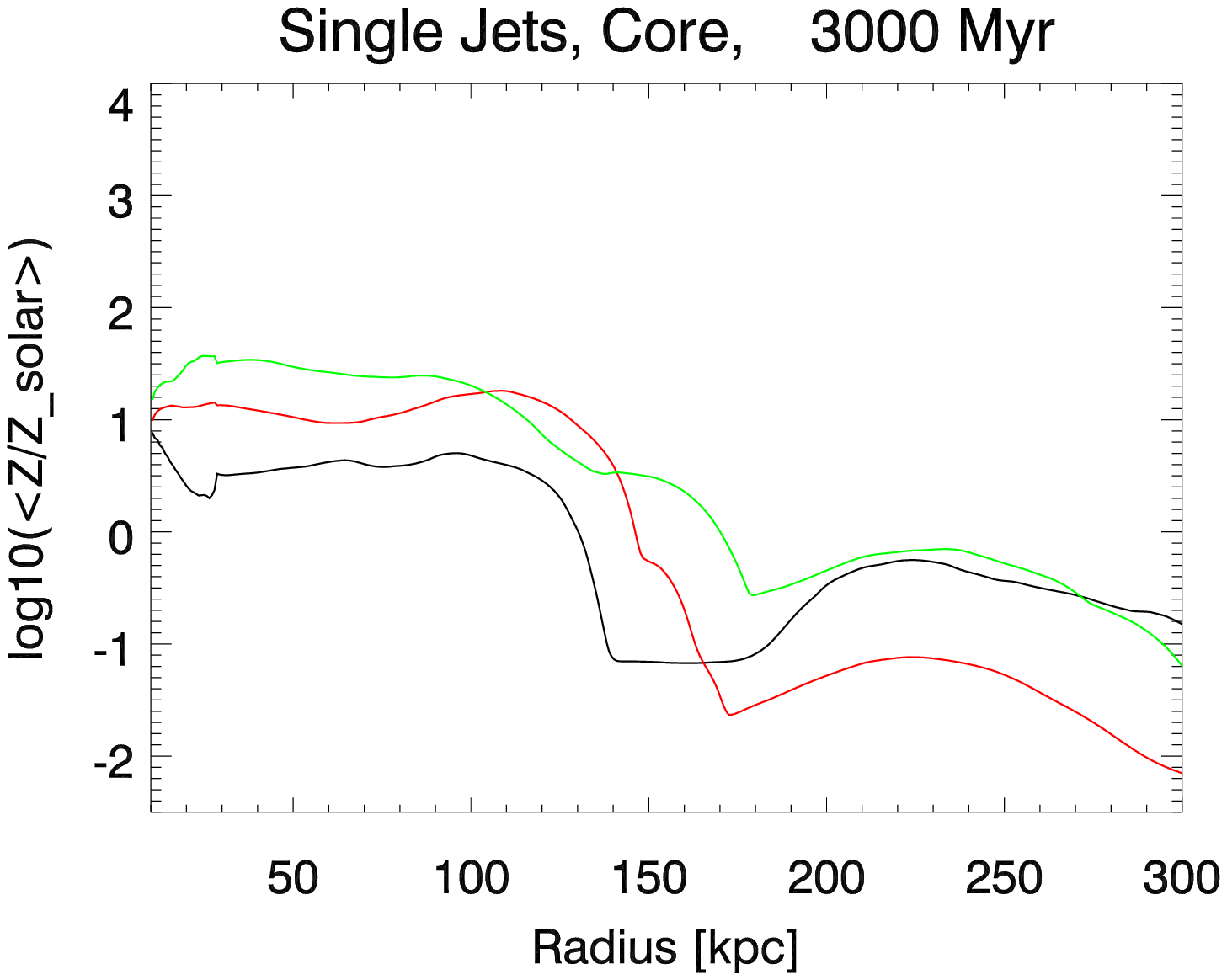} 
	\hspace*{\columnsep}%
  \includegraphics[bb=65 20 483 355, scale=0.34, clip=]{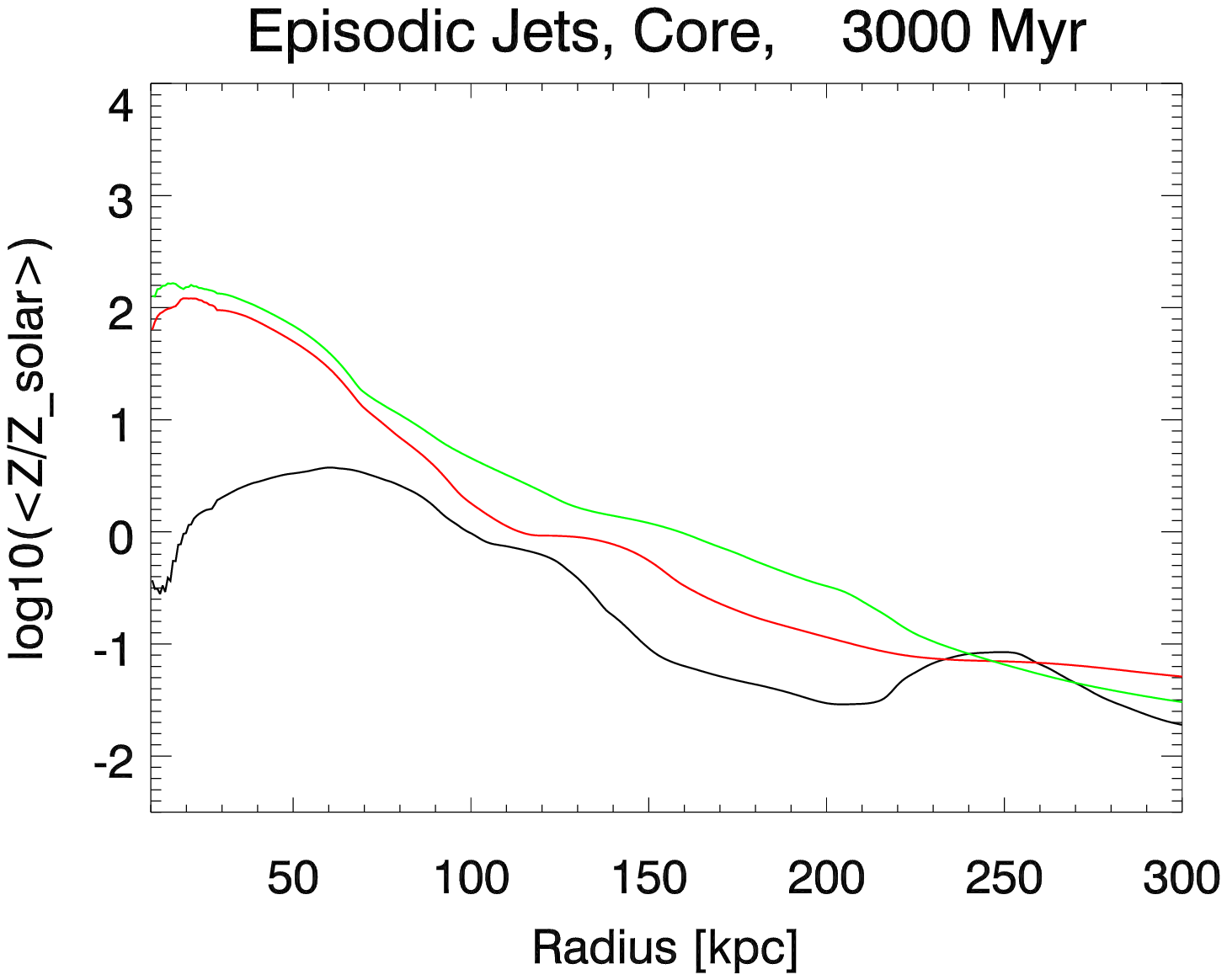} \\ 
  \caption{A zoom into the central 300\,kpc of Figure~\ref{fig2}.}
\label{fig3}
\end{figure*}

\begin{figure*}
  \vspace*{0pt}
  \includegraphics[bb=65 20 483 355, scale=0.34, clip=]{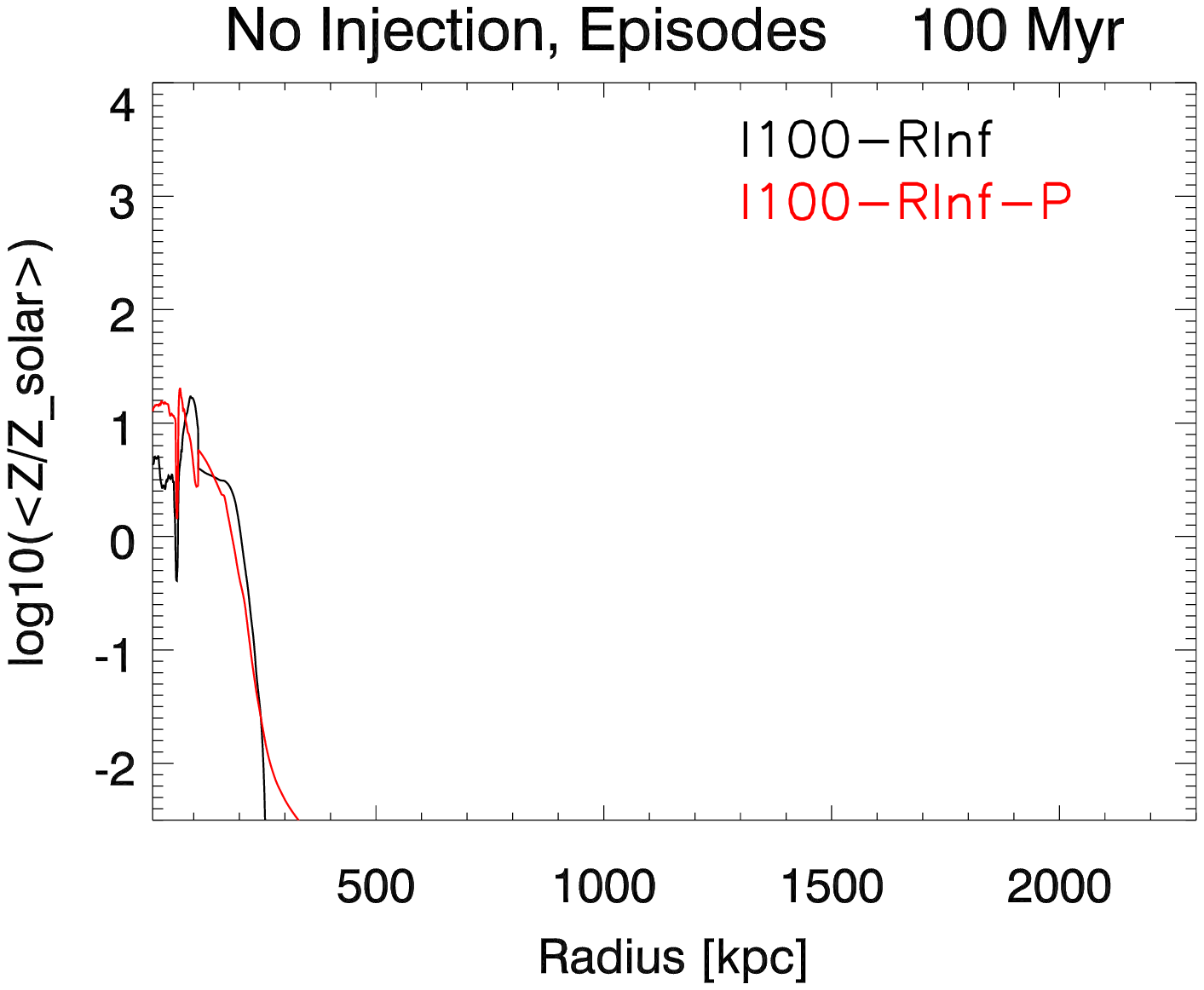} 
	\hspace*{\columnsep}%
  \includegraphics[bb=65 20 483 355, scale=0.34, clip=]{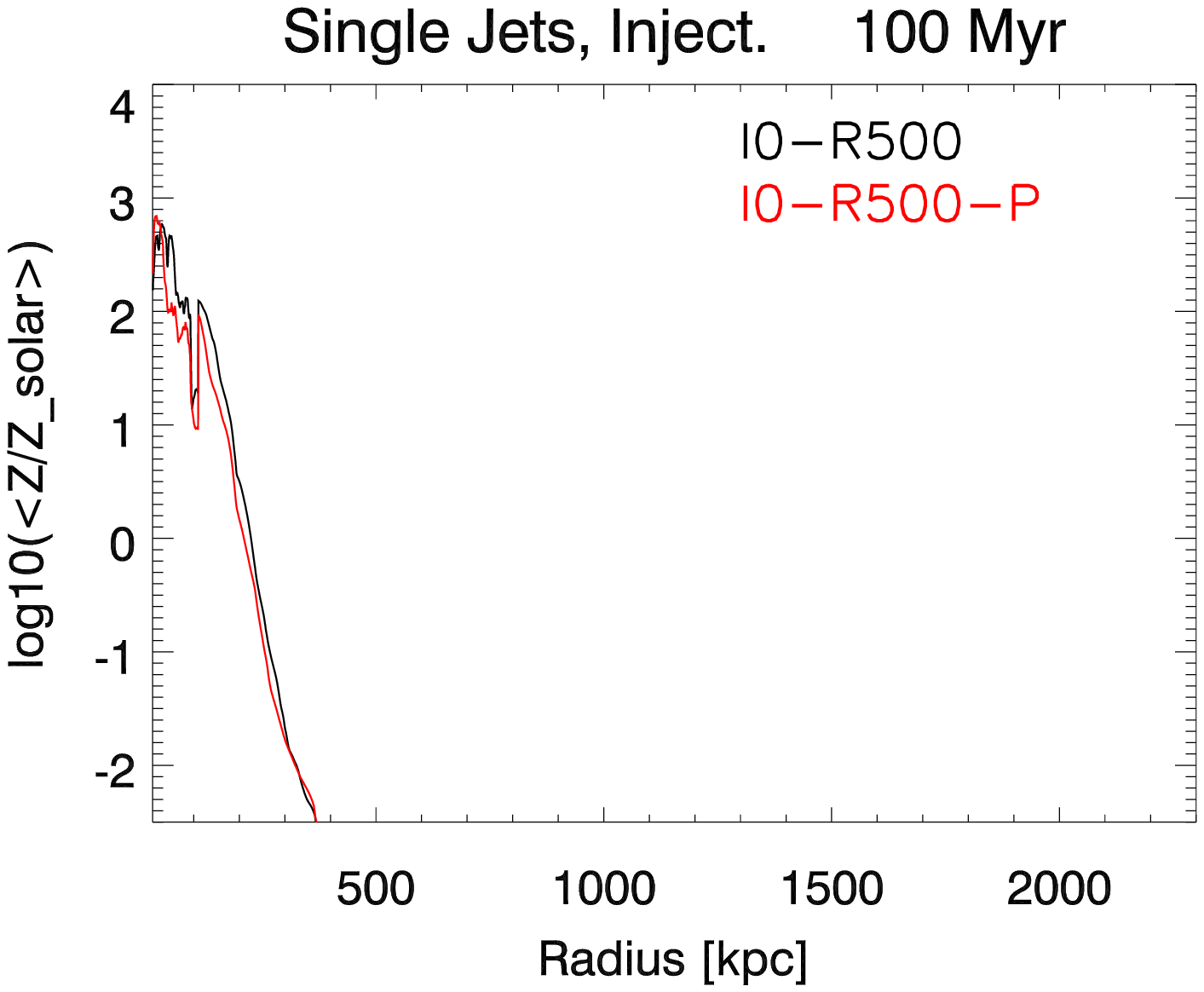} 
	\hspace*{\columnsep}%
  \includegraphics[bb=65 20 483 355, scale=0.34, clip=]{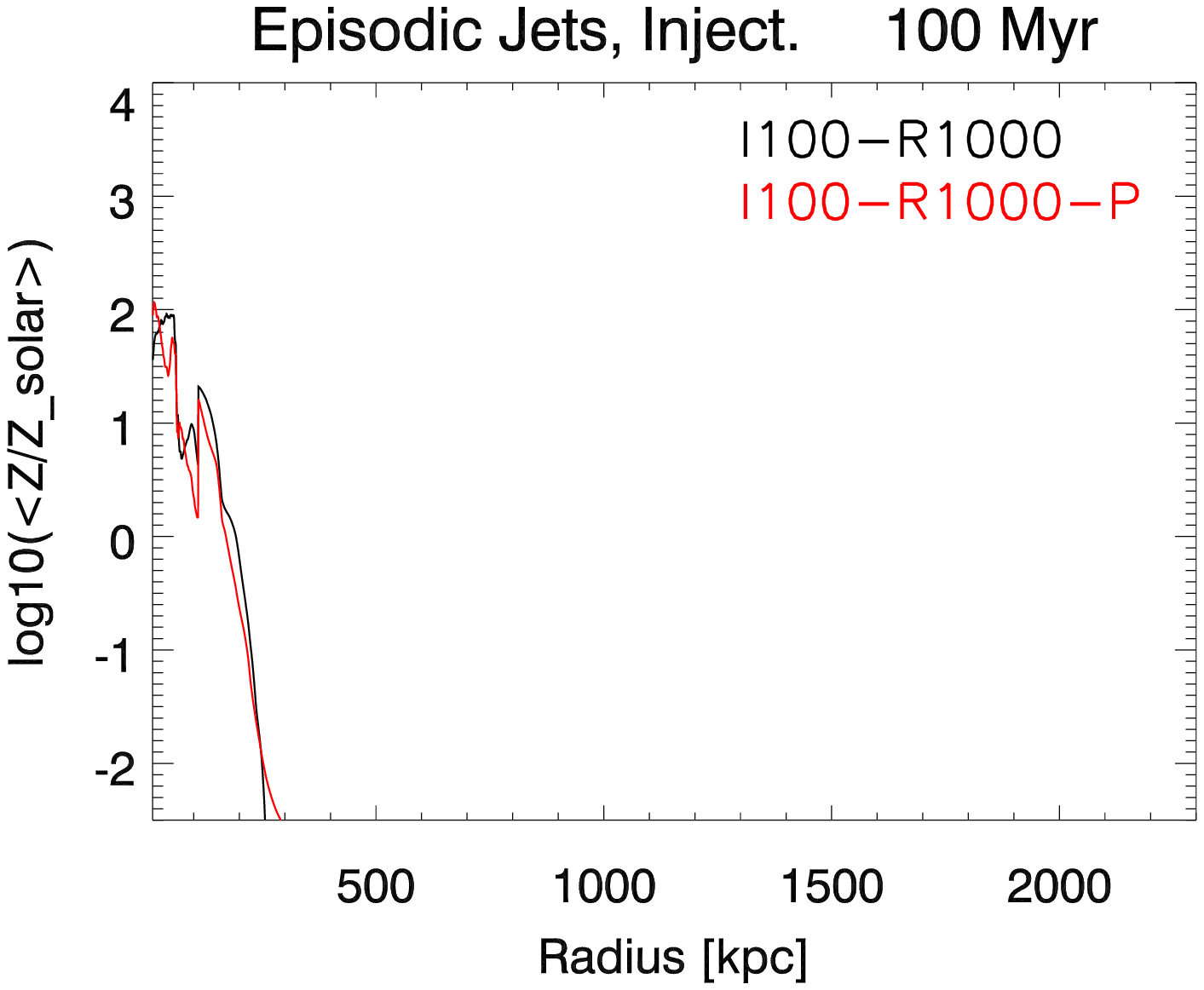} \\
  \includegraphics[bb=65 20 483 355, scale=0.34, clip=]{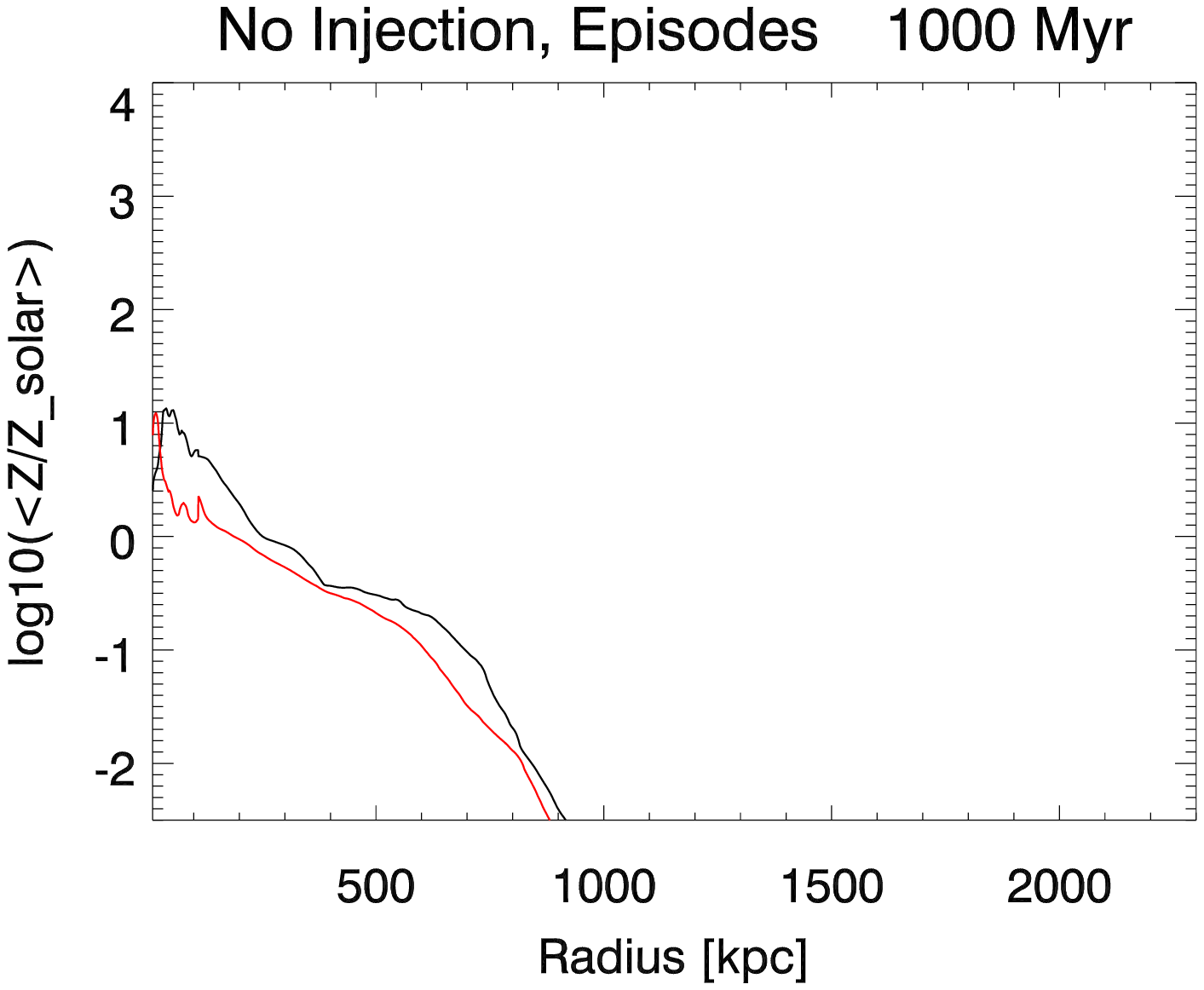} 
	\hspace*{\columnsep}%
  \includegraphics[bb=65 20 483 355, scale=0.34, clip=]{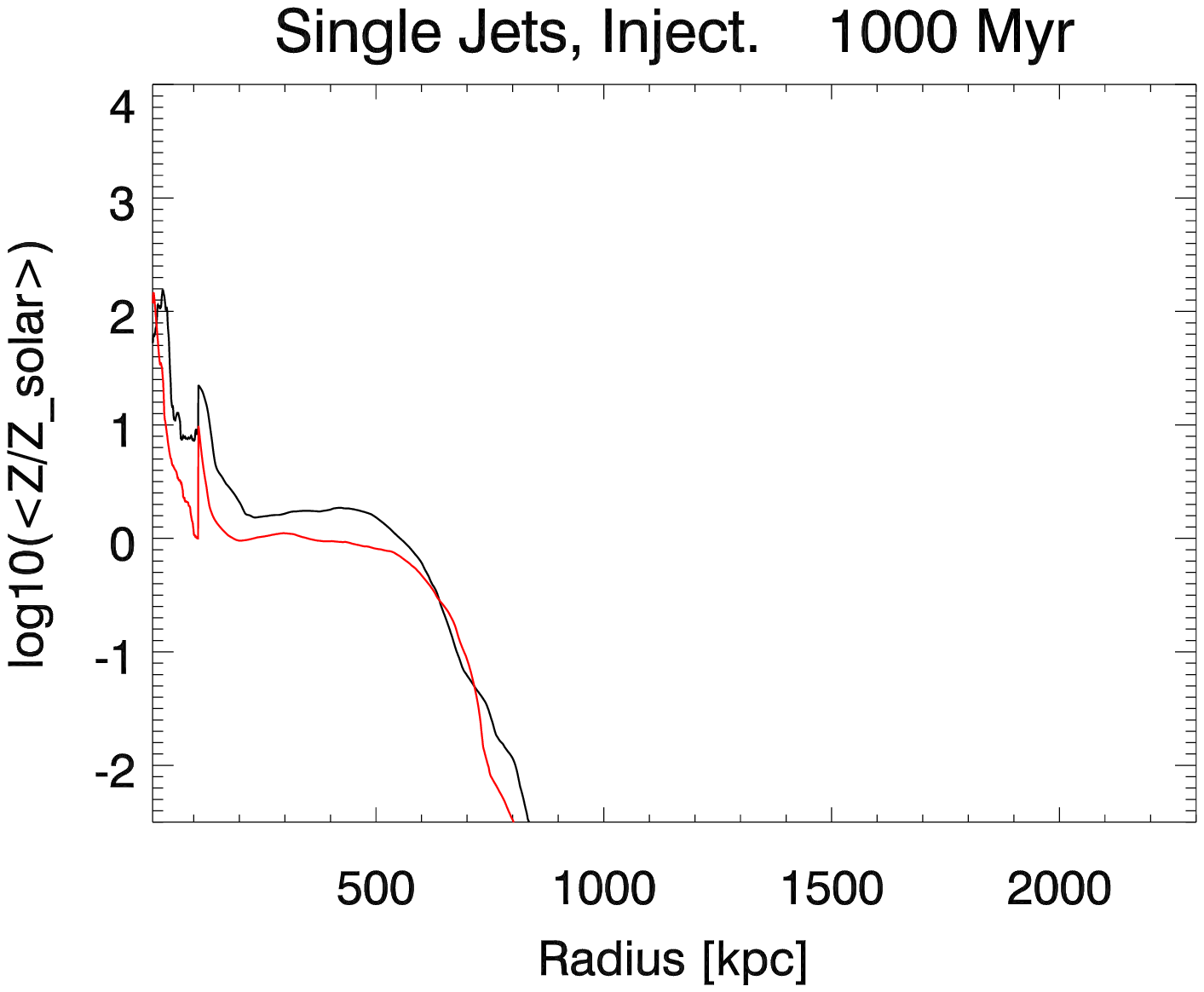} 
	\hspace*{\columnsep}%
  \includegraphics[bb=65 20 483 355, scale=0.34, clip=]{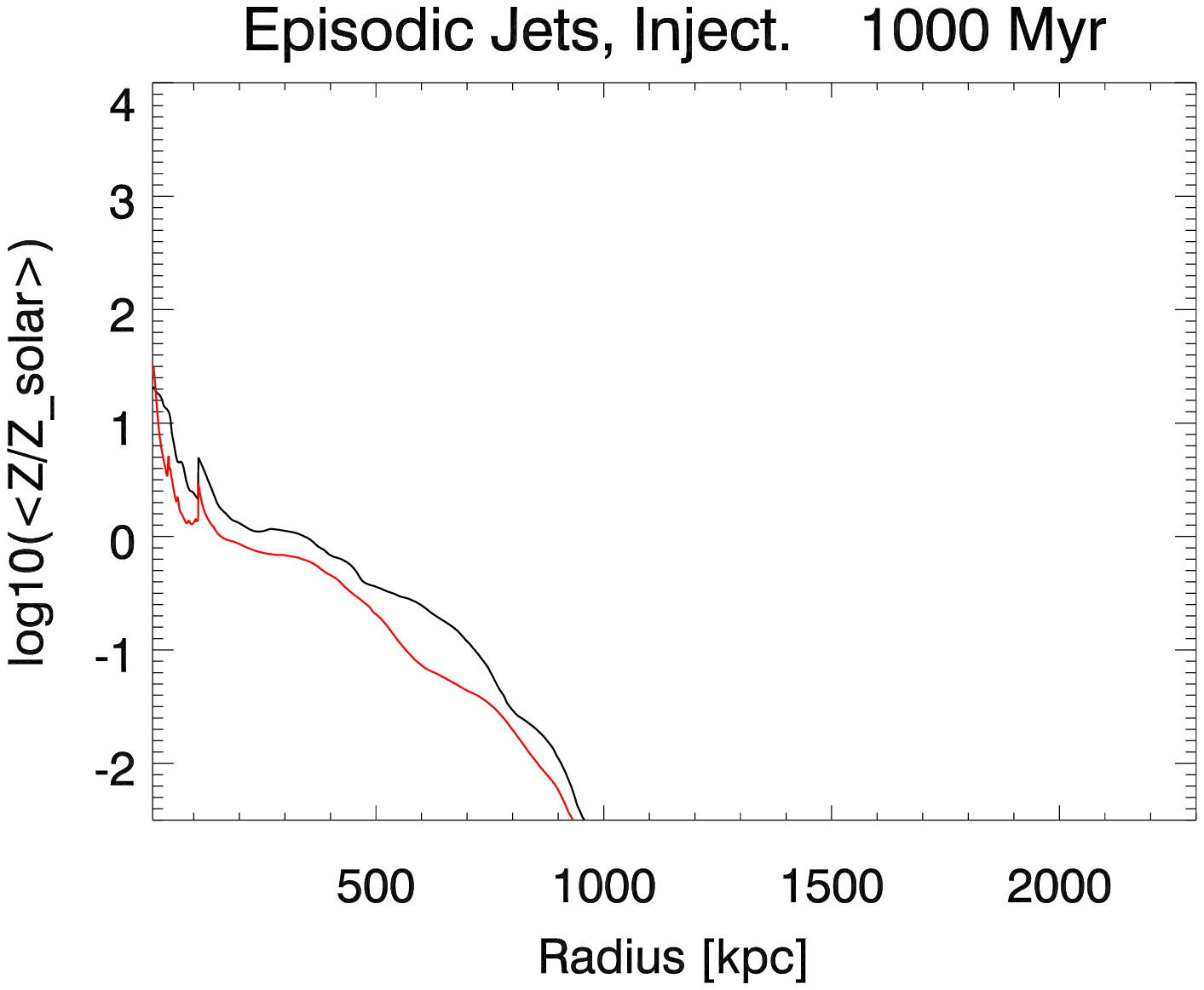} \\
  \includegraphics[bb=65 20 483 355, scale=0.34, clip=]{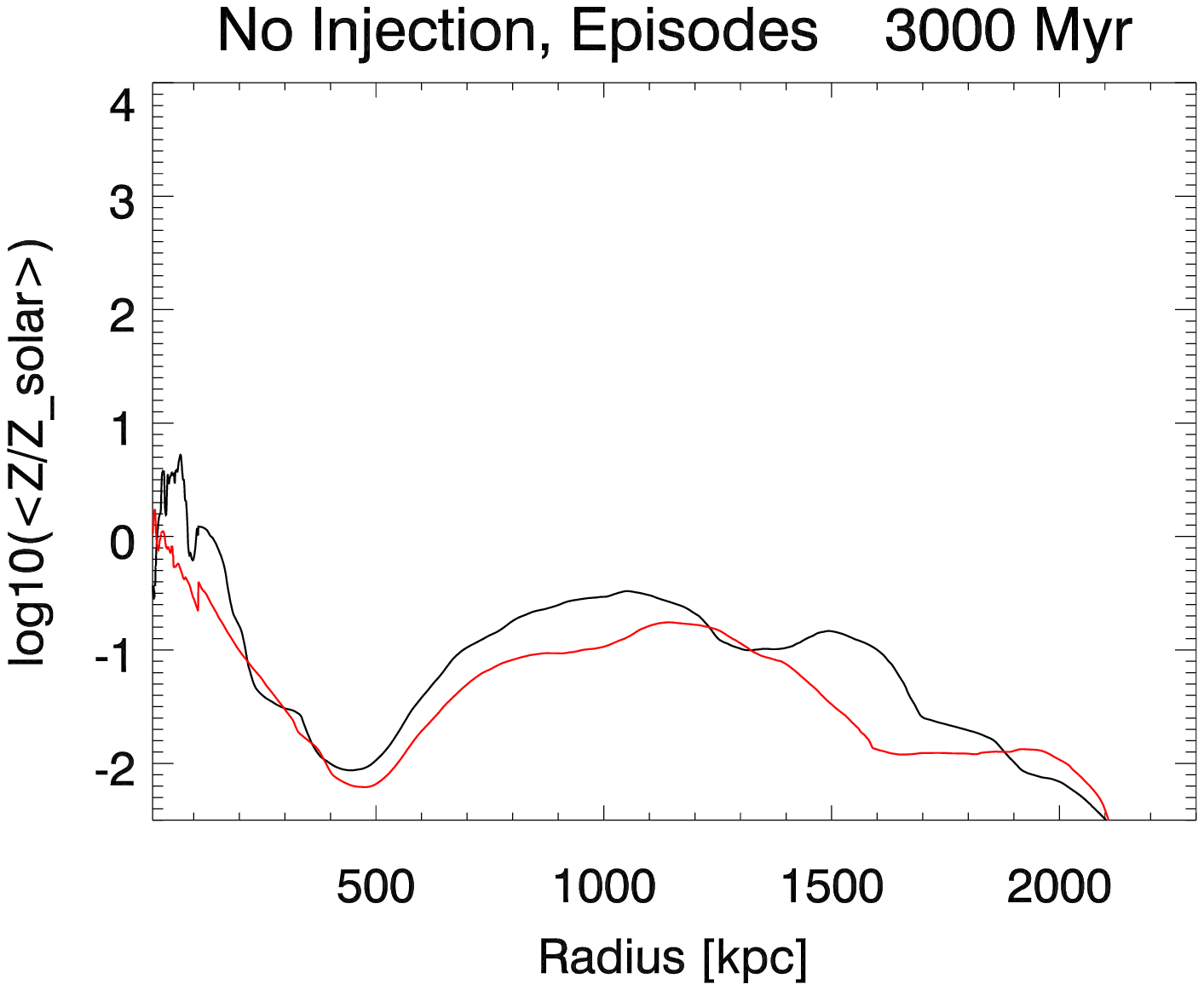} 
	\hspace*{\columnsep}%
  \includegraphics[bb=65 20 483 355, scale=0.34, clip=]{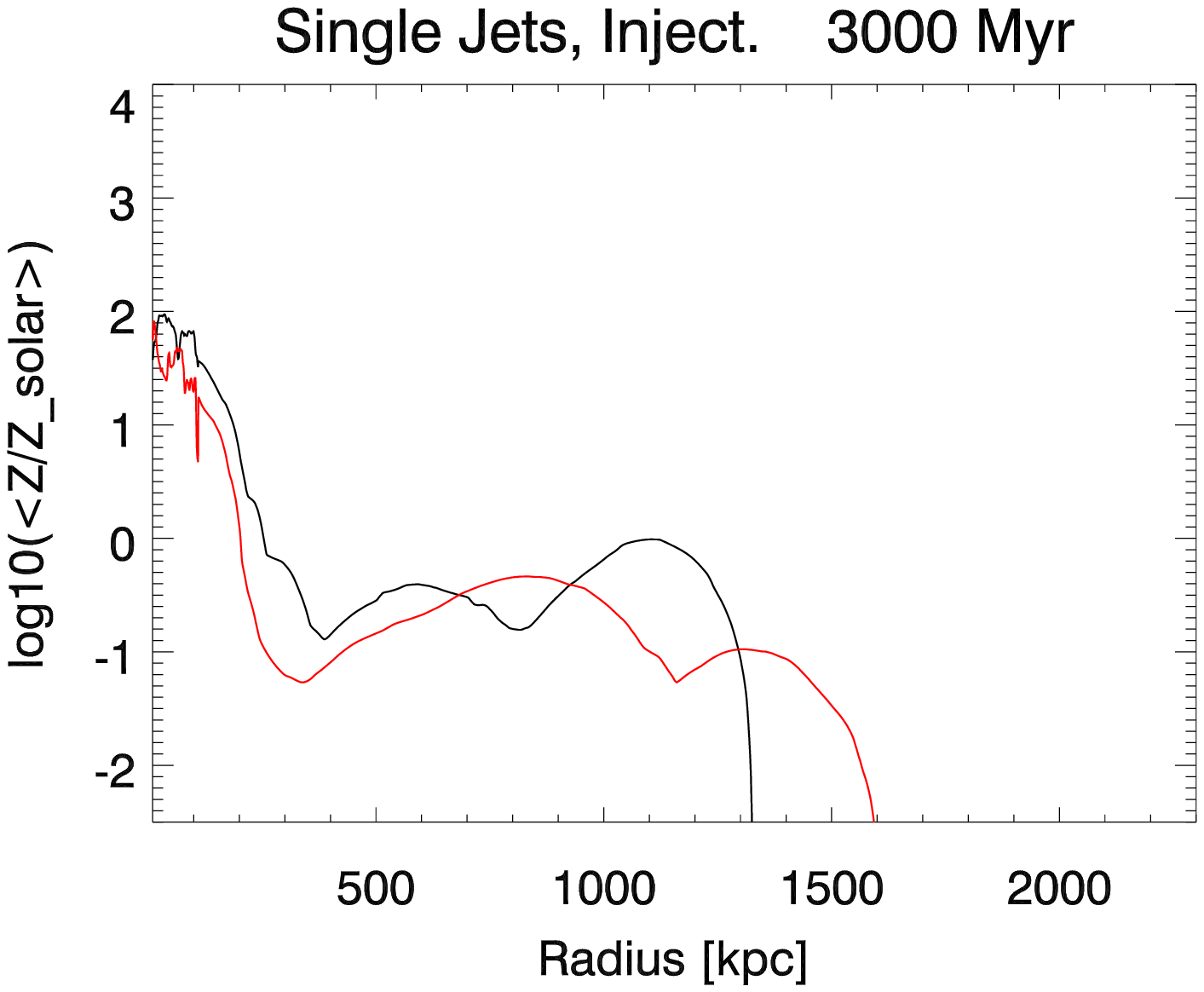} 
	\hspace*{\columnsep}%
  \includegraphics[bb=65 20 483 355, scale=0.34, clip=]{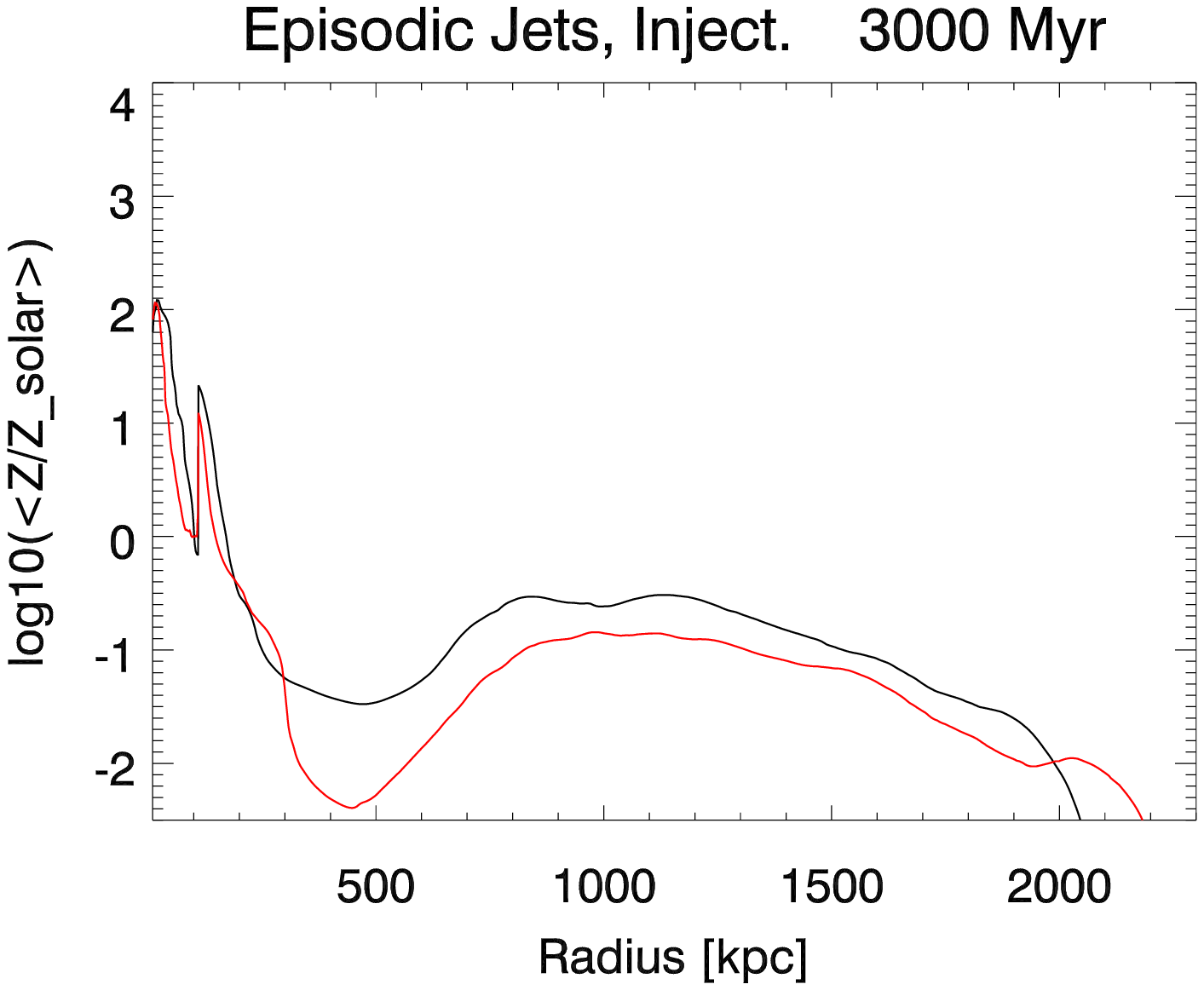} 
  \caption{Time evolution of the logarithm of the mean relative metallicities 
showing the differences 
	between Runs with a flat (black) and a Plumer (red) metal distributions.
The dynamical evolution of the ICM metallicity is independent of the metal 
distribution details at the central dominant galaxy.}
\label{fig4}
\end{figure*}

Figure~\ref{fig2} shows the (logarithm) azimuthal average of the relative metallicity 
of the ICM gas density, as functions of radial distance from the cluster's core at 
different epochs. The relative metallicity in a given shell of thickness $dr$ is 
calculated as:

\begin{equation}
\frac{Z(r)}{Z_0} = 
\frac{1}{Z_0} \frac{\int \rho_m(r',\theta) r'^2 sin(\theta) d\theta dr'}
{\int \rho_g(r',\theta) r'^2 sin(\theta) d\theta dr'},
\label{plots}
\end{equation}
where $Z_0$ is an arbitrary normalisation factor which, in Section~\ref{data} below,
we use to compare our simulated metallicity gradients with the observations from 
\citet{leccardi08}. 
The left most column in Figure~\ref{fig2} (hereafter 2L) shows Runs without 
any metal injection and with different jet active phases. The middle column (2M) 
shows Runs with a single 30\,Myr outburst and different metal distributions and 
injection rates. The right most column (2R) displays Runs with the same episodic 
jets, but different injection rates (see Table~1). 
Furthermore, Figures~\ref{fig3} and~\ref{fig4} present plots with a similar format, 
arranged by columns (and we address them with the same short notation for the left
middle and right columns). Figure~\ref{fig3} shows details of the central region of all 
the plots in Figure~\ref{fig2} while Figure~\ref{fig4} compares runs with the same 
parameters except for the galaxy metal distribution. It is important to note that 
the total AGN energy input to the ICM is the same in all cases.

\begin{figure*}
  \vspace*{0pt}
  \includegraphics[bb=55 20 480 355, scale=0.34, clip=]{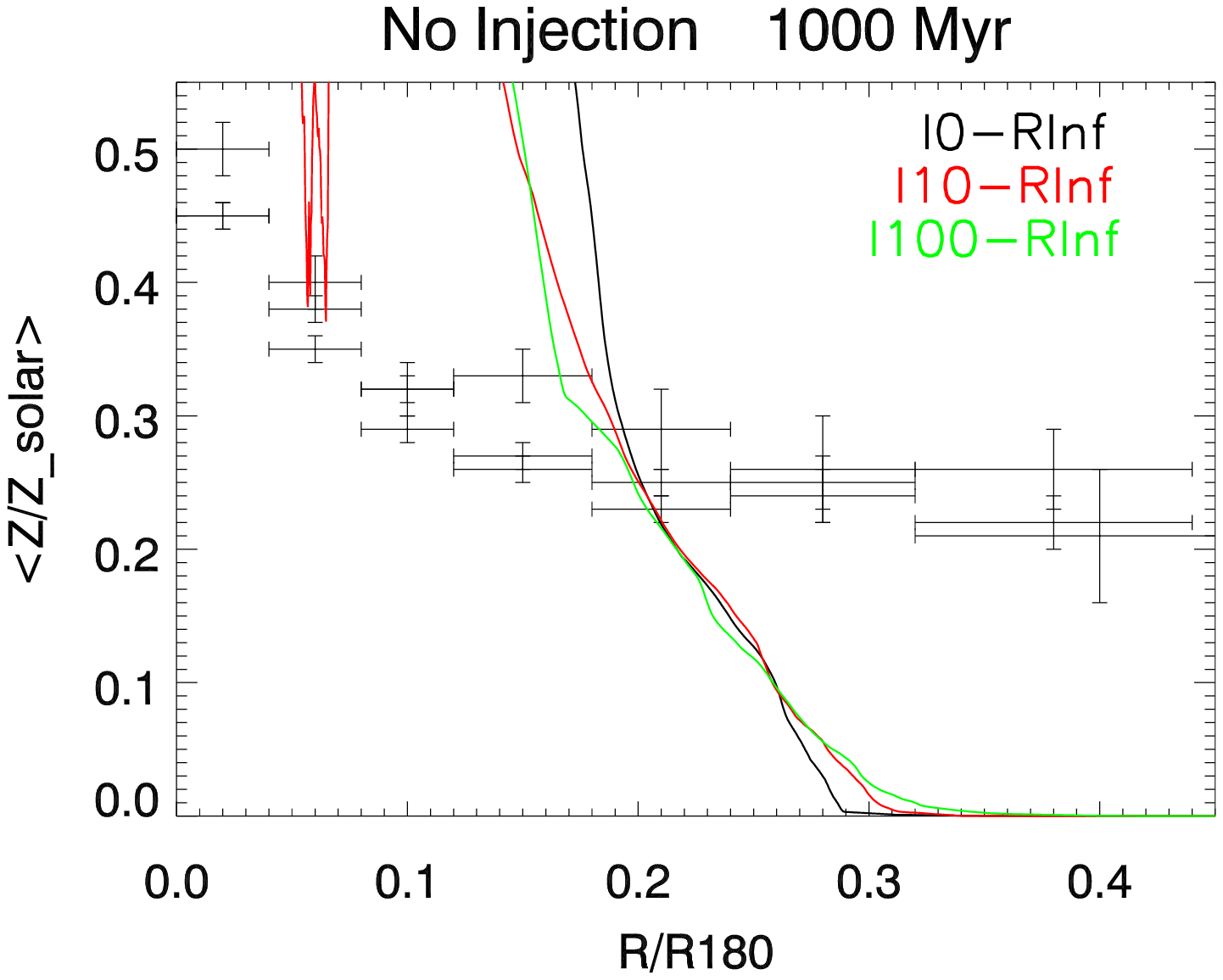} 
	\hspace*{\columnsep}%
  \includegraphics[bb=55 20 480 355, scale=0.34, clip=]{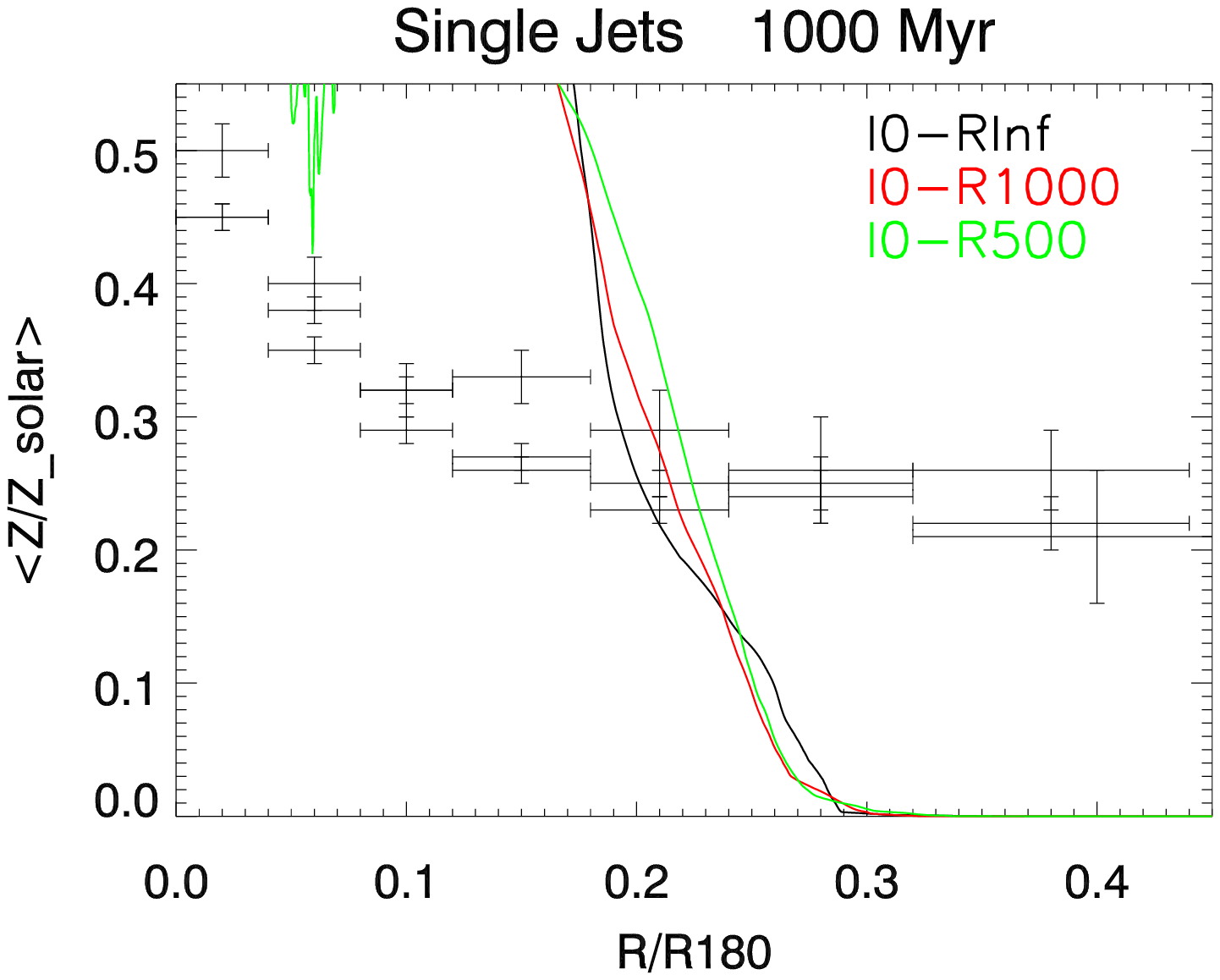} 
	\hspace*{\columnsep}%
  \includegraphics[bb=55 20 480 355, scale=0.34, clip=]{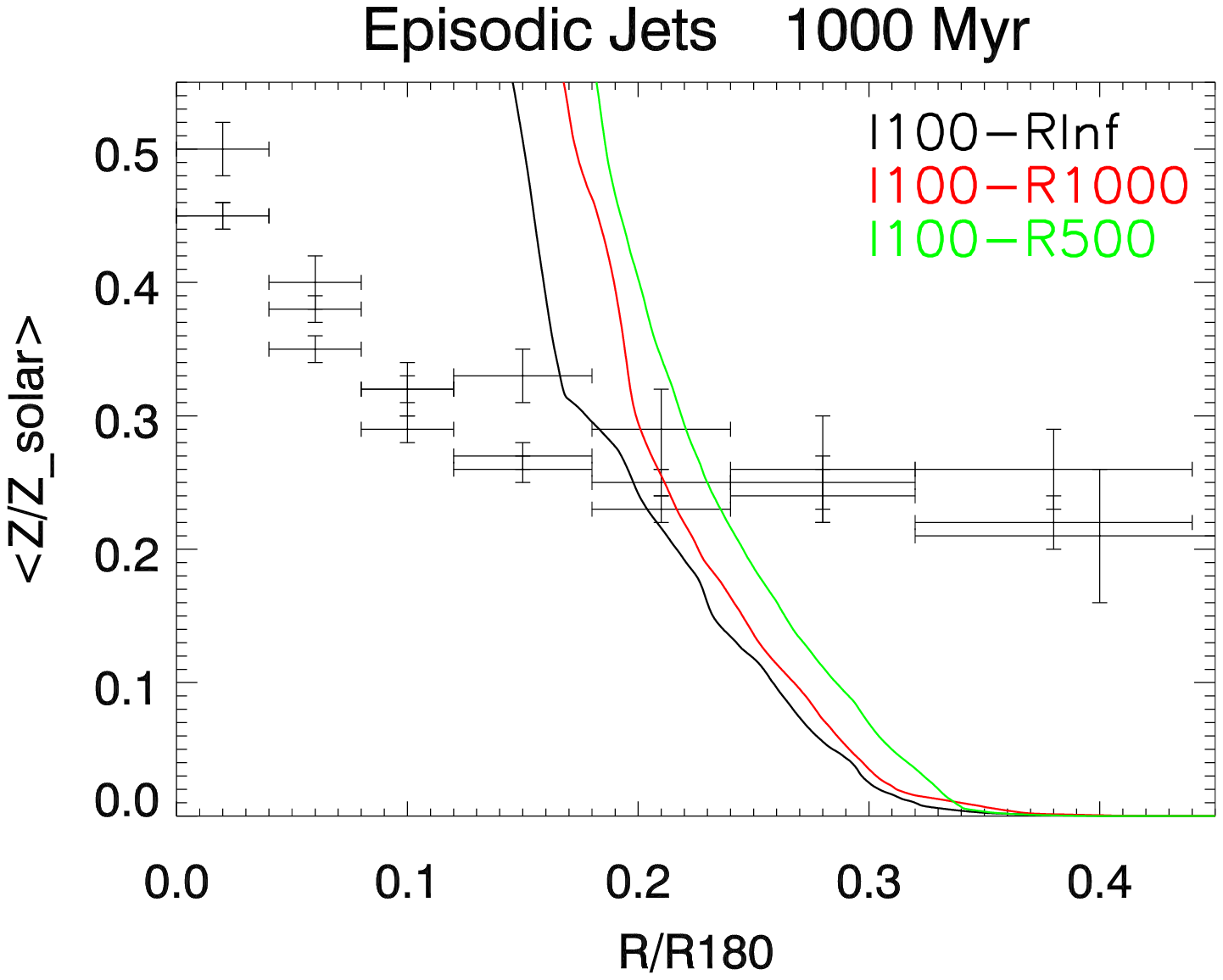} \\
  \includegraphics[bb=55 20 480 355, scale=0.34, clip=]{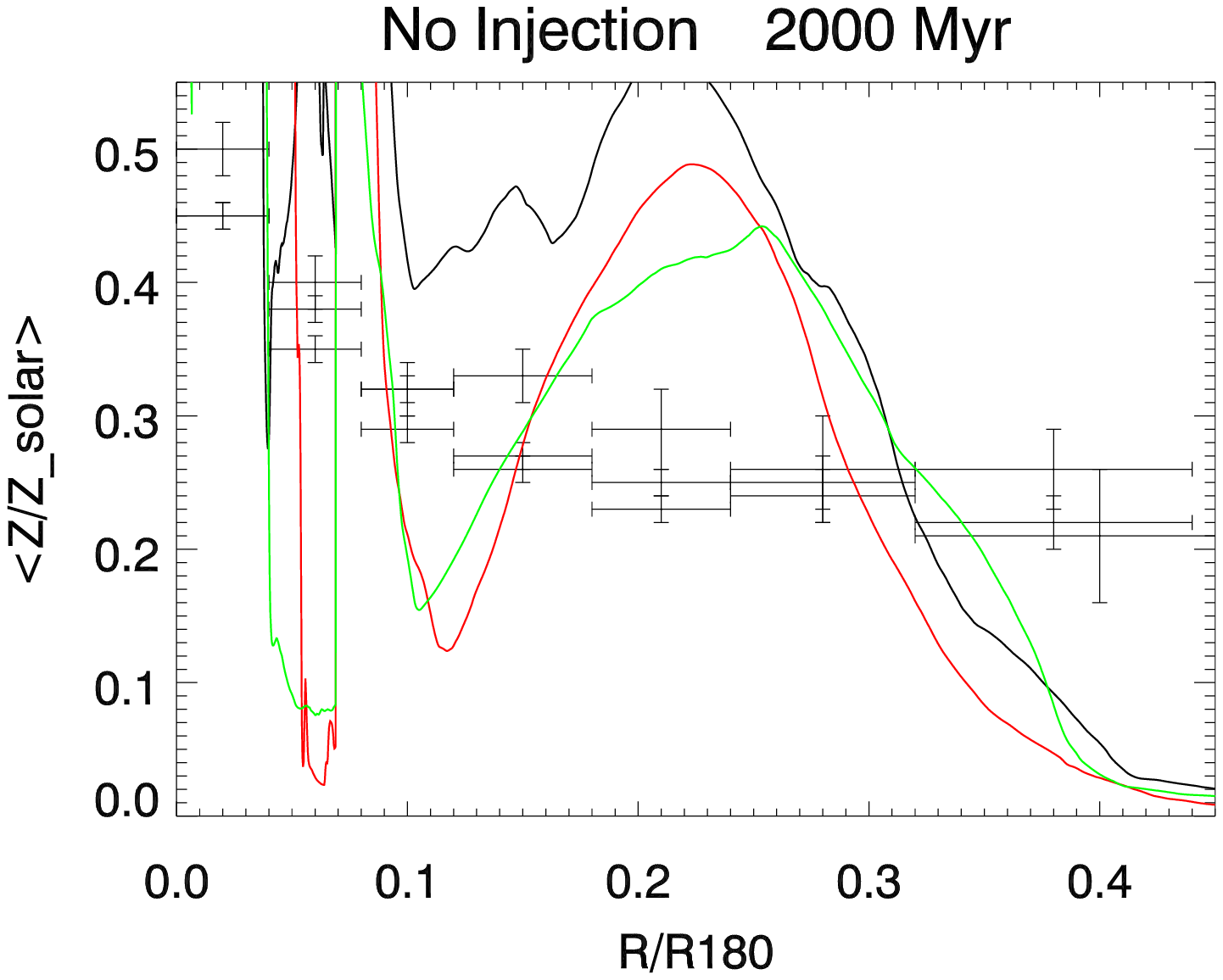} 
	\hspace*{\columnsep}%
  \includegraphics[bb=55 20 480 355, scale=0.34, clip=]{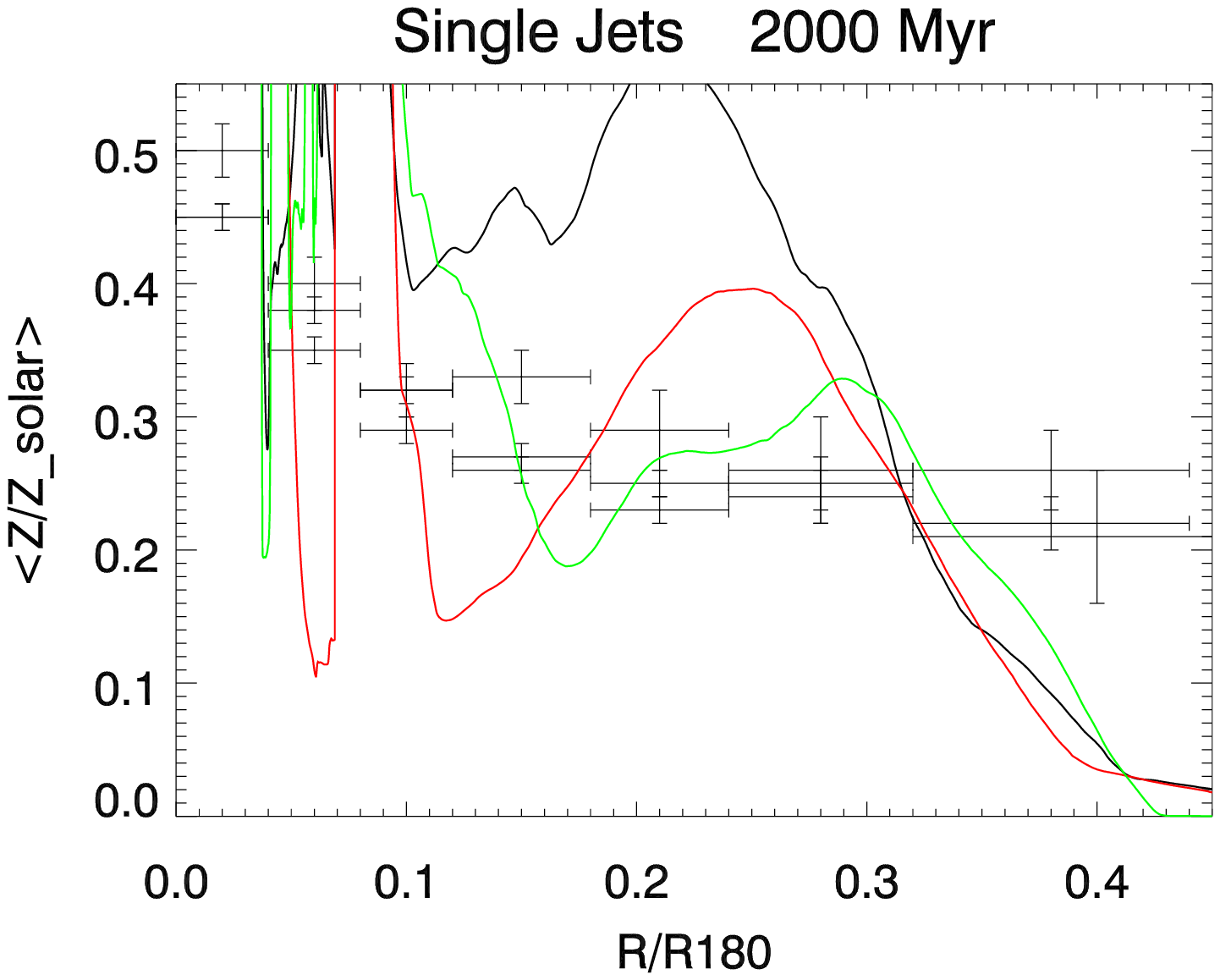} 
	\hspace*{\columnsep}%
  \includegraphics[bb=55 20 480 355, scale=0.34, clip=]{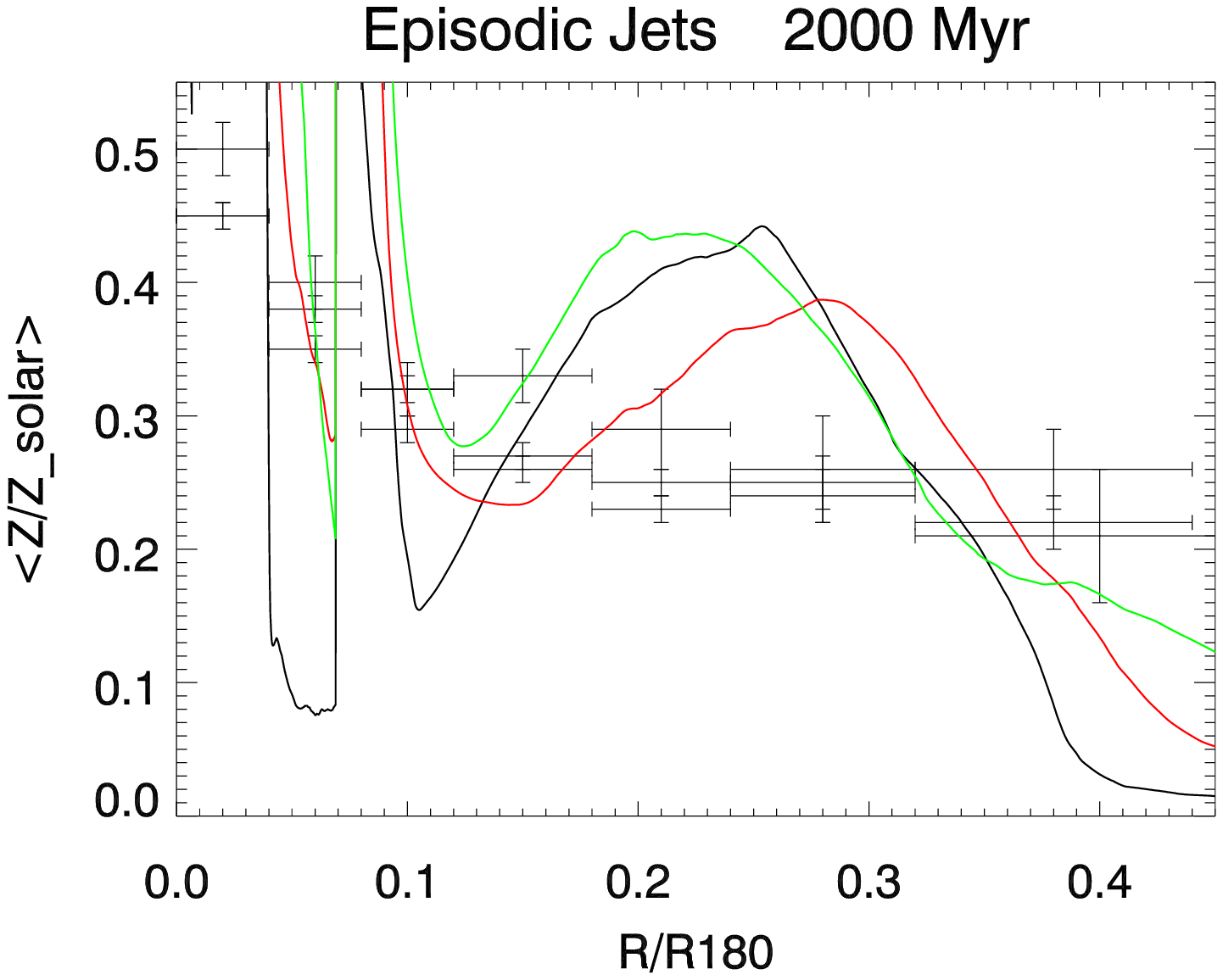} \\
  \includegraphics[bb=55 20 480 355, scale=0.34, clip=]{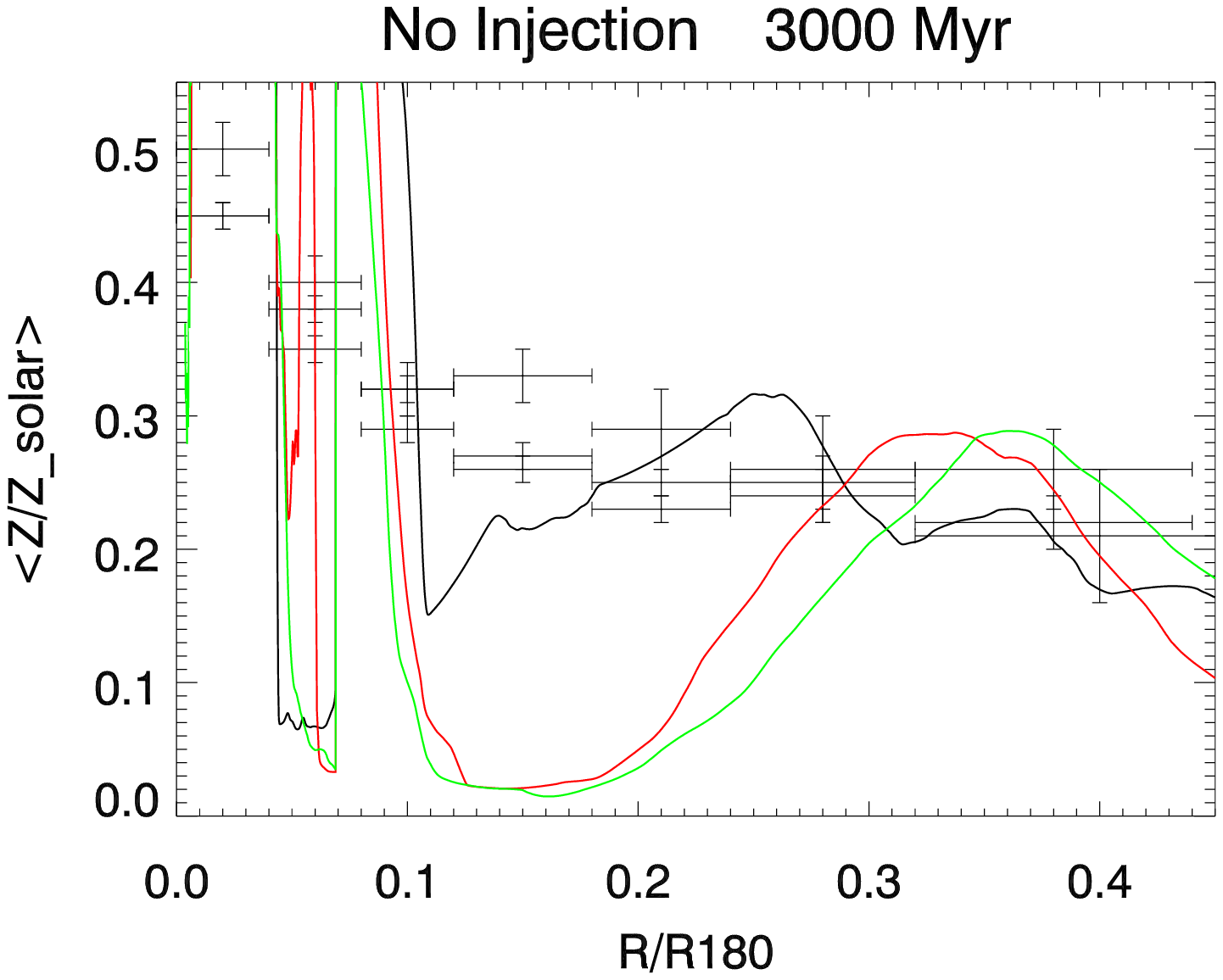} 
	\hspace*{\columnsep}%
  \includegraphics[bb=55 20 480 355, scale=0.34, clip=]{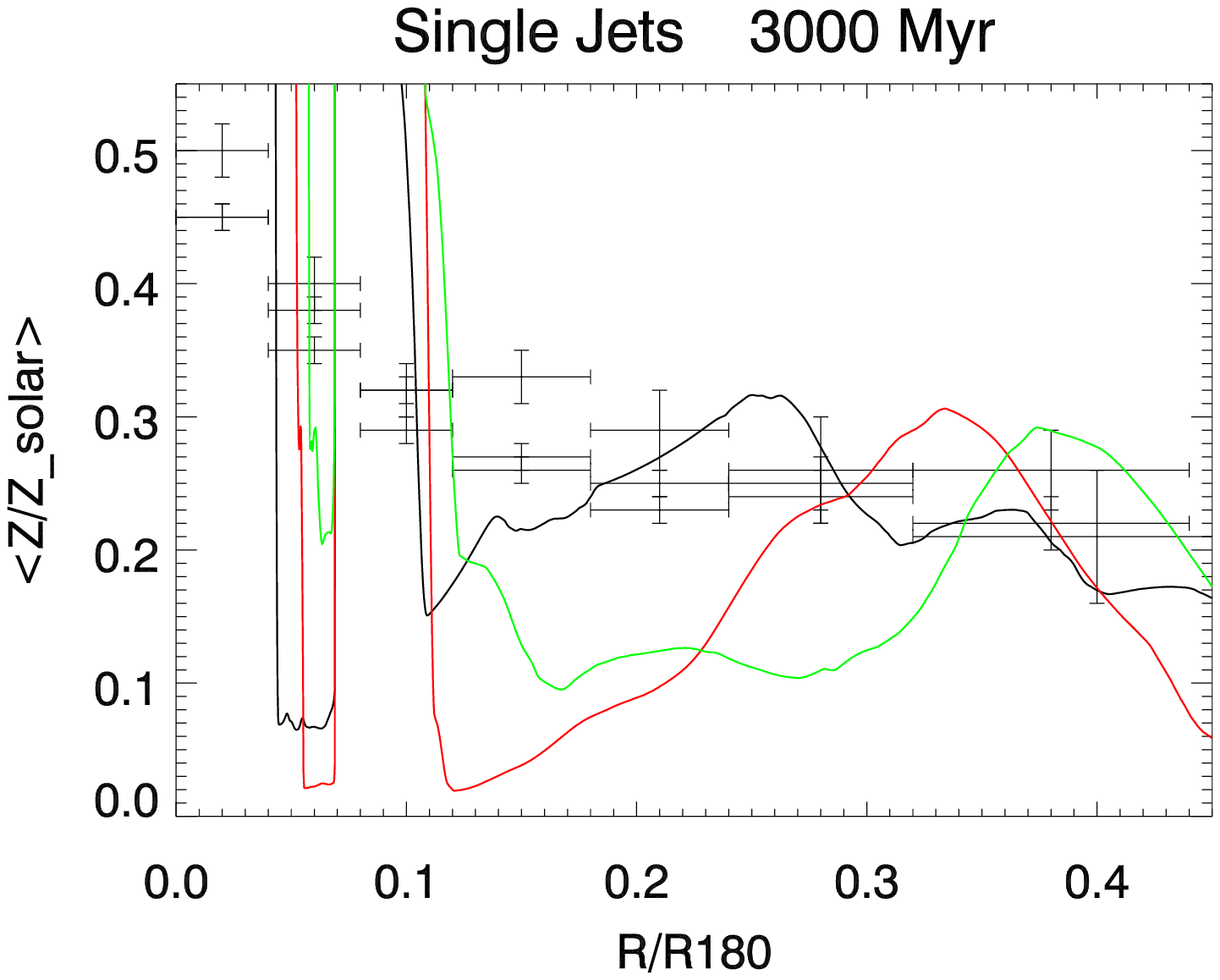} 
	\hspace*{\columnsep}%
  \includegraphics[bb=55 20 480 355, scale=0.34, clip=]{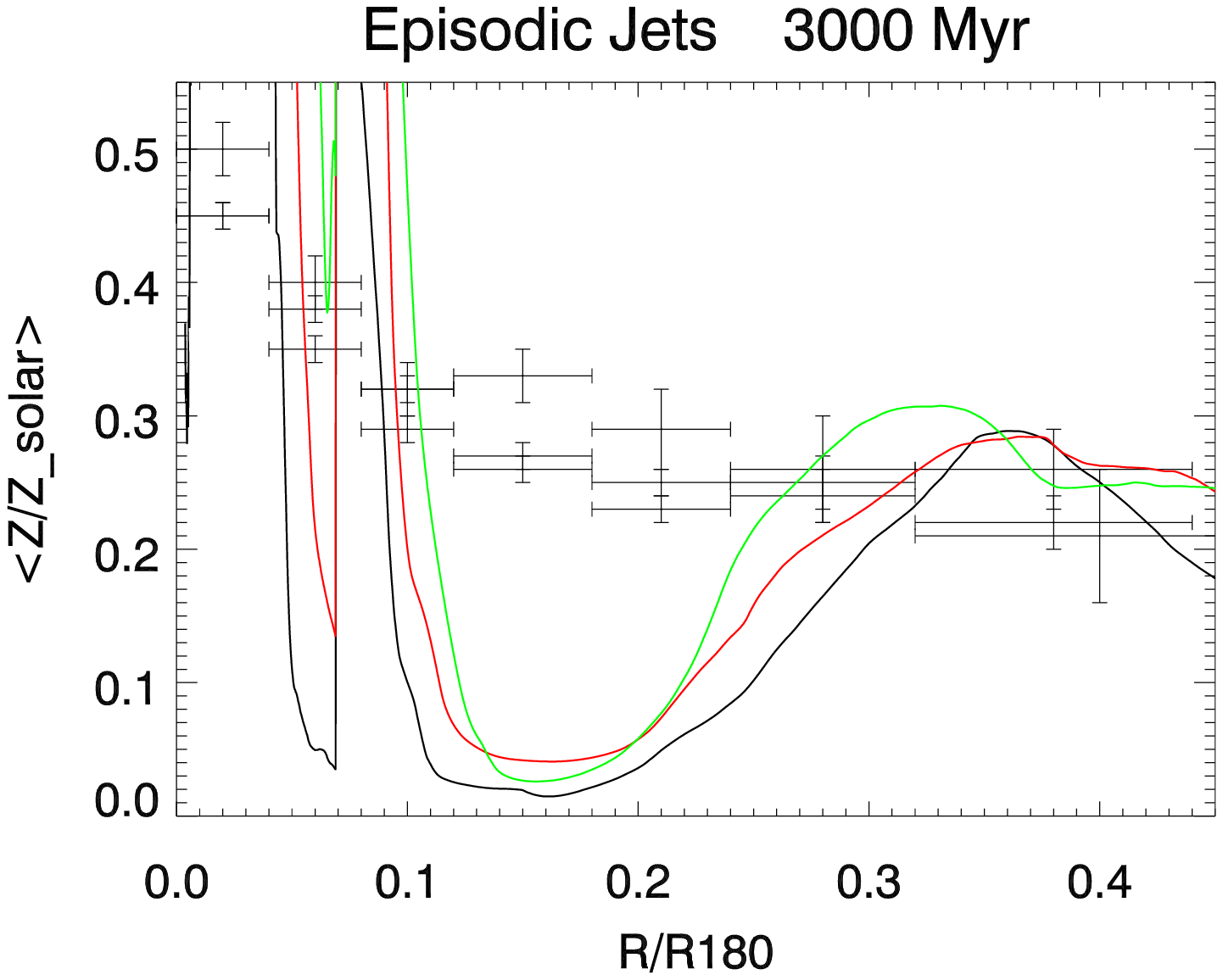} 
  \caption{General evolution of the mean relative metallicities of our simulations
overlaid on the ICM radial metallicity profiles from the data composition (Figure~5) by 
\citet{leccardi08}.}
\label{fig5}
\end{figure*}

Figure~\ref{fig2} shows that although powerful AGN jets are transitory phenomena, 
they produce important dynamical and chemical effects on the ICM on megaparsec scales 
for a few gigayears. The large-scale dynamical evolution of the ICM and its 
metals is similar for all our simulations (see Figures~\ref{fig2}, \ref{fig3} 
and~\ref{fig4}). The initial metal distribution of the central galaxy is rapidly
lost and the very steep and negative gradient of the metallicity rise slowly.
The central region, \hbox{r\,$\la$\,250\,kpc}, remains quite steep specially
if metal injection is present. From 2\,Gyr onwards regions with flat gradients
develop for \hbox{r\,$\ga$\,500\,kpc} and for some central regions as well 
(see Figures~\ref{fig3} and~\ref{fig5}). 

The number of jet episodes has an effect from about 2\,Gyr onwards, mainly within 
the regions \hbox{r\,$\in$\,(200,750)\,kpc} and 
\hbox{r\,$\in$\,(1500,2150)\,kpc}.
The metallicity profiles in Figure~\ref{fig2} show that after evolving for
3\,Gyr, the ICM metals advected by single outburst jets (single jets) 
reach \hbox{radii\,$\sim$\,1800\,kpc} from the cluster's core. However, if 
intermittent jets are present, metals \hbox{reach~$\sim$\,300\,kpc} further.
On the other hand, metallicity plots, more significantly from systems with intermittent jets, 
show reductions in metallicity in the region \hbox{r\,$\in$\,(200,750)\,kpc}
of an order of magnitude (see Figures~\ref{fig2} and~\ref{fig5}).
Similar depressions were found by \citet{heath07} in their simulations but it is difficult 
to find any traces of such metal distribution gradients with current 
X-ray spectroscopic observations of clusters. In our simulations, gas and 
metals are continuously injected into the central dominant galaxy (see 
equation~(\ref{update2})) and thus the AGN relic bubbles inflated by episodic jets 
carry with them more metals and gas than bubbles formed by single jets do. 
The ICM gas in the surroundings will occupy the space left 
by the bubbles when they rise, producing central inflows in the cluster. These inflows 
are larger for episodic jets.
The combined effects of these inflows (replenish central regions with metal-poor gas)
and the rising bubbles (moving hot gas and metals away form the core) 
result in an intermediate intermediate region in the cluster, with low metallicity. 
Furthermore, during the intermittent jets' active phases every outburst 
drives an expanding shock wave leaving turbulent flows 
in its wake. Shocks of secondary outbursts 
sweepout turbulent gas formed by previous outbursts. Thus, the hydrodynamical 
structures produced are more complex when intermittent jets are present, in agreement with
the radio observations of intermittent jets of \citet{brocksopp07}.
These physical conditions facilitate the mixing of gases with different metal concentrations, 
resulting in smoother metallicity profiles as shown in the last row of 
Figures~\ref{fig2}, \ref{fig3} and~\ref{fig5}. 

The metal injection has effects on the general metal abundances present in the ICM,
specially at the core (see~2M, 2R,~3M, 3R~5M and~5R).
The faster the metal replenishment time, t$_c$, the higher the metal abundance. 
Figure~\ref{fig3} shows that the metallicity profiles at radii\,$\la$75\,kpc are 
dominated by the central galaxy. 
On the other hand, an important metallicity drop is evident at the core of the cluster 
between~600\,Myr and~1\,Gyr, for runs with metal injection (see~2M, 2R,~3M and~3R). 
This loss and redistribution of metals takes place the moment the equatorial part of
the jets' cocoon collapses, and the formation of the buoyant bubbles of hot gas. 
In this process some small proportion of metals are lost in our simulations
through the inner computational boundary -- while others, are lifted to larger 
radii to form the bubbles.
However, by comparing Runs with and without injection in Figures~\ref{fig2} 
and~\ref{fig3}, from 1\,Gyr onwards, it seems that the rapid central metallicity 
changes mentioned above (between~600\,Myr and~1\,Gyr) have no important effects 
on the general metallicity evolution of the ICM.

The metal distribution details of the central galaxy (either flat or Plumer, see 
Section~\ref{conserv}) have no effect on the dynamical evolution of the ICM 
metallicity on large scales, as shown in Figure~\ref{fig4}. The discrepancies in 
4M at 3\,Gyr (and 2M as well) are only due to numerical advection.

\subsection{Comparison with Radio and X-Ray observations}
\label{data}

As discussed above, intermittent jets drive turbulence and mixing
in the ICM. Direct evidence of jets with three outburst episodes
was first found in the FR~II radio galaxy \hbox{B0925+420} by \citet{brocksopp07}. 
The ages of the lobes of this radio source (\hbox{$\sim$\,a~few\,Myr}) are comparable 
to the intermittancy timescales we use in our simulations. 
Furthermore, there is observational evidence of turbulence and diffusion in the ICM,
e.g., see \citealt{rebusco06} and references there in. However, a quantitative study of
the ``nested cocoons'' and the ICM turbulence and diffusion is beyond the scope of this
letter.

In Figure~\ref{fig5} we compare the mean metallicity profiles from the X-Ray 
data presented in \citet{leccardi08} and references there in, 
to our simulations at different
epochs. It is evident that for \hbox{radii\,$\la$\,0.12\,R$_{180}$} our simulations present 
pronounced metallicity gradients that do not follow the general trends present at larger 
radii. In our simulations this inner region is dominated by the metals of the central 
galaxy (see Figures~\ref{fig2} and~\ref{fig3}). 
We do not expect our simulations to reproduce the metallicity gradients on small scales,
where the profiles are likely to be shaped by
less-energetic phenomena
than AGN activity such as galactic winds.
Away from the core however some of our 
simulations develop metallicity gradients comparable, over 
radii from~0.1 \hbox{to~0.4\,R/R$_{180}$},
to those in the data (see Figure~\ref{fig5}). Table~2 summarises the radial ranges over 
which the mean metallicity of our simulations 
is in agreement with experimental data. 
The best fit is obtained with \hbox{Run~I0-RInf}, with a single jet outburst 
and no metal injection.

\begin{table}
\centering
 \begin{minipage}{\columnwidth}
  \caption{Ranges of agreement between our simulations and the observations from 
\citet{leccardi08}, at 3\,Gyr.}
  \begin{tabular}{@{}lc@{}}
  \hline
   Simulation 	&Approximate radial 	\\	
   name		&range [R$_{180}$]	\\	
 \hline
 I0-RInf 	  &(\,0.18\,,\,0.45\,)	\\ 
 I0-R1000    	  &(\,0.26\,,\,0.40\,)	\\ 
 I0-R500    	  &(\,0.32\,,\,0.45\,)	\\ 
 I10-RInf	  &(\,0.27\,,\,0.42\,)	\\ 
 I100-RInf	  &(\,0.31\,,\,0.45\,)	\\ 
 I100-R1000 	  &(\,0.29\,,\,0.45\,)	\\ 
 I100-R500 	  &(\,0.25\,,\,0.45\,)	\\ 
\hline
\end{tabular}
\end{minipage}
\label{table2}
\end{table}

\section{Conclusions}
\label{conclu}

We present 2D axisymmetric hydrodynamical simulations of the metal advection in the ICM,
driven by the jets of a radio-loud AGN. 
We implement the ICM with a cooling flow, and the central cD~galaxy 
with a metal distribution and continuous metal formation, and powerful intermittent jets.
The jets produce the early and late features typical of canonical models, simulations
and observations of FR~II objects (e.g., see
\citealt{churazov01,alexander06,heath07,brocksopp07}). 
We model the metals using tracer fields and explore the consequences of varying the 
jet intermittency, the galaxy metal distribution profile and the metal formation rates.
We have found the following.

Although powerful AGN jets are transitory phenomena they produce important 
dynamical and chemical effects on the ICM on megaparsec~scales for a few gigayears.

The convective flows formed long after the AGN active phases drag $\sim$95\% of the 
initial metals in the central galaxy to distances \hbox{$\ga$\,1.5\,Mpc}.

We implement intermittent jets with the same total energy input to the ICM.
The general metallicity evolution for all our simulations is very similar on 
large-scales: \hbox{after~2\,Gyr}, approximately flat metallicity gradients are
formed from \hbox{radii~$\sim$\,700} \hbox{to~$\sim$\,1300\,kpc}. Metallicity
drops are formed (found by \citet{heath07} as well) of factors \hbox{of~$\sim$\,15} 
at \hbox{r\,$\in$\,(200,750)\,kpc}. 
These features might be correlated to the mass of the gas and metal advected by 
the relic bubbles, but high resolution X-ray observations are needed to see
traces of such metallicity gradients. 
Furthermore, intermittent jets distribute metals $\sim$\,300\,kpc further than single
outburst jets do, and also facilitate the mixing 
of gases with different metal concentrations and induce turbulence in the ICM.

The metal injection has effects on the general metal abundances present in the ICM:
the faster the metal replenishment time the higher the metal abundance.
For \hbox{radii\,$\la$\,0.12\,R$_{180}$} 
our simulations are dominated by the metals of the central galaxy and do not 
follow the general trends present at larger radii on either our simulations or
the X-Ray observations. Since metal production and jets are realistic in active 
galaxies, metalliciy gradients in the very central regions of cool core clusters
are likely to be shaped by less-energetic phenomena. On the other hand, away from 
the core, some of our simulations (specially the one with a single jet outburst and no 
metal injection) develop metallicity gradients in agreement with, over particular radial ranges, 
the observations of \citet{leccardi08} and references there in.

We model massive central galaxies with the same total metal mass in all cases.
The metal distribution details of the central galaxy have no effect on the dynamical 
evolution of the ICM metals.

\section*{Acknowledgements}

The software used in these investigations was in part developed by 
the DOE-supported ASC / Alliance Center for Astrophysical Thermonuclear 
Flashes at the University of Chicago. MHE acknowledges support from
CONACyT (196898/217314).




\bsp

\label{lastpage}
 
\end{document}